\documentclass[a4paper,11pt]{article}

\usepackage{amsmath}
\usepackage{amssymb}
\usepackage[nosort]{cite}
\usepackage{graphicx}   
\usepackage{color} 
\usepackage{bbm}
\usepackage[margin=2cm]{caption}
\usepackage[latin1]{inputenc} 
\usepackage[T1]{fontenc}
\usepackage[bookmarks=true,hyperfigures=true]{hyperref}
 \usepackage[a4paper,text={470pt,700pt},centering]{geometry}

\makeatletter
\let\old@makecaption=\@makecaption
\def\@makecaption{\small\old@makecaption}
\makeatother

\makeatletter
\newlength{\apb@width}
\newcommand{\autoparbox}[2][c]{\settowidth{\apb@width}{#2}\parbox[#1]{\apb@width}{#2}}
\newcommand{\includegraphicsbox}[2][]{\autoparbox{\includegraphics[#1]{#2}}}
\makeatother

\providecommand{\hypersetup}[1]{}

\providecommand{\pdfbookmark}[3][]{}
\hypersetup{plainpages=false}
\hypersetup{pdfpagemode=UseOutlines}
\hypersetup{bookmarksnumbered=true}
\hypersetup{bookmarksopen=true}
\hypersetup{pdfstartview=FitH}
\hypersetup{colorlinks=false}
\hypersetup{citebordercolor={.5 1 .5}}
\hypersetup{urlbordercolor={.5 1 1}}
\hypersetup{linkbordercolor={1 .7 .7}}
\hypersetup{citebordercolor={.6 .9 .6}}
\hypersetup{urlbordercolor={.7 .8 1}}
\hypersetup{linkbordercolor={1 .7 .7}}
\hypersetup{pdfborder={0 0 1 [3]}}

\numberwithin{equation}{section}

\newcommand{\unit}{\mathbbm{1}}

\newcommand{\ket}[1]{\mathopen{|}#1\mathclose{\rangle}}
\newcommand{\bra}[1]{\mathopen{\langle}#1\mathclose{|}}
\newcommand{\braket}[2]{\mathopen{\langle}#1|#2\mathclose{\rangle}}
\newcommand{\Tr}{\text{Tr}}

\ifx\genfrac\sdflkaj

\else

\fi

\newcommand{\nn}{\nonumber}

\makeatletter
\def\mr@ignsp#1 {\ifx\:#1\@empty\else #1\expandafter\mr@ignsp\fi}%
\newcommand{\multiref}[1]{\begingroup
\xdef\mr@no@sparg{\expandafter\mr@ignsp#1 \: }%
\def\mr@comma{}%
\@for\mr@refs:=\mr@no@sparg\do{\mr@comma\def\mr@comma{,}\ref{\mr@refs}}%
\endgroup}
\makeatother
\newcommand{\hypref}[2]{\ifx\href\asklfhas #2\else\href{#1}{#2}\fi}

\newcommand{\secref}[1]{Sec.~\multiref{#1}}

\newcommand{\appref}[1]{App.~\multiref{#1}}

\newcommand{\tabref}[1]{Tab.~\multiref{#1}}

\newcommand{\figref}[1]{Fig.~\multiref{#1}}
\renewcommand{\eqref}[1]{(\multiref{#1})}

\def\[{\begin{equation}}
\def\]{\end{equation}}
\def\<{\begin{eqnarray}}
\def\>{\end{eqnarray}}
\newcommand{\beq}{\begin{equation}}
\newcommand{\eeq}{\end{equation}}
\newcommand{\beqa}{\begin{eqnarray}}
\newcommand{\eeqa}{\end{eqnarray}}

\definecolor{purple}{rgb}{0.5,0,0.5}

\usepackage{enumitem}

\begin{document}
\thispagestyle{empty}

\begin{flushright}\footnotesize
	\texttt{SAGEX-22-22\\
		TCDMATH 22-01
	}
\end{flushright}
\vspace{1cm}

\begin{center}%
	{\Large\textbf{
		Chaotic spin chains in AdS/CFT}\par}%
	
	\vspace{15mm}
	
	\textrm{
		Tristan McLoughlin$^*$ and Anne Spiering$^\dagger$ }\vspace{8mm} \\
	\textit{$*$
		School of Mathematics \& Hamilton Mathematics Institute \\
		Trinity College Dublin, Ireland} \\
		\texttt{\small tristan@maths.tcd.ie\phantom{\ldots}} \\ \vspace{4mm}
		
		\textit{$\dagger$ Niels Bohr Institute, 
		 University of Copenhagen
		 \\ Blegdamsvej 17, 2100 Copenhagen \O, Denmark
	} \\
	\texttt{\small anne.spiering@nbi.ku.dk\phantom{\ldots}} \\ \vspace{4mm}

	\par\vspace{14mm}
	
	\textbf{Abstract} \vspace{5mm}
	
	\begin{minipage}{14cm}
		\parindent=2em
		We consider the spectrum of anomalous dimensions in planar $\mathcal{N}=4$ supersymmetric Yang-Mills theory and its $\mathcal{N}=1$ super-conformal Leigh-Strassler deformations. 
				The two-loop truncation of the integrable $\mathcal{N}=4$ dilatation operator in the SU$(2)$ sector, which is a next-to-nearest-neighbour deformation of the XXX spin chain, is not strictly integrable at finite coupling and we show that it indeed has Wigner-Dyson level statistics.  However, we find that it is only weakly chaotic in the sense that the cross-over to chaotic dynamics is slower than for generic chaotic systems.

			\parindent=2em
	For the Leigh-Strassler deformed theory with generic parameters, we show that the one-loop dilatation operator in the SU$(3)$ sector is chaotic, with a spectrum that is well described by GUE Random Matrix Theory. For the imaginary-$\beta$ deformation, the statistics are GOE and the transition from the integrable limit is that of a generic system. This provides a weak-coupling analogue of the chaotic dynamics seen for classical strings in the dual background. 
			
	\parindent=2em
		We further study the spin chains in the semi-classical limit described by generalised Landau-Lifshitz models, which are also known to describe large-angular-momentum string solutions in the dual theory.  We show that for the higher-derivative theory following from the two-loop $\mathcal{N}=4$ SU$(2)$ spin chain, the maximal Lyapunov exponent is close to zero, consistent with the absence of chaotic dynamics. For the imaginary-$\beta$ SU$(3)$ theory, the resulting Landau-Lifshitz model has classically chaotic dynamics at finite values of the deformation parameter. 
	\end{minipage}
	
\end{center}

\newpage
\tableofcontents

\vspace{10mm}
\hrule
\vspace{5mm}

\section{Introduction}
The realisation that in planar $\mathcal{N}=4$ supersymmetric Yang-Mills (SYM) theory, the dilatation operator acting on single-trace operators can be mapped to a long-range integrable spin chain \cite{Minahan:2002ve, Beisert:2003tq, Beisert:2003yb}, has led to an impressive range of non-perturbative results, see \cite{Beisert:2010jr, Arutyunov:2009ga, Bombardelli_2016, Gromov:2017blm} for reviews. These results make fundamental use of the holographic duality \cite{Maldacena:1997re} which relates the anomalous dimensions of single-trace operators to the energies of non-interacting closed strings in an AdS$_5\times$S$^5$ background. While many of these results have been extended to other theories, such as to deformations of $\mathcal{N}=4$ SYM or to theories which share the feature of being highly supersymmetric like ABJM theory \cite{Aharony:2008ug}, such theories are very special and in general one will not have at hand the tools provided by integrability. 

For classical strings moving in more general geometries there is in fact significant evidence of chaotic dynamics. 
By computing the power spectrum, Lyapunov characteristic exponents and Poincar\'e sections, in \cite{PandoZayas:2010xpn} it was shown  that the motion of classical circular strings in the AdS$_5$-Schwarzschild background is chaotic.
This was subsequently extended to a wide range of backgrounds such as the AdS soliton background \cite{Basu:2011dg},  AdS$_5 \times$T$^{1,1}$ \cite{Basu:2011di}, other AdS$_5 \times$ Einstein$_5$ spaces \cite{Basu:2011fw} and p-brane backgrounds \cite{Stepanchuk:2012xi, Chervonyi:2013eja, Asano:2016qsv}, see also \cite{Basu:2012ae, Giataganas:2014hma, Asano:2015qwa, Hashimoto:2016wme, Basu:2016zkr, Ishii:2016rlk, Giataganas:2017guj, Roychowdhury:2017vdo, deBoer:2017xdk, Nunez:2018ags, Filippas:2019ihy, Rigatos:2020igd, Hashimoto:2018fkb, Nunez:2018qcj, Akutagawa:2019awh, Ma:2019ewq, Filippas:2019bht, Pal:2021fzb, Roychowdhury:2021jqt, Hashimoto:2016dfz,Cubrovic:2019qee, Giataganas:2021ghs}. Moreover, the string motion in the gravity dual of $\beta$-deformed $\mathcal{N}=4$ SYM theory was shown to be chaotic for imaginary values of $\beta$ \cite{Giataganas:2013dha}. In this work we provide a weak-coupling analogue of this classical string chaos by considering the spectral statistics of the perturbative $\beta$-deformed dilatation operator in the dual gauge theory.

On the gauge-theory side, chaotic behaviour in classical Yang-Mills theory has been long known, see \cite{Matinyan:1981dj, Matinyan:1986hc,Biro:1994bi} for reviews of early work. It was shown that upon restricting to homogeneous field configurations or by dimensionally reducing  on $S^3$, the resulting Hamiltonian has non-integrable dynamics. This can be related to the computation of anomalous dimensions as, by using a conformal transformation, the dilatation generator of a conformal theory defined on $\mathbb{R}^{1,3}$ becomes the Hamiltonian for the theory on $\mathbb{R}\times S^3$. In the case of $\mathcal{N}=4$ SYM, reducing the theory on $S^3$ produces the BMN matrix theory \cite{Berenstein:2002jq} which is known to have classically chaotic dynamics \cite{Asano:2015eha}. More generally, Yang-Mills is a theory of many interacting degrees of freedom and so we expect it to demonstrate the property of quantum many-body chaos. Spectral statistics can provide a key signature of quantum chaos in many-body systems, see \cite{zelevinsky1996nuclear, DAlessio:2015qtq} for reviews. In particular, the BGS conjecture \cite{Bohigas:1983er} states that quantum-chaotic systems have spectra whose fluctuations are described by Random Matrix Theory (RMT). 
An analysis of the spectrum of the planar   
dilatation operator of $\mathcal{N}=4$ SU$(N)$ SYM theory expanded in the 't Hooft coupling $\lambda$ was performed in \cite{McLoughlin:2020siu} and \cite{McLoughlin:2020zew}, and it was shown that, for rank-one sectors of the one-loop planar theory, the spacing between adjacent levels is described by the Poisson distribution as is characteristic of integrable systems \cite{berry1977level}.
However, when non-planar corrections are included, the spectrum becomes chaotic and the level spacings closely follow the Wigner-Dyson distribution for the Gaussian Orthogonal Ensemble (GOE). Additional analysis of spectral observables and multi-trace eigenstates provided further evidence that, at finite values of $N$, the theory is described by GOE RMT. For operators whose dimensions scale as $N^2$, and so are sensitive to non-planar effects, the theory has also been shown to exhibit fast scrambling \cite{deMelloKoch:2020jmf}.

In this work, we examine a different set of questions and consider quantum-chaotic behaviour that also occurs at weak coupling but in the planar limit. First we consider the two-loop planar dilatation operator of $\mathcal{N}=4$ SYM theory restricted to a closed SU$(2)$ sector. The theory is explicitly integrable at one-loop order, where it corresponds to the $XXX_{1/2}$-Heisenberg spin chain, and is believed to be integrable also at finite coupling. At loop order $k$, the spin chain  has interactions involving $k+1$ adjacent spins and so the all-order chain has infinite interaction range. When truncated to just the two-loop terms involving next-to-nearest neighbour interactions, the spin chain is only perturbatively integrable and higher charges are only conserved to leading order in the coupling. We investigate the level statistics of this two-loop spin chain at finite coupling and find that it is indeed described by GOE RMT.
The two-loop deformation is nonetheless non-generic and only gives rise to a weakly chaotic system in the sense that the transition between the integrable and chaotic regimes is slower and only occurs for larger values of the deformation than in the generic, strongly chaotic case. 
This is consistent with the work \cite{Szasz-Schagrin:2021pqg} which showed that deformations of the XXZ spin chain constructed via the boost method of \cite{Bargheer:2008jt, Bargheer:2009xy} are weakly chaotic and the fact that the two-loop dilatation operator is similarly constructed. An interesting corollary of this analysis is that the SU$(2)$ preserving, next-to-nearest deformation of the XXX spin chain is only weakly chaotic. We find similar results for twisted versions of the theory which describe a rank-one sector of the integrable $\beta$-deformed theory.

To understand quantum chaos in more generic planar theories, we consider the  one-loop dilatation operator in the SU$(3)$ sector of the marginally deformed $\mathcal{N}=4$ SYM theory preserving $\mathcal{N}=1$ supersymmetry \cite{Leigh:1995ep}. This theory is parametrised by two additional complex parameters, $h$ and $q$. There are specific values for which the theory is integrable \cite{Bundzik:2005zg}, see also \cite{Roiban:2003dw, Berenstein:2004ys, Freyhult:2005ws, Beisert:2005if}, and where the level statistics are Poisson. In general this is not the case and we show that for generic values the spectrum of the theory is described by Gaussian Unitary Ensemble (GUE) RMT. In the case $h=0$ and $q\in \mathbb{R}$, which corresponds to imaginary values of $\beta$ as studied in \cite{Giataganas:2013dha}, it is again GOE. 
By examining the transition from integrable points to the chaotic regime, we see that these deformations are strongly chaotic, thus suggesting a weak-coupling analogue of classical string chaos. Interestingly, quantum chaos with GOE statistics has previously been seen for strings in space-times dual to confining theories \cite{PandoZayas:2012ig}.

To make the connection between spin-chain quantum chaos and classical string chaos more manifest, we analyse the large-length limit of the spin chain. In this limit the low-energy states of the Heisenberg spin chain are known to have a semi-classical description  in terms of a two-dimensional Landau-Lifshitz  (LL) type sigma model \cite{Fradkin:1991nr}. The same sigma model action can be found from a large angular momentum, or fast-string, limit of the AdS$_5 \times$S$^5$ world-sheet theory \cite{Kruczenski:2003gt} which thus provides a bridge between the two sides of the duality. This identification has been extended to larger sectors \cite{Hernandez:2004uw, Stefanski:2004cw,Kruczenski:2004cn, Hernandez:2004kr, Stefanski:2005tr}, including a PSU$(2,2|4)$ LL model describing the complete one-loop ${\cal N}=4$ SYM dilatation generator \cite{Stefanski:2007dp}. The semi-classical limit of the higher-loop, longer-range Hamiltonians gives, after integrating out the short wavelength modes, a generalised LL action with higher-derivative interactions. The two-loop effective LL action with four-derivative terms was found from the spin chain \cite{Kruczenski:2004kw} and matches the corresponding limit of the string theory. However, in general the LL action following from the spin chain and string theory disagree. The allowed terms in a generalised LL action can be combined with arbitrary coefficients which then parametrise the space of possible theories. The agreement at one- and two-loops between the gauge-theory LL model valid at weak coupling and the string LL model valid at strong coupling is presumably a result of a non-renormalisation theorem which fails for the $\lambda^3$, six-derivative terms \cite{Ryzhov:2004nz, Tseytlin:2004xa,Minahan:2005qj, Minahan:2005mx, Tirziu:2006ve}. Nonetheless, even where the two models disagree, having a common sigma-model description is extremely useful for understanding the duality. We show that while the two-loop LL model with higher derivative interactions has more erratic behaviour, the dynamics do not appear to be strongly chaotic. In particular, we compute the Lyapunov characteristic exponents for a specific family of solutions, and show that they are not clearly larger than zero. This is consistent with the weak chaotic behaviour of the spin chain and the classical integrability of the string theory. 

LL models for deformed backgrounds have also been studied and in \cite{Frolov:2005ty} the LL model corresponding to the $\beta$-deformed theory, i.e.\ the marginally deformed theory with $h=0$ and $q=\rm{exp}(i\beta)$ with $\beta\in \mathbb{R}$, was considered. In the gauge theory this corresponds to considering the long-wavelength limit of the twisted spin chain, while on the string side it describes fast strings moving in the dual geometry constructed in  \cite{Lunin:2005jy} by using the TsT-transformation. In this case the classical string theory is integrable \cite{Frolov:2005dj} and there are all-order twisted Bethe equations for the spectrum \cite{Beisert:2005if}, see also \cite{Gromov:2010dy, deLeeuw:2012hp, Kazakov:2015efa,Zoubos:2010kh, vanTongeren:2013gva} for reviews. The more general dual background for $q=\rm{exp}(\kappa+i \beta)$ was also constructed in \cite{Lunin:2005jy} by taking additional S-duality transformations. In this case the geometry is not an exact string solution and there are likely $\alpha'$ corrections. Moreover, the string theory is no longer integrable and as mentioned has been shown explicitly to be chaotic \cite{Giataganas:2013dha}. One can still take the fast string limit of the world-sheet theory and study the long-wavelength limit of the one-loop dilatation operator. This was done in \cite{Frolov:2005iq} for a holomorphic SU$(3)$ sector where, after a careful accounting of the relevant BPS states, the same deformed LL model was found in both cases. In this work we study the dynamics of this deformed LL model and show, by computing the Lyapunov exponents for specific solutions, that in case of real $q=\rm{exp}(\kappa)$ it has classically chaotic dynamics. This demonstrates that there is classical chaos in the semi-classical limit of the spin chain consistent with our quantum spin-chain results and also in agreement with the classical chaos found in the string theory. We further study the transition from integrable to chaotic dynamics as $\kappa$ increases from zero. We find that, for all the configurations we consider, there are finite values of $\kappa$ for which the largest Lyapunov exponent remains zero. Such behaviour is familiar in classically chaotic systems but is surprising in the quantum theory where we expect a sharp transition to chaotic dynamics in the infinite volume limit and is likely due to the fact that the limit considers only the low-energy states.

\section{Weakly chaotic SU(2) spin chains}
In $\mathcal{N}=4$ SYM theory, the planar dilatation operator acting on single-trace gauge-invariant operators composed of two types of complex scalars, $X$ and $Z$, can be written as an SU$(2)$ invariant spin-chain Hamiltonian. The map between gauge-theory operators and spin-chain states is
\<
\Tr(ZXZZX\dots )\quad\widehat{=}\quad \ket{\uparrow\downarrow\uparrow \uparrow\downarrow \dots}
\>
with each $Z$ replaced by $\ket{\uparrow}$, each $X$ by $\ket{\downarrow}$, and where the cyclicity of the trace must be imposed as an additional condition on the spin-chain state.  This corresponds to restricting to states that are invariant under cyclic shifts of the lattice sites or, equivalently, that have zero lattice momentum. 
The perturbative dilatation operator acting on such operators at the origin can be computed as an expansion in the rescaled 't Hooft coupling $g^2$,
\<
\mathfrak{D}=\sum_{k=0}^{\infty} g^{2k}\mathfrak D^{(2k)}~~~\text{with}~~~~g^2=\frac{\lambda}{16\pi^2}~.
\>
In the planar limit, the spectrum of the dilatation operator can be reproduced by a spin-chain Hamiltonian with periodic boundary conditions which is written as a sum of terms each acting on contiguous sets of spin-chain sites
\<
\mathfrak{D}^{(2k)}=\sum_{i=1}^L \mathfrak{D}_i^{(2k)}~,
\>
where $L$ corresponds to the length of the spin chain. 
The leading term, $\mathfrak{D}^{(0)}_i=\unit_{i}$, is simply the identity operator and counts the number of fields/spins and gives the classical scaling dimension. The one-loop and two-loop terms \cite{Minahan:2002ve,Beisert:2003tq} are 
\<
\mathfrak{D}^{(2)}_i=2(\unit_{i,i+1}-\mathbb{P}_{i,i+1})~,
~~\mathfrak{D}^{(4)}_i=-8\unit_{i,i+1}+12\mathbb{P}_{i,i+1}-2(\mathbb{P}_{i,i+1}
\mathbb{P}_{i+1,i+2}+\mathbb{P}_{i+1,i+2}\mathbb{P}_{i,i+1})~,
\>
where $\mathbb{P}_{a,b}$ is the permutation operator acting on the $a$-th and $b$-th sites. Writing the permutation operator in terms of the usual spin operators, $\mathbf{S}=\tfrac{1}{2}\vec{\sigma}$,
\<
\mathbb{P}_{a,b}=\tfrac{1}{2}(\unit_{a,b}+4 \mathbf{S}_a\cdot \mathbf{S}_b)~,
\>
we can write the one- and two-loop terms as a special case of the Heisenberg spin-chain Hamiltonian  with next-to-nearest-neighbour (NNN) corrections
\<
	\label{eq:NNNHam}
H=\epsilon_0 \sum_{i=1}^L  \unit_{i,i+1}+\lambda_1 \sum_{i=1}^L \mathbf{S}_i\cdot \mathbf{S}_{i+1} +\lambda_2 \sum_{i=1}^L \mathbf{S}_i\cdot \mathbf{S}_{i+2}~,
\>
 with the choice
\<
\label{eq:2looppar}
\lambda_1=-4\left(1-4g^2\right)~,~~~\lambda_2=-4g^2~,~~~\epsilon_0 =-\frac{1}{4}(\lambda_1+\lambda_2)=1-3g^2~.
\>
It is believed, and there is extensive evidence, that planar $\mathcal{N}=4$ SYM theory is integrable at all orders in $\lambda$ (see \cite{Beisert:2010jr,Arutyunov:2009ga,Bombardelli_2016} for reviews). Specifically, the perturbative $L\to \infty$ spin-chain spectrum can be reproduced by asymptotic Bethe equations which have a finite radius of convergence, $|g|<1/4$ \cite{Beisert:2006ez}. 
On the other hand, while the unperturbed XXX spin chain is integrable, it is known that the XXX spin chain with NNN interactions
\<
H=H_{XXX}+\delta H_{NNN}~,
\>
e.g.\ \eqref{eq:NNNHam} with
$\epsilon_0=0$, $\lambda_1=1$ and $\lambda_2=\delta$, is chaotic \cite{hsu1993level, poilblanc1993poisson}. 

Since the all-order planar dilatation operator is integrable, as is the one-loop term on its own, it is interesting to see the appearance of chaos when we truncate to include only the two-loop terms.
 In particular we are interested in the behaviour of the theory as we vary the ratio of the strength of the next-to-nearest-neighbour term to the nearest-neighbour term which controls the strength of integrability breaking. We can write this ratio in terms of the planar gauge coupling as
\<
\label{eq:effgc}
 \delta=\frac{g^2}{1-4g^2}~,
\>
where we see that $\delta$ increases monotonically from $g^2=0$ until $g^2=1/4$ where it becomes negative and then asymptotically approaches $\delta=-1/4$. We thus expect that the system becomes chaotic as we increase the coupling $g^2$ from $0$ to $1/4$ and then become somewhat less chaotic. In order to explore this behaviour, we numerically compute the statistical properties of the energy levels in the following. 

\subsection{Level-spacing statistics} 
For quantum-chaotic systems it is generally believed\footnote{For example, see the books \cite{mehta2004random,porter1965statistical} or the reviews \cite{zelevinsky1996nuclear,d2016quantum}.} that, while  the overall dependence of the spectral density on the energy is specific to a given system, the fluctuations of energy levels about this average trend are universal and described by Random Matrix Theory. This is often referred to as the BGS conjecture \cite{Bohigas:1983er}. For RMT, the distribution of spacings, $s$, between adjacent levels is well-described by the Wigner-Dyson distribution
\<
\label{eq:WDdist}
P_{WD}(s)=A(\alpha) s^\alpha e^{-B(\alpha)s^2}~,
~~~ A(\alpha)=2 \frac{\Gamma(1+\tfrac{\alpha}{2})^{1+\alpha}}{\Gamma(\tfrac{1+\alpha}{2})^{2+\alpha}}~,~~~
B(\alpha)=\frac{\Gamma(1+\tfrac{\alpha}{2})^{2}}{\Gamma(\tfrac{1+\alpha}{2})^{2}}~,
\> 
where the parameter $\alpha$ depends on the Gaussian ensemble: $\alpha=1$ for the orthogonal, $\alpha=2$ for the unitary and $\alpha=4$ for the symplectic ensemble. The choice of the appropriate ensemble for a given Hamiltonian is determined by the presence or absence of discrete symmetries, and in particular the structure of the correlations between energy levels is influenced by
time-reversal symmetry. In general the time-reversal operator is an anti-unitary operator which can be written in the form $\mathcal{T}=KC$, where $K$ is a particular unitary operator and $C$ takes the complex conjugate of any operator or state it acts upon. 
It is known, see e.g. \cite{mehta2004random}, that for systems which are time-reversal symmetric and additionally rotationally invariant, there is a choice of states for which the Hamiltonian is real and symmetric and thus the appropriate ensemble is the orthogonal one.
On the other hand, for integrable systems energy levels are essentially uncorrelated and it has been shown that in a variety of integrable systems, e.g. \cite{berry1977level, poilblanc1993poisson, hsu1993level}, the level-spacing distribution is Poisson 
\<
P_{P}(s)=e^{-s}~.
\>
 
Before computing the level-spacing distribution for various parameter configurations of the Hamiltonian  \eqref{eq:NNNHam}-\eqref{eq:2looppar}, we must first de-symmetrise the spectrum by focussing on a subspace of states all of which have the same values of the global charges. States with different charges cannot mix and so the Hamiltonian is block diagonal, leaving eigenvalues in different sectors uncorrelated, and thus a level-statistics analysis is only meaningful within individual blocks.
For the Hamiltonian \eqref{eq:NNNHam}-\eqref{eq:2looppar} de-symmetrisation requires considering chains with fixed length $L$, and fixed total spin $\text{S}^z=\sum_{i=1}^L \text{S}_i^z$ or equivalently fixed impurity number $M=L/2-\text{S}^z$. As the Hamiltonian is translation invariant, we must also fix the total momentum $P$ and to connect with the cyclicity of the trace defining gauge-invariant operators we choose $P=0$. Furthermore, there is a parity charge corresponding to the symmetry transformation which reverses the ordering of spins in a spin-chain state,
 e.g. 
\<
\mathcal{P}:\ket{\downarrow \uparrow \uparrow \downarrow \downarrow \uparrow}\to \ket{\uparrow \downarrow\downarrow \uparrow \uparrow \downarrow}~,
\>
 and here we consider the larger sector of positive-parity states.  Finally, as there is a global SU$(2)$ symmetry, we only consider lowest-weight states satisfying S$^{-} \ket{\Psi}=0$. 
 
After computing the energy eigenvalues, $\{e_i\}$, for a particular de-symmetrised subspace, we must unfold the spectrum to remove the overall energy dependence and fix the average spacing to be one. This we do by splitting the cumulative level number
\begin{equation}
n(e)= \sum_{i=1}^m \Theta(e-e_i)\,,
\label{eq:staircase}
\end{equation}
where $m$ is the total number of states in the sector, into an average and fluctuation part
\begin{equation}
n(e)= n_\mathrm{av}(e) + n_\mathrm{fl}(e)\,.
\end{equation}
The average part is determined by first ordering the energy levels, then discarding some (usually around 5-10$\%$) of the low- and high-energy levels, and then fitting the spectrum to a relatively high-order (usually degree 15) polynomial which defines $n_\mathrm{av}$. 
The unfolded spectrum is then given by
\begin{equation}
\varepsilon_i= n_\mathrm{av}(e_i)\,,
\end{equation}
and captures the physics of the spectral fluctuations. Finally, we compute the level spacings $s_i=\varepsilon_{i+1}-\varepsilon_i$, divide the spacings range into bins (usually 15-20) and approximate the distribution, $P(s)ds$, by counting the number of spacings in each bin. The details of the unfolding and binning procedure can in certain cases have significant effects, so we choose values where the results do not vary strongly for small changes in the parameters. 

\begin{figure}
	\centering
	$
	\includegraphicsbox[scale=0.45]{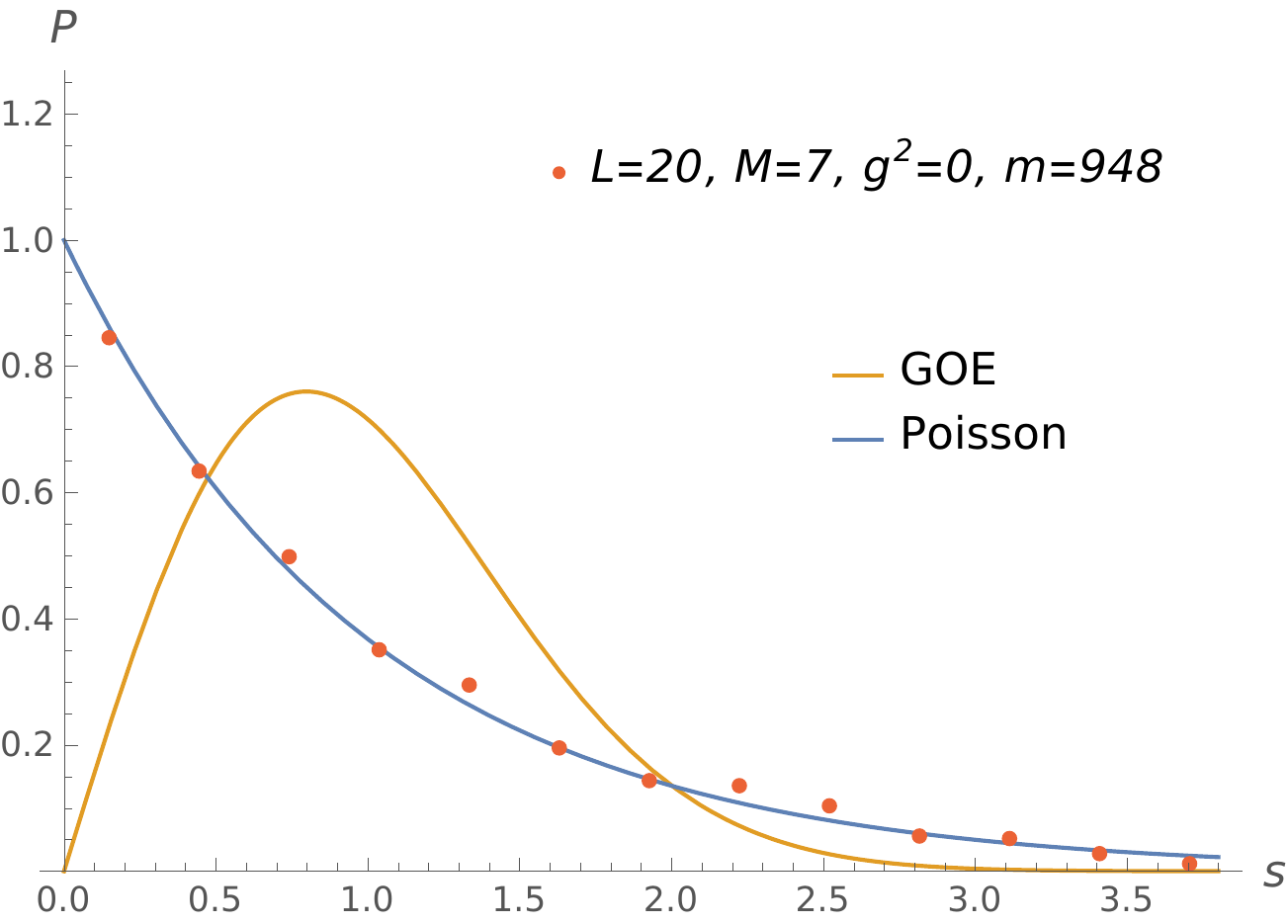}$ 
	$
	\includegraphicsbox[scale=0.45]{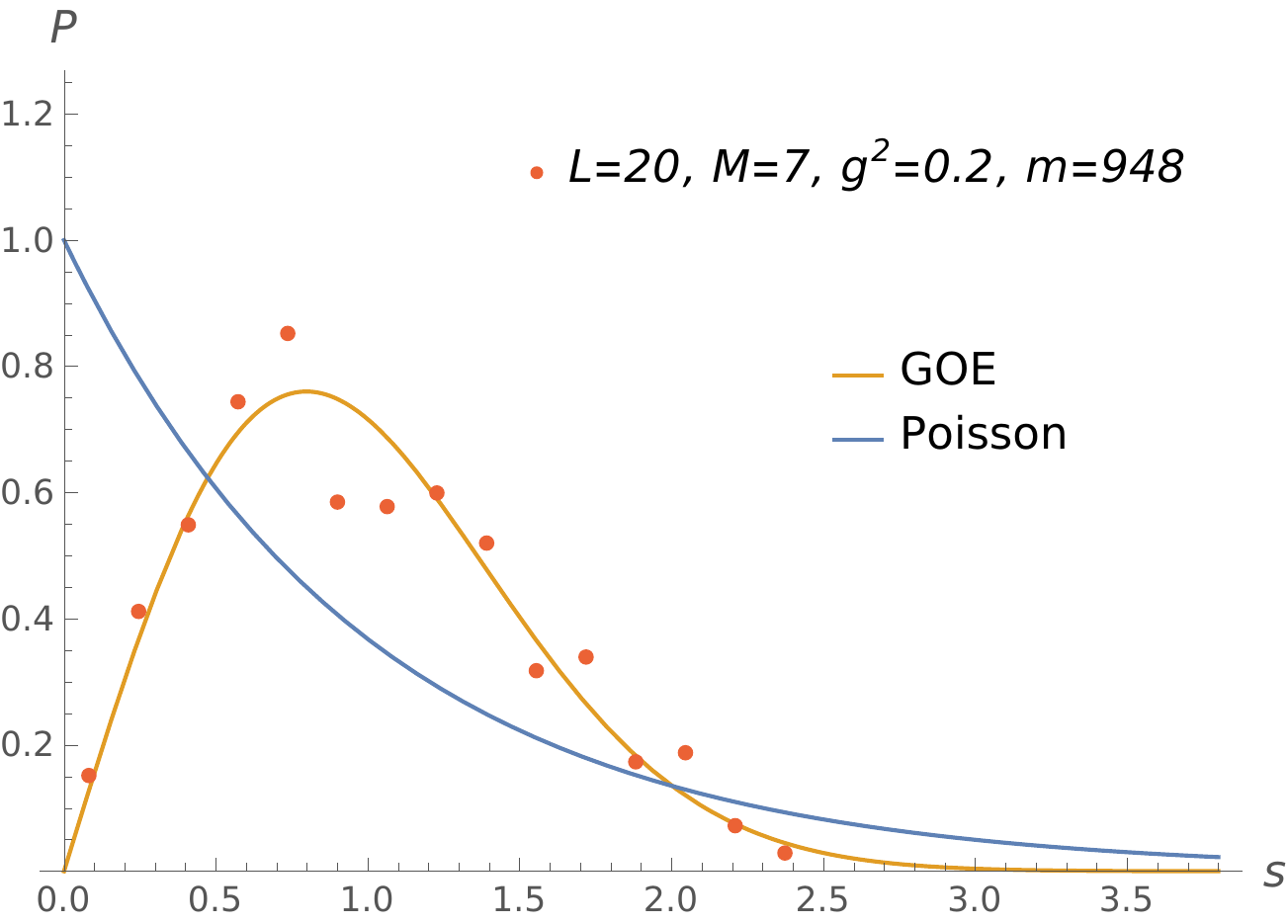}$ \\\vspace{0.3cm}
		$
	\includegraphicsbox[scale=0.45]{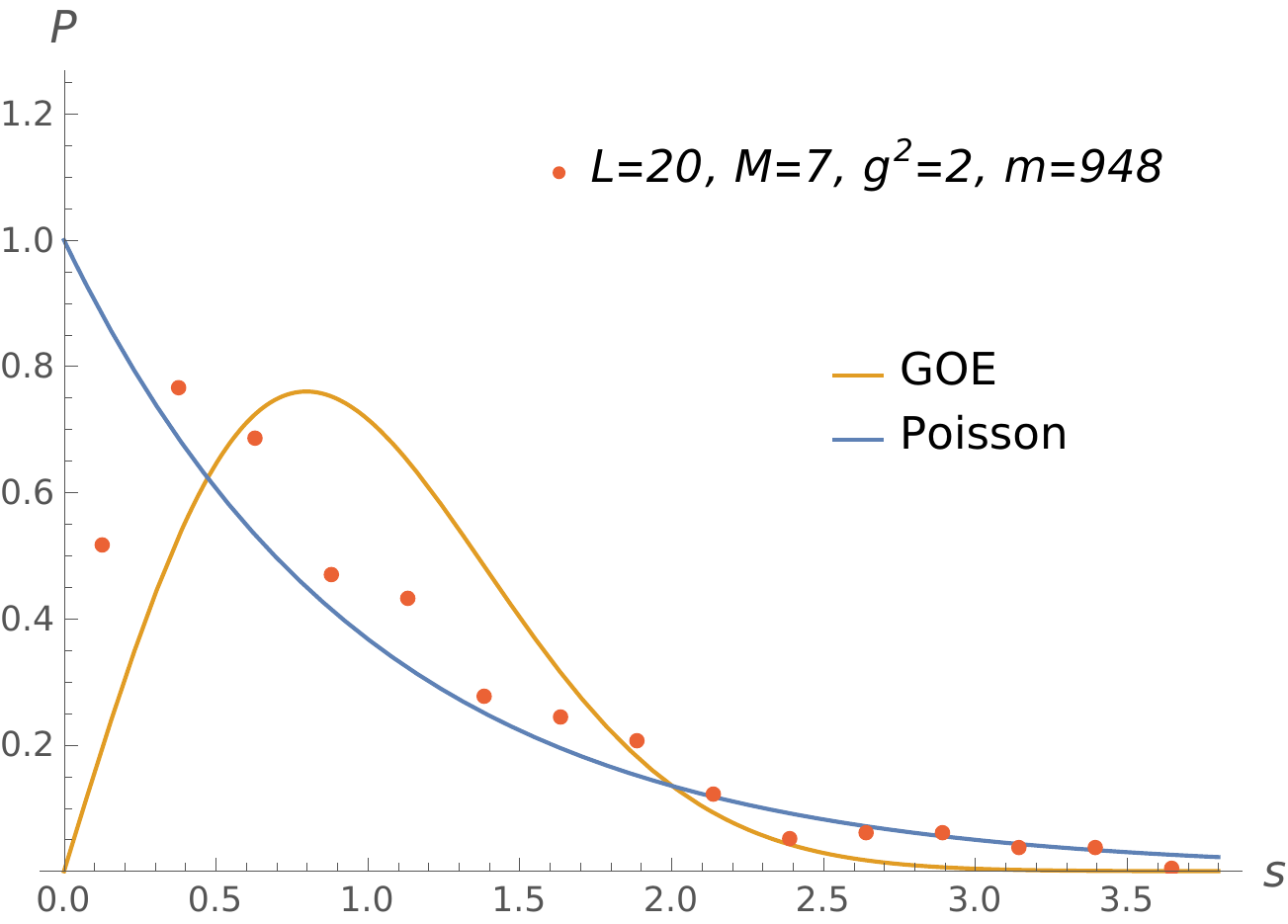}$	$
	\includegraphicsbox[scale=0.45]{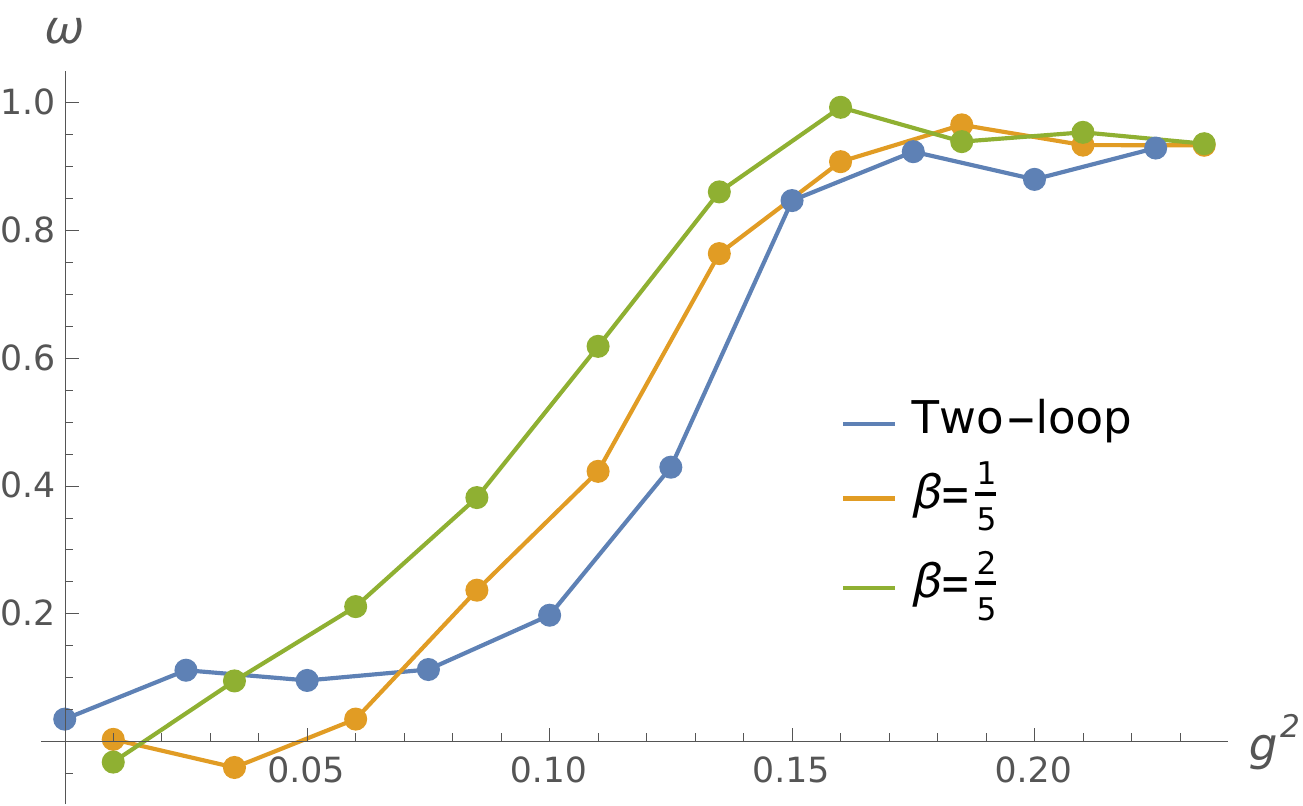}$ 
	\caption{Distribution of level spacings of the two-loop spin chain for positive-parity primary operators with $L=20$ and $M=7$ for $g^2=0,0.2,2$ (top left, right and bottom left). The Wigner-Dyson distribution with $\alpha=1$ and the Poisson distribution are shown for comparison. Bottom right:  Brody-distribution parameter $\omega$ as a function of $g^2$ for zero-momentum states with $L=20$, $M=7$ and for the twisted deformation with $\beta=1/5$ and $\beta=2/5$.}
	\label{fig:LStwoloopsu2}
\end{figure}
We show the results of these numerical calculations for the sector of positive-parity primary states with $L=20$ and $M=7$ in Fig.\ \ref{fig:LStwoloopsu2}, where the  Poisson distribution for $g^2=0$ and the GOE Wigner-Dyson distribution for $g^2=0.2$ is manifest. More quantitatively, we can define a generalised Brody distribution, 
\<
\label{eq:gBrody}
P_B(s)=A_B(\alpha, \omega) s^{\alpha \omega } e^{-B_B(\alpha, \omega) s^{1+\omega}} 
~,~~~ 
\>
with
\<
A_B(\alpha,\omega)=(1+\omega) \frac{\Gamma(\tfrac{2+\alpha \omega }{1+\omega})^{1+\alpha \omega }}{\Gamma(\tfrac{1+\alpha \omega }{1+\omega})^{2+\alpha \omega}}~,~~~
B_B(\alpha, \omega)=\frac{\Gamma(\tfrac{2+\alpha \omega }{1+\omega })^{1+\omega}}{\Gamma(\tfrac{1+\alpha \omega }{1+\omega })^{1+\omega}}~,
\>
which interpolates between the Wigner-Dyson distribution \eqref{eq:WDdist} for arbitrary $\alpha$ at $\omega=1$ and the Poisson distribution at $\omega=0$. For $g^2=0$ and assuming $\alpha=1$, we find the best fit is for a value of $\omega$ close to zero, in particular $\omega=0.03$, i.e.\ the nearest-neighbour levels are uncorrelated which is consistent with the integrability of the $g^2=0$ spin-chain model. For $g^2=0.2$ and assuming $\alpha=1$, we find the best fit is for a value of $\omega$ close to unity, $\omega=0.88$. If we assume $\omega=1$ instead, the best fit is for $\alpha=0.82$, and fitting both parameters at the same time we find $(\alpha=1.1, \omega=0.81)$ which is close to the GOE values and consistent with a quantum-chaotic nature of this spin-chain system. We can formally continue the coupling beyond $g^2=1/4$ such that the effective coupling $\delta$ becomes negative and we see that in this case the distribution becomes intermediate between Poisson and Wigner-Dyson with a best fit of $\omega=0.33$ when assuming $\alpha=1$. 

That we find a GOE spectrum is consistent with the symmetries of the Hamiltonian in \eqref{eq:NNNHam}. General spin-chain Hamiltonians with SU$(2)$ symmetry and arbitrary interaction range
\cite{Roiban:2003dw, Berenstein:2004ys}
\<
\label{eq:XXXlongrange}
H_{\rm{SU}(2),s}=\sum_{i=1}^L \big( \unit_{i,i+s}-\vec{\sigma}_i \cdot \vec{\sigma}_{i+s}\big)~
\>
are invariant under the action of the time-reversal operator $\mathcal{T}=KC$,
\<
H=\mathcal{T}H\mathcal{T}^{-1}~,
\> 
where $K$ is the unitary operator
\<
K=e^{\tfrac{i\pi}{2}\sum_k \sigma^y_k}=\prod_{k=1}^L i\sigma^y_k
\>
which, up to a minus sign, simply flips each spin. This symmetry guarantees that the  Hamiltonian is real and symmetric, which argues that the spectrum of the XXX spin chain with a long-range deformation of the type \eqref{eq:XXXlongrange} should be described by the orthogonal ensemble.
However, because the symmetry involves a spin flip, the action of the time-reversal operator will relate states in different sectors (except for the case where the $M=L/2$) and thus does not quite provide the explanation for GOE statistics within a sector of fixed $M$. This is important, as in order to see the Wigner-Dyson statistics, we worked within such sectors of fixed $M$. Fortunately, there is an additional symmetry: If we take $K$ to be the spin-chain parity operator $\mathcal{P}$ which takes $\sigma^A_i \sigma^B_{i+s}\to \sigma^B_i \sigma^A_{i+s}$, the combined operator $\mathcal{P}C$ acts within each of the fixed $M$ sectors and leaves the spin-chain Hamiltonian invariant. Now by working with eigenvectors of $\mathcal{P}$, which is also necessary to fully desymmetrise the spectrum, we can choose a basis of states such that $\mathcal{P}=+1$ and the Hamiltonian is real and symmetric when restricted to these sectors. This construction can be generalised to a time-reversal symmetry of the one-loop non-planar dilatation operator of $\mathcal{N}=4$ SYM theory by interpreting $\mathcal{P}$ as the operator which reverses the order of fields within individual traces of gauge-invariant operators and so explains the GOE statistics seen in \cite{McLoughlin:2020siu,McLoughlin:2020zew}.

\subsection{Transition from integrable to chaotic dynamics}
The smooth transition from Poisson to Wigner-Dyson statistics is quite generic for finite-size systems, e.g.\ \cite{Rabson, santos2010onset, modak2014universal, modak2014finite, bulchandani2021onset}, and is naturally explained in the Hamiltonian system \eqref{eq:NNNHam}-\eqref{eq:2looppar} by the integrable structure no longer being exact at finite $g^2$.  At the same time, deformations which preserve integrability to leading order in the deformation parameter and only break it at higher orders are special and have been shown  to be ``weak'' \cite{Szasz-Schagrin:2021pqg} in the sense that the onset of chaos is slower, i.e.\ the cross-over from Poisson to Wigner-Dyson occurs at larger couplings than for generic deformations. 
In \cite{Szasz-Schagrin:2021pqg}, the authors considered the XXZ spin chain with an integrability-breaking current deformation which gives rise to long-range interactions. These current deformations were generated by the spin-chain boost method \cite{Bargheer:2008jt, Bargheer:2009xy} which was originally introduced in the context of the integrable planar $\mathcal{N}=4$ long-range spin chains. This method guarantees that the deformation can be completed into a long-range integrable theory non-perturbatively in the deformation parameter. This similarity suggests that the two-loop planar dilatation correction is weak in the same sense, and we will explore this in the following. 

We consider the transition from integrable to chaotic level statistics for the two-loop Hamiltonian \eqref{eq:NNNHam} with parameters \eqref{eq:2looppar} as we vary $g^2$ from $0$ to $1/4$ and the same transition for a Hamiltonian where a twist parameter $\beta\in\mathbb{R}$ is introduced into the NNN-deformation as
\<
\label{eq:twist}
\sum_{i=1}^L \mathbf{S}_i\cdot \mathbf{S}_{i+2}\rightarrow \sum_{i=1}^L (2 e^{2i \beta}\text{S}^-_i\text{S}^+_{i+2}+2e^{-2i \beta}\text{S}^+_i \text{S}^-_{i+2}+ \text{S}^z_i \text{S}^z_{i+2})~
\>
with $\text{S}^\pm=\tfrac{1}{2}(\text{S}^x\pm i \text{S}^y)$. Note that here we are not twisting the leading NN term, and thus we cannot assume that this deformation corresponds to an integrable deformation for generic values of $\beta$ even at $\mathcal{O}(g^2)$ and so it is not weak in the above sense.

For a range of values of $\beta$ and $g^2 \in [0,0.25]$ we compute the level-spacing distributions and, assuming $\alpha=1$, find the values of $\omega$ that give the best fits of $P_B$ to the data. 
As can be seen from the bottom right plot in Fig.\ \ref{fig:LStwoloopsu2}, the undeformed two-loop spin chain transitions from the integrable, $\omega=0$, case to the chaotic, $\omega=1$, distribution consistently slower than the generic twisted deformation which is compatible with it being a weak deformation. It is worth noting that although we are using the parameters \eqref{eq:2looppar}, this Hamiltonian is for intents and purposes the XXX spin chain with a NNN interaction. It is somewhat surprising to find that this well-known chaotic system is weak in any sense\footnote{We are grateful to B. Pozsgay, D. Sz\'asz and G. Takacs for very useful discussions on this point.}. However, there were some previous suggestions that this is indeed the case, e.g.\ the deformation was noted by \cite{hsu1993level} to be weaker than the interchain perturbation. More recently, \cite{malikis2020quasi, Kurlov:2021pxl} found that for the XXX chain with NNN deformation there are  quasi-conserved charges . This is to be contrasted with the next-to-NNN (range-four) deformation for which there are no quasi-conserved charges at all.  

One subtlety with our comparison of the SU$(2)$-invariant chain and the twisted chain is that the twisted deformation, in addition to breaking integrability, breaks the global SU$(2)$ and parity symmetries. As a result it mixes larger subsectors (for $L=20$ and $M=7$ the dimension is $m=3876$) than the undeformed theory and so the comparison is not between equivalent sectors of eigenstates. When $g^2$ is taken to $0$ in the twisted case, the symmetries are restored and the subspaces with different charges  decouple. The combined distribution of the corresponding sectors is neither Poisson nor Wigner-Dyson as it has many more spacings near $0$. This explains the negative values of $\omega$ near $g^2=0$ for the twisted deformation. It is thus interesting to turn to the $\beta$-deformed gauge theory which is described by a twisted spin chain even at leading order and where the NNN deformation does not change the global symmetries.

\subsection{Marginally deformed SU(2) spin chain} 
	 
	 \label{sec:su2def}
The $\beta$-deformed $\mathcal{N}=4$ SYM theory is a Leigh-Strassler deformation \cite{Leigh:1995ep} preserving $\mathcal{N}=1$ supersymmetry, but breaking the SO$(6)$ $R$-symmetry of $\mathcal{N}=4$ SYM to U$(1)^3$. In terms of $\mathcal{N}=1$ superfields, $\Phi_i$ with $i=0,1,2$, the superpotential can be written as 
\<
W=g' ~ \Tr[e^{-i\beta}\Phi_0 \Phi_1 \Phi_2 -e^{i\beta} \Phi_1 \Phi_0 \Phi_2]~,
\>
where in general the parameters $g'$ and $\beta$ are allowed to be complex, though in this section we restrict them to be real. Additionally, requiring conformal invariance provides a constraint relating $g'$, $\beta$, the gauge-coupling constant $g$ and the rank $N$ of the gauge group, as well as constraining the gauge group to SU$(N)$ rather than U$(N)$.
The Lagrangian may be obtained from the undeformed Lagrangian by replacing all products of fields by a Moyal-like star-product corresponding to a non-commutativity in the internal SU$(4)$ R-symmetry directions \cite{Lunin:2005jy}. As a result of Filk's theorem \cite{FILK199653}, planar Feynman diagrams in the deformed theory are generally the same as those in $\mathcal{N}=4$ SYM theory up to an overall phase depending on the charges of the external fields\footnote{There are important caveats to this statement for short operators \cite{Fokken:2013mza}. In particular, at a fixed loop order $l$, we must restrict our attention to operators of length $L$ greater than the interaction range of connected subdiagrams. For the computation of the planar dilatation operator this implies that we consider operators with $L>l+1$.}. Consequently, the perturbative dilatation operator is just a twisted version of the $\mathcal{N}=4$ spin chain and the all-loop spectrum of asymptotically long operators is given by twisted Bethe equations \cite{Beisert:2005if, Ahn:2010ws}.   

If we again focus on an SU$(2)_\beta$ sector comprising operators made of two complex scalars, $X$ and $Z$, the resulting spin-chain Hamiltonian can again be written in terms of spin operators acting on each site, $\mathbf{S}_i$. The two-loop Hamiltonian was constructed in \cite{Frolov:2005ty} by means of a unitary transformation 
\<
U(\beta)=\text{exp}\left(i\beta\sum_{k=1}^L k \text{S}^z_k\right)
\>
which acts on the spin operators as 
\<
U(\beta) \text{S}^{\pm}_k U^\dagger(\beta)=e^{\pm i k \beta} \text{S}^{\pm}_k~, ~~~
U(\beta) \text{S}^{z}_k U^\dagger(\beta)=\text{S}^z_k
\>
and so we can find the SU$(2)_\beta$ Hamiltonian from the $\mathcal{N}=4$ spin chain by making the replacement
\<
\mathbf{S}_a\cdot \mathbf{S}_b \rightarrow \left(\mathbf{S}_a\cdot \mathbf{S}_b\right)_\beta =2 e^{+i\beta(b-a)} \text{S}_a^- \text{S}_b^+ +2 e^{-i\beta(b-a)} \text{S}^+_a \text{S}^-_b +\text{S}^z_a \text{S}^z_b~.
\> 
The unitary transformation corresponds to a twist of the boundary conditions, so the spectrum remains integrable for finite $\beta$ and $\lambda$ and is given by the same Bethe equations but with an additional boundary factor. 

For later convenience we define a Hamiltonian with different twists in the NN and NNN contributions
\<
\label{eq:HamNNdef}
H_{\beta}=\epsilon_0 \sum_{i=1}^L \unit_{i,{i+1}} +\lambda_1 \sum_{i=1}^L  \left(\mathbf{S}_i\cdot \mathbf{S}_{i+1}\right)_\beta+\lambda_2 \sum_{i=1}^L  \left(\mathbf{S}_i\cdot \mathbf{S}_{i+2}\right)_{\beta_2}~,
\>
where the two-loop planar dilatation operator of the $\beta$-deformed theory corresponds to the parameters \eqref{eq:2looppar} and $\beta_2=\beta$. Further, with this choice of parameters, there is a long-range integrable completion of this Hamiltonian and so the two-loop deformation, while not integrable at finite $g$, should be weakly chaotic in the sense of \cite{Szasz-Schagrin:2021pqg}.

\begin{figure}
	\centering
	$
	\includegraphicsbox[scale=0.45]{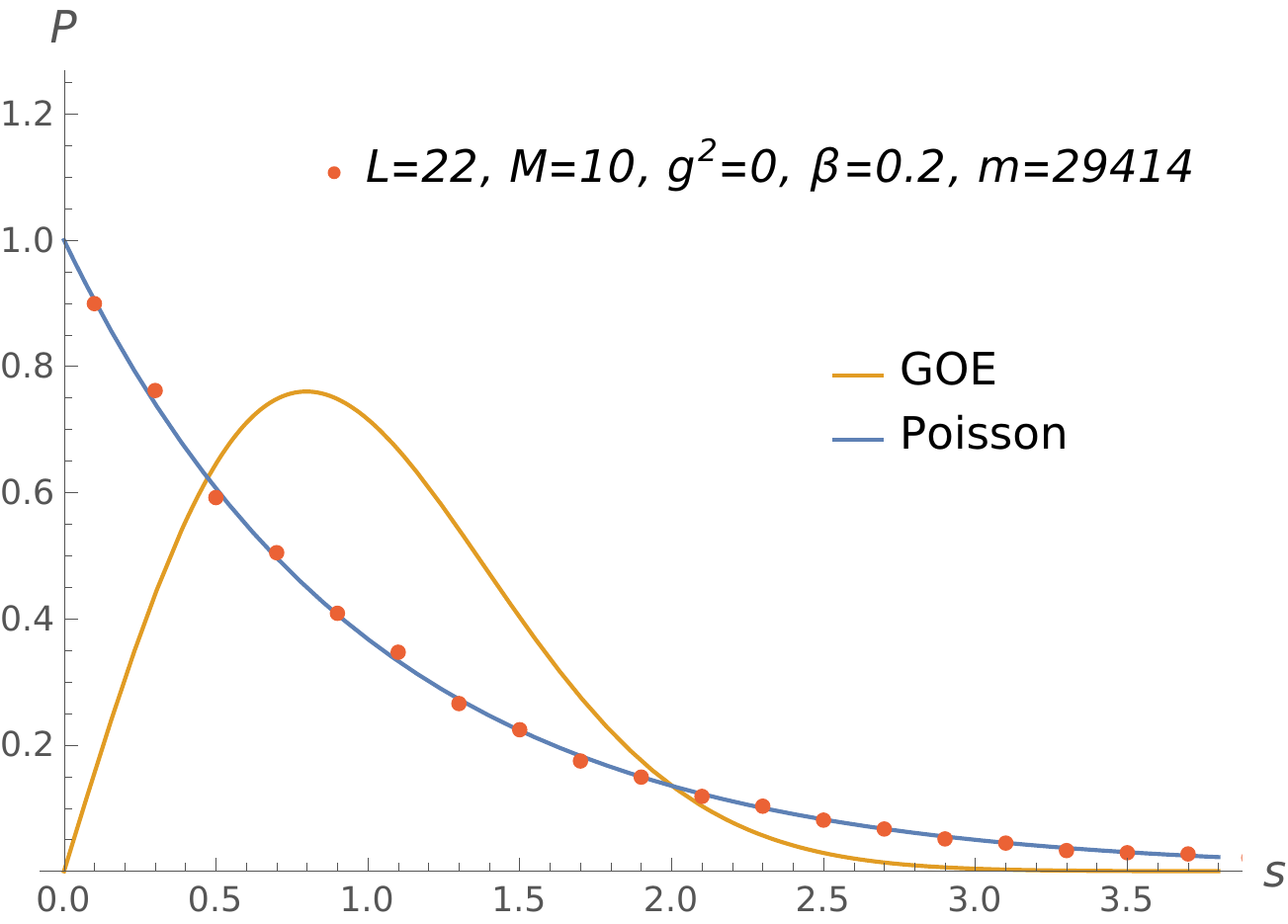}$ 
	$
	\includegraphicsbox[scale=0.45]{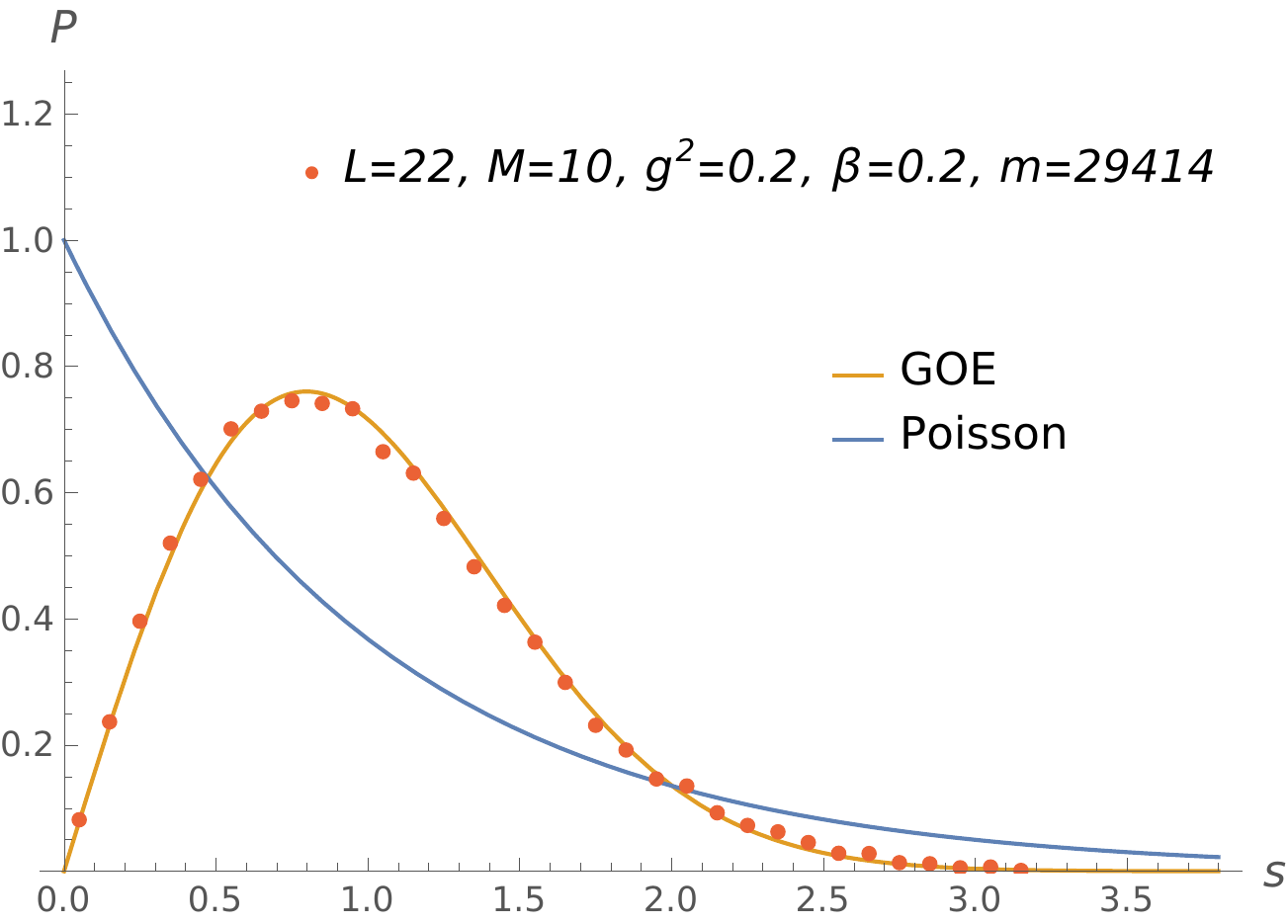}$ \\\vspace{0.3cm}
	$
	\includegraphicsbox[scale=0.45]{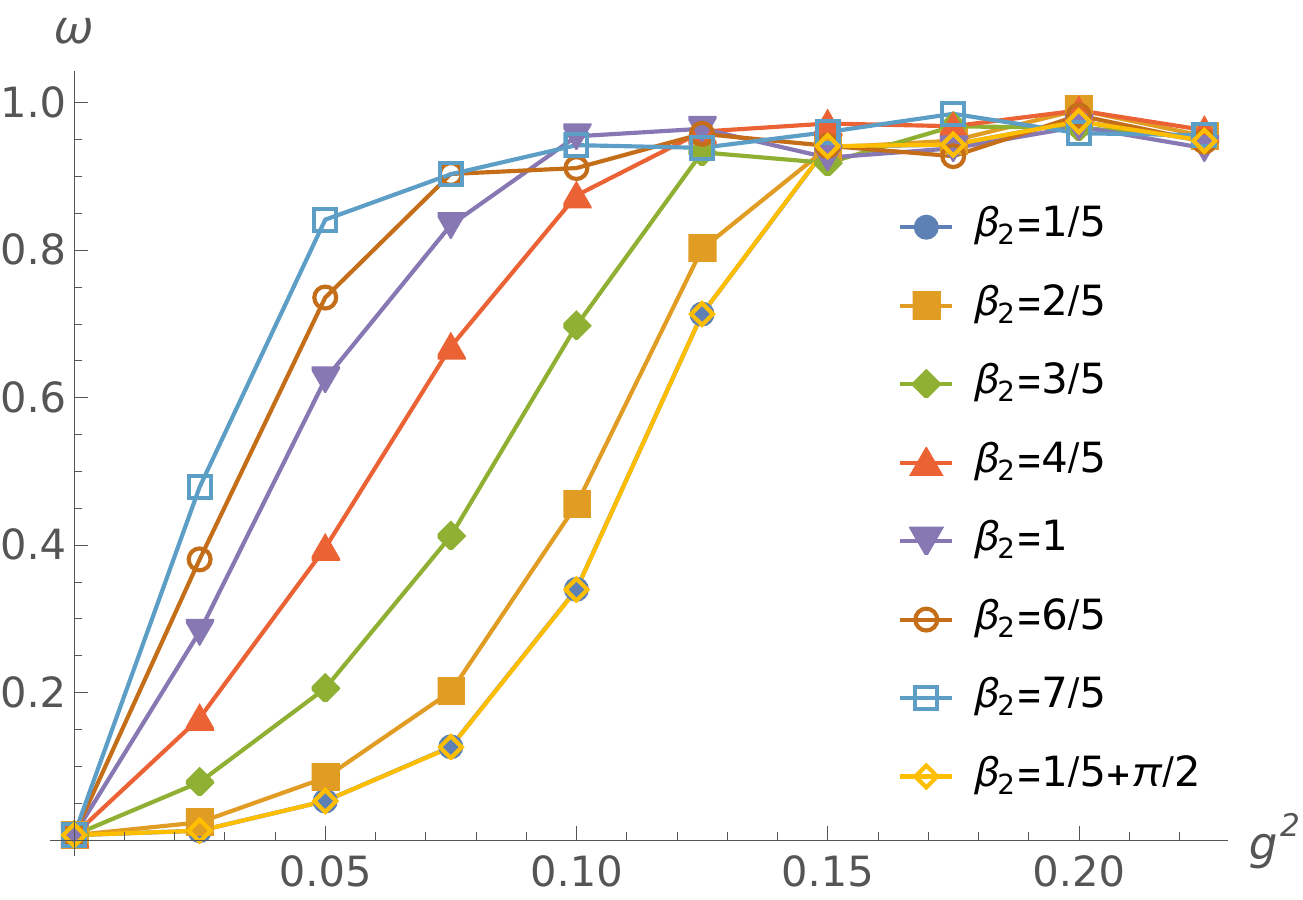}$	$
	\includegraphicsbox[scale=0.45]{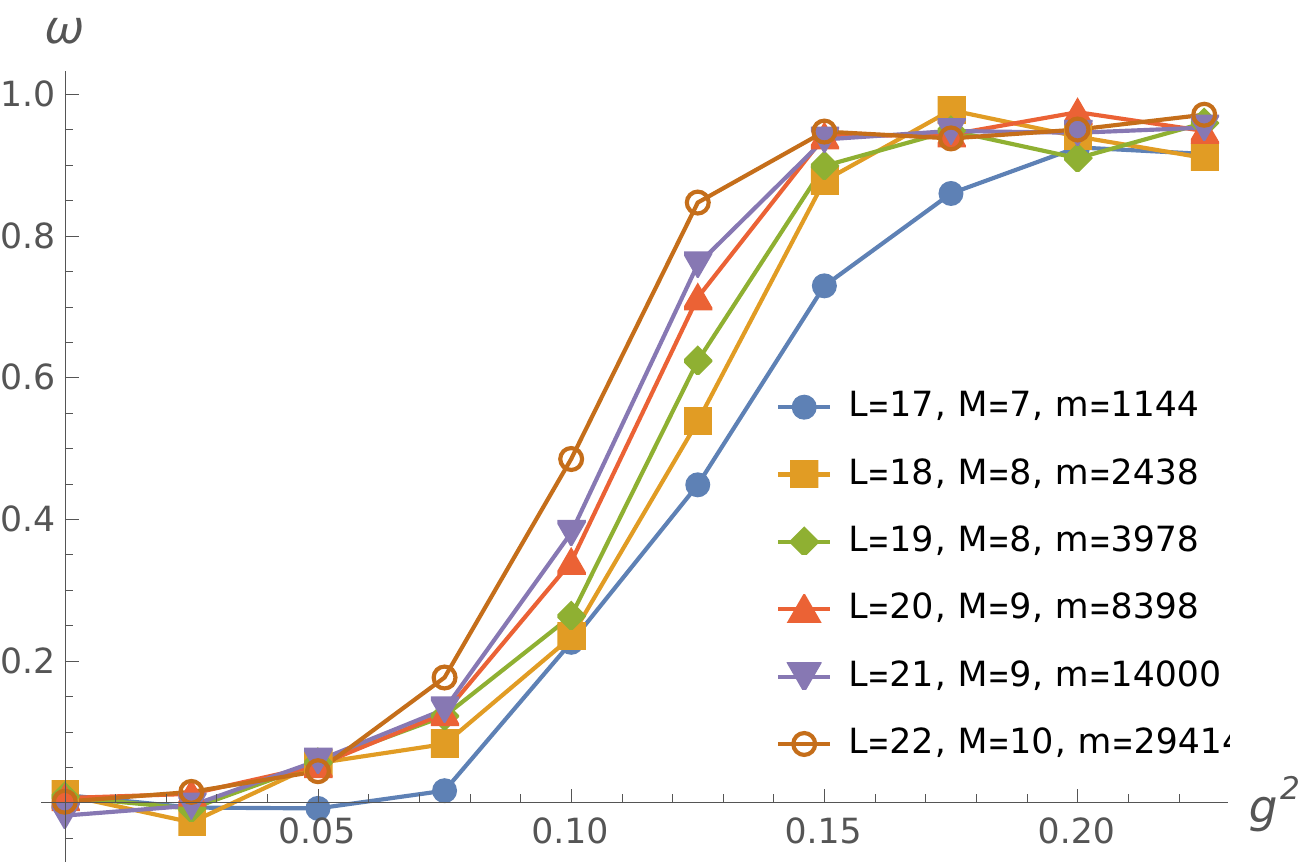}$ 
	\caption{Top row: Distribution of level spacings of the deformed two-loop spin chain  for  $L=22$, $M=10$ zero-momentum operators with $\beta=1/5$ and $g^2=0$ and $0.2$. Bottom left:  Brody-distribution parameter $\omega$ as a function of $g^2$ for zero-momentum states with $L=20$, $M=9$ for $\beta=1/5$ and $\beta_2\in [\tfrac{1}{5},\tfrac{1}{5}+\tfrac{\pi}{2}]$.
		Bottom right: Brody-distribution parameter $\omega$ as a function of $g^2$ for spin chains of varying length, $L\in[16,22]$, with $\beta=\beta_2=1/5$.}
	\label{fig:LStwoloopsu2beta}
\end{figure}
This can be seen in the numerical analysis of the level spacings in Fig.~\ref{fig:LStwoloopsu2beta}. The analysis is essentially the same as in the undeformed case with the simplification that, as the twist breaks the global SU$(2)$ and parity symmetries, translation invariance is the only symmetry and thus we de-symmetrise by restricting to zero-momentum operators. When the coupling $g^2$ is set to $0$, the Hamiltonian is just that of the twisted XXX spin chain and the distribution is Poisson, cf.\ Fig.\ \ref{fig:LStwoloopsu2beta} (top left). For sufficiently large values of $g^2$ the theory becomes chaotic and the distribution is well described by the GOE Wigner-Dyson distribution as can be seen in Fig.\ \ref{fig:LStwoloopsu2beta} (top right). That the distribution is orthogonal is again consistent with the discrete symmetries as the previous notion of  time-reversal symmetry extends to this case even though the rotational invariance is broken: For a spin chain with a Hamiltonian composed of terms of the form
\<
H_{D,s}=\sum_{i=1}^L \big[ \unit_{i,i+s}-\sigma_i^z \sigma_{i+s}^z-2{ e^{ i s \beta}}\sigma_i^- \sigma_{i+s}^+ -2 {e^{-i s \beta}}\sigma_i^+ \sigma_{i+s}^-\big]~,
\>
the previous combinations of complex conjugation and either spin-flip or parity is a symmetry of the Hamiltonian. This implies that the statistics of the twisted spin chain with non-nearest-neighbour interactions are also GOE.  

When $\beta_2\neq \beta$ there is presumably no completion into a long-range integrable theory and so the deformation should be strongly chaotic. In Fig.\ \ref{fig:LStwoloopsu2beta} (bottom left) we compare different values of $\beta_2$ by computing the level-spacing distribution for a range of values $g^2\in [0,0.25]$ and then finding the value of $\omega$ for which $P_B$ with $\alpha=1$ best fits the data. It can be seen that  the transition from $\omega=0$ to $\omega=1$ is slowest for $\beta_2=\beta$ with the transition being quicker for increasing values of $\beta-\beta_2$. It is not shown in the figure but the curve for $\beta_2=0$ is very close to that of $\beta_2=2/5$ and so the dependence appears to be on $|\beta-\beta_2|$ mod $\pi$. Additional curves for  $\beta_2$ increasing beyond those of $\beta_2=1/5+\pi/2$ move closer to the original curve, and are coincident when $\beta_2=\beta+\pi$, and so the dependence on $g^2$ appears to reach a limiting curve as $\beta_2-\beta=\pi/2$. This provides an interesting example of a deformation which smoothly interpolates from weakly to strongly chaotic. 

The authors of \cite{Szasz-Schagrin:2021pqg} also demonstrated that the cross-over behaviour of weak deformations has a different finite-size scaling than seen in strongly chaotic systems. As the size of the system increases, the value of the cross-over coupling, $g_c$, at which the system transitions from integrable to chaotic level statistics generally decreases. In \cite{modak2014finite}, the authors found numerical evidence that $g_c$ scales as a power of the system size, with the exponent being universal though possibly depending on the RMT ensemble relevant to the chaotic regime and so on the discrete symmetries of the model. For interacting models the transition from Poisson to GOE Wigner-Dyson distributions has a cross-over coupling which scales as $L^{-3}$. The same result was found in \cite{Szasz-Schagrin:2021pqg} for the XXZ spin chain with a NNN-deformation which is not SU$(2)$ invariant. In contrast, for the weakly-chaotic current deformation the cross-over was found to scale as $L^{-2}$. The scaling of the cross-over coupling thus provides another test for weakly chaotic systems.  

\begin{figure}
	\centering
	$
	\includegraphicsbox[scale=0.7]{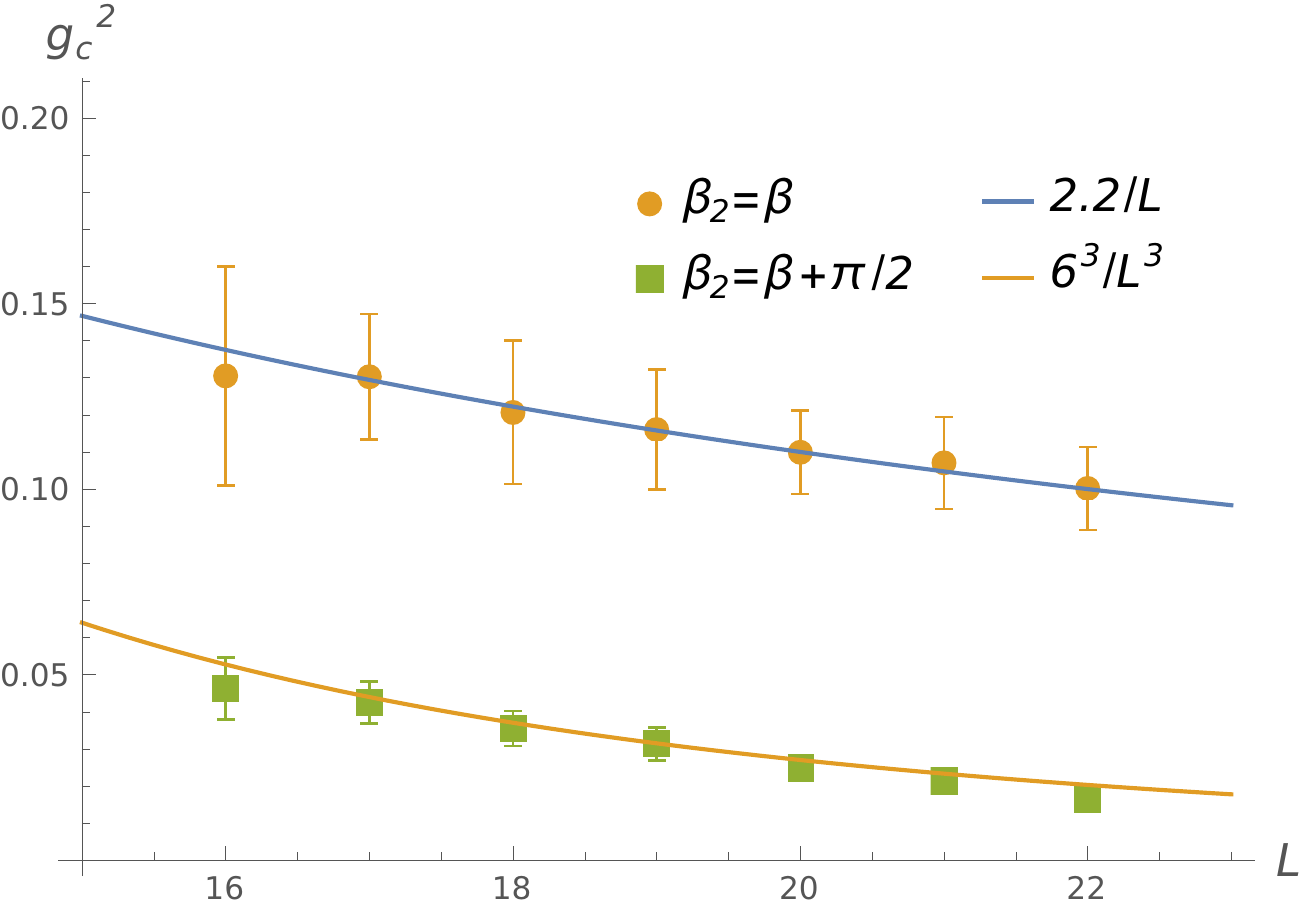}$  
	\caption{
		Cross-over coupling for the twisted two-loop spin chain with $L\in[16,22]$ for both $\beta=\beta_2=1/5$ and $\beta=1/5$, $\beta_2=\beta+\pi/2$. The curves $\sim 1/L$ and $\sim 1/L^3$ are shown for comparison.
	}
	\label{fig:CrOvsu2beta}
\end{figure}

Different authors use different definitions of the cross-over, but the results should be independent of specific definitions. In this work we define $g_c^2$ to be the coupling for which the level spacing is best fit by the Brody distribution with $\omega=1/2$ and $\alpha=1$\footnote{The peak of the Brody distribution at $\omega=1/2$ is $s=1/3^{3/2}\Gamma[5/3]\simeq 0.53> 1/\sqrt{2\pi}$ which is larger but close to the peak value used in \cite{Szasz-Schagrin:2021pqg}.}. We numerically determine the cross-over coupling for a given $L$ and $M$ by computing $\omega$ for a range of couplings  and then finding the parameters $a$ and $b$ for which this data is best fit by the function $\tfrac{1}{2}(\tanh(a x-b)+1)$ with the cross-over coupling being given by $b/a$. The system size, in terms of the dimension of the Hilbert space of each zero-momentum subsector, is a function of both $L$ and $M$ and in principle we might hold either fixed and vary the other but this may not be the most physical comparison. We compute the cross-over coupling for different values of $L$ but hold S$^z/L$ as close to constant as possible. This is similar to \cite{modak2014finite} which considered chains at half filling, S$^z=0$, and \cite{Szasz-Schagrin:2021pqg} which also considered fixed total spin S$^z=1$, though we include chains of odd length in our computations. In Fig.\ \ref{fig:CrOvsu2beta} we plot the estimated cross-over coupling for the Hamiltonian \eqref{eq:HamNNdef} of the weakly, $\beta_2=\beta$, and strongly chaotic, $\beta_2=\beta+\pi/2$, cases for chains with $L\in[16,22]$ with S$^z/L\simeq 7/16$. For the strongly chaotic case we find that the dependence on $L$ is $g^2_c=N_c/L^{b_c}$ with 
\<
b_c=3.3 \pm 0.2~,~~~N_c=(5\pm 3)\times 10^2~,
\>
while for the weak case we find 
\<
b_c=0.91\pm 0.06~,~~~ N_c=1.7\pm 0.3~.
\>
We have quoted errors for the coefficients but these should not be taken too seriously -- they are simply the standard errors of the fit to the conjectured form of $g_c$ and are likely a significant underestimate.
The parameters for the strongly chaotic case are consistent with previous results of $b_c=3$. For the weak case we find $b_c\simeq 1$ which differs from $b_c=2$ found in \cite{Szasz-Schagrin:2021pqg}. This may be due to the different global symmetries of the two models. While neither model has a global SU$(2)$ symmetry, the anisotropy of XXZ model considered in \cite{Szasz-Schagrin:2021pqg} is a much more significant breaking than the twist we have used. Nonetheless, these results are compatible with the weakly chaotic case having a scaling exponent smaller than in the strongly chaotic case. 

\section{Marginally deformed SU(3) spin chain} 
\label{sec:su3}
We now extend our considerations to larger sectors of the field theory, and specifically the SU$(3)$ subsector of $\mathcal{N}=4$ SYM theory. It consists of single-trace operators made from holomorphic combinations of three complex scalars $\phi_i$, with $i=0,1,2$, which can be combined with fermionic partners into the components of three $\mathcal{N}=1$ chiral superfields $\Phi_i$.  As before, we consider the effect of marginal deformations on the spectrum of anomalous dimensions for such operators and study the spectral statistics in the context of chaos and integrability, however restricting to one-loop order in this section. The general Leigh-Strassler deformation is given in terms of two complex parameters, $h$ and $q$, which we parametrise as $q=\text{exp}(i\beta+\kappa)$\footnote{We absorb a factor of 2 into the definition of $\beta$ compared to \secref{sec:su2def} to simplify our subsequent formulas.}, with $\beta, \kappa \in \mathbbm{R}$, and its superpotential is
\<
W=g''\Tr[\Phi_0 \Phi_1 \Phi_2 -q \Phi_1 \Phi_0 \Phi_2 +\frac{h}{3} (\Phi_0^3 +\Phi_1^3+\Phi_2^3)]~.
\>
At one-loop order, the SU$(3)$ sector of chiral operators $\mathcal{O}=\psi^{i_1 \dots i_L}\Tr(\phi_{i_1}\dots \phi_{i_L})$ is closed and one can extract the corresponding planar dilatation operator. We follow the notation of \cite{Bundzik:2005zg} and write it as an SU$(3)$ spin-chain Hamiltonian with nearest-neighbour interactions. The $i$-th spin-chain site, where $i=1, \dots, L$, can be in one of three states
\<
\ket{M_i}~,~~~ M_i=0,1,2,
\>
and the Hamiltonian can be written as $
H_{MD}=\sum_{i=1}^L H_{i, i+1}$
with
\<
\label{eq:Hsu3}
H_{i ,i+1}&=&\sum_{\ell=0}^2 \Big[E^{\ell,\ell}_{i} E_{i+1}^{\ell+1,\ell+1}-q E^{\ell+1,\ell}_{i}E_{i+1}^{\ell,\ell+1}
-\bar{q} E^{\ell,\ell+1}_{i}E_{i+1}^{\ell+1,\ell}
\nn\\
& &+q \bar{q} E^{\ell+1,\ell+1}_{i}E_{i+1}^{\ell,\ell}-q \bar{h} E^{\ell+1,\ell+2}_{i}E_{i+1}^{\ell,\ell+2}-\bar{q} h E^{\ell+2,\ell+1}_{i}E_{i+1}^{\ell+2,\ell}\nn\\
& &+h E^{\ell+2,\ell}_{i}E_{i+1}^{\ell+2,\ell+1}\
+\bar{h} E^{\ell,\ell+2}_{i}E_{i+1}^{\ell+1,\ell+2}
+h \bar{h} E^{\ell,\ell}_{i}E_{i+1}^{\ell,\ell}\Big]~,
\>
where the operators $E_i^{m,n}$ act on the $i$-th site as
\<
E^{m,n}\ket{r}=\delta^{nr}\ket{m}~.
\>
The authors of \cite{Bundzik:2005zg} considered values of $h$ and $q$ for which this spin chain is integrable and found $R$-matrices satisfying the Yang-Baxter equation for the cases
\<
\label{eq:qhInt}
(q, h)=\big\{(e^{i \beta},0), 
	(0,1/\bar{h}), ( (1+\rho)e^{\frac{2\pi i m}{3}}, \rho e^{\frac{2\pi i n}{3}}),(-e^{\frac{2\pi i m}{3}},e^{\frac{2\pi i n}{3}})\big\}~,
\>
all of which consist of Hopf twists of the undeformed theory.

We are interested in the general case of arbitrary complex parameters $h$ and $q$, and in understanding the level statistics of the corresponding spin-chain spectra. As in the SU$(2)$ sector we begin with a study of the model's symmetries to properly de-symmetrise the spectrum by restricting to subsectors of states with the same global quantum numbers. We consider closed spin chains of fixed length, $L$, with periodic boundary conditions. As the Hamiltonian and boundary conditions are then translation invariant, the states are labelled by the total momentum $P$ and, motivated by the connection to single-trace operators, we focus on spin chains with $P=0$.

As can be seen from the form of the individual terms in the Hamiltonian \eqref{eq:Hsu3}, the Hilbert space splits into three sectors, each having a fixed value of $M=\sum_{i=1}^L M_i$ counted modulo $3$. The Hamiltonian additionally has a manifest symmetry under the cyclic flavour transformation 
\<
0\to 1 \to 2 \to 0
\label{eq:flcyc}
\>
and for $L \neq 0$ mod $3$ the three sectors are mapped into each other by this transformation. By direct diagonalisation for small $L\neq 0$ mod $3$ (we considered $L \leq 13$), we find that the Hilbert space generically further splits into two subsectors which can be seen in the eigenvectors having many zeroes. This is due to the states possessing a charge associated with a combined spin-chain parity transformation $\mathcal{P}$ and a discrete flavour transformation $\mathcal{C}$. The flavour transformation $\mathcal{C}$ sends all sites with $M_i=v+1$ to $v+2$ and those with $M_i=v+2$ to $v+1$ where $v$ is given by $L v=M$ (mod $3$). For example in the $(L=4,M=0)$ sector, the flavour transformation exchanges $\ket{1}$ and $\ket{2}$ and so the state
 \<
\ket{0111}-\ket{0222}
 \>
 has negative charge under the generalised parity transformation $\mathcal{CP}$, while all other states in this sector have positive parity. As a result, this state does not mix with the others and is in a distinct superselection sector. The eigenvectors carry a definite value of this generalised parity and so organise into two orthogonal subspaces 
 
The situation is somewhat different for chains where $L=0$ (mod $3$) as states of fixed $M$ are still invariant under the cyclic flavour transformation. The eigenvectors are labelled by the corresponding charge, $Z=\{1, e^{2\pi i/3}, e^{4\pi i/3}\}$, and the space breaks into three further subsectors. For $M=1$ and $M=2$ we de-symmetrise by considering sectors of definite $Z$. 
 For the case $M=0$ the spectrum is in fact degenerate with, for example, the $(L=3,M=0)$ states
 \<\ket{000}-\ket{111}~~~\text{and}~~~\ket{000}-\ket{222}
 \>
 both having the eigenvalue $3 h \bar{h}$. In these cases it is possible to define the generalised parity transformation using any of the three flavour transformations $v \leftrightarrow v+1$ $v=0,1,2$ and this symmetry separates the spectrum into two, with the degenerate states in different subspaces. Nevertheless, this does not fully de-symmetrise the spectrum and as a result we do not consider $L=0$ (mod 3) $M=0$ sectors in this work. 
 
Finally, we also study the system with $h=0$, which includes the integrable case $q=e^{i \beta}$ with $\beta\in \mathbb{R}$ but also non-integrable cases where $q$ is not a pure phase. In this case the Hamiltonian not only  commutes with the total value of $M$, but also the number of each individual flavours is conserved and so we must work with states of fixed length $L$ and fixed number of sites, $R$, with $\ket{1}$, and, $S$, with $\ket{2}$. 
 
 \subsection{Level-spacing statistics}
 
\begin{figure}
	\centering
	$
	\includegraphicsbox[scale=0.5]{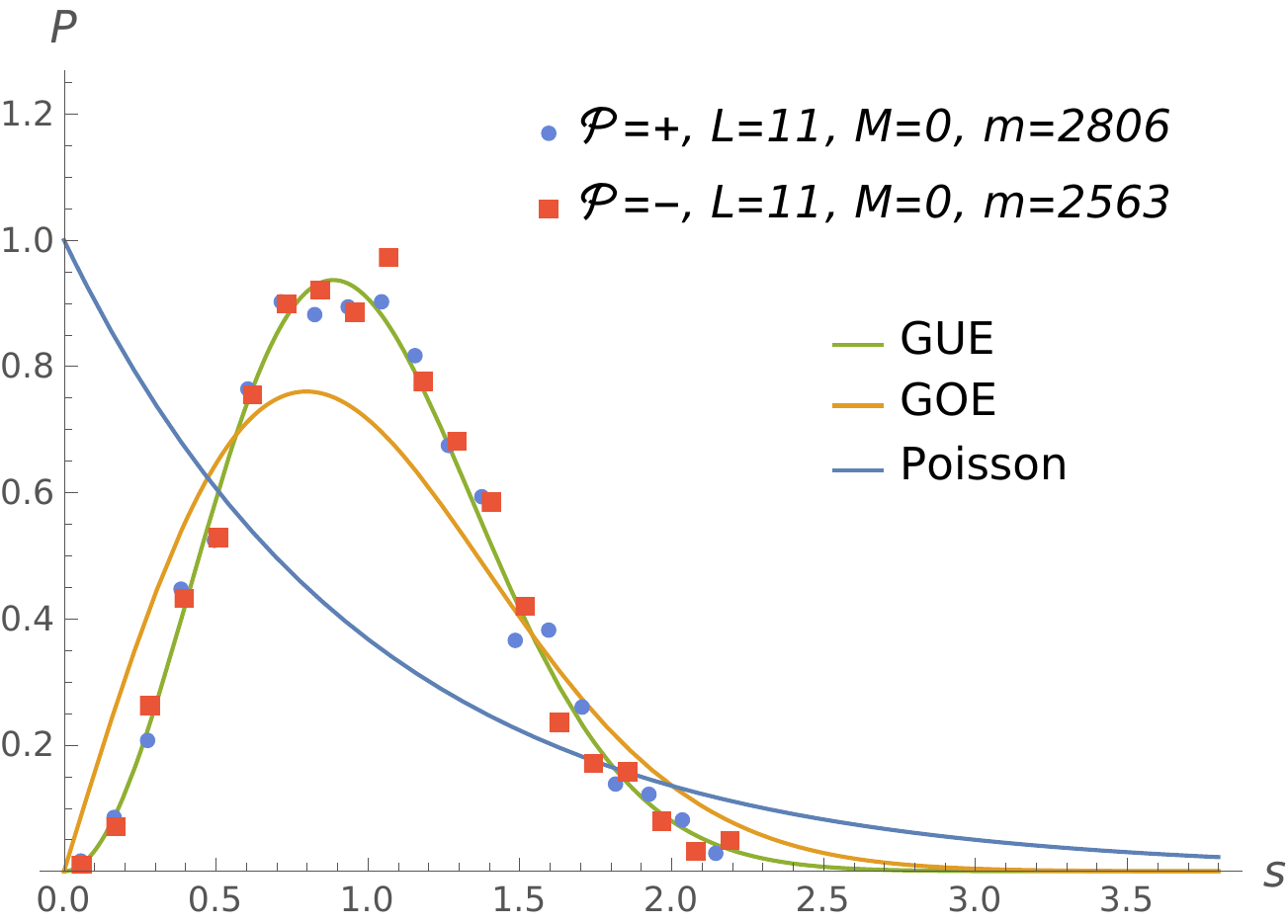}$ 
	$
	\includegraphicsbox[scale=0.5]{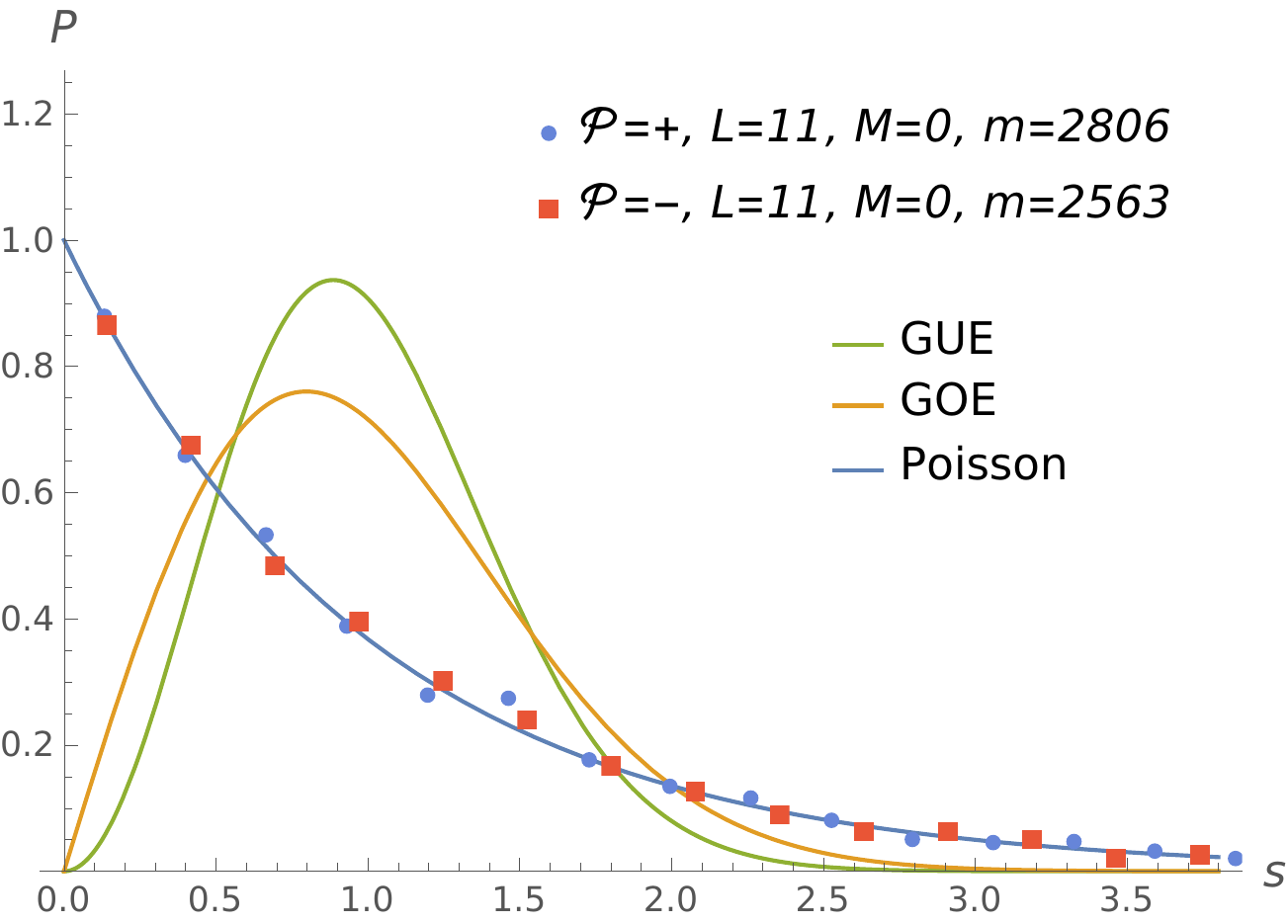}$ 
		$
	\includegraphicsbox[scale=0.5]{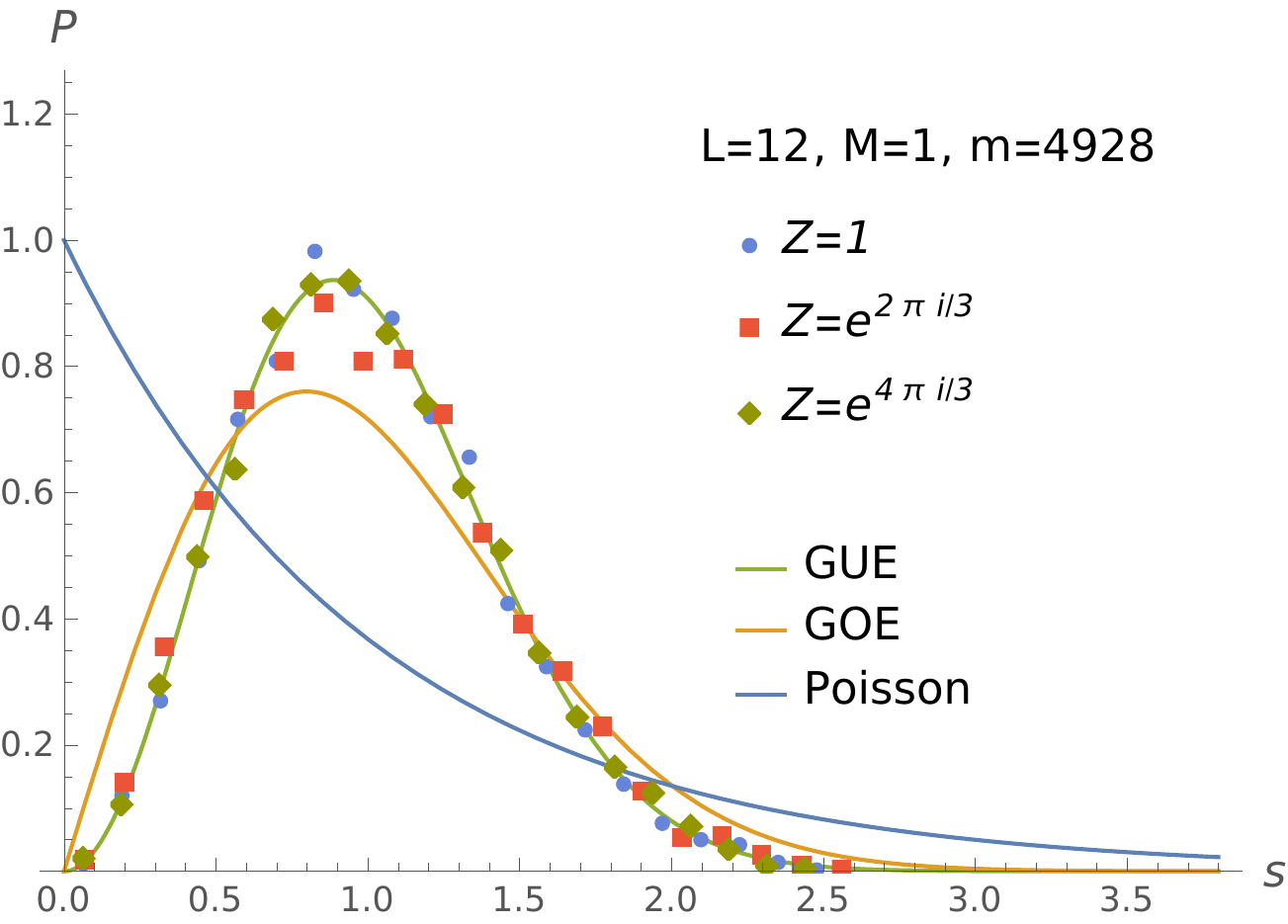}$ 
	$
	\includegraphicsbox[scale=0.45]{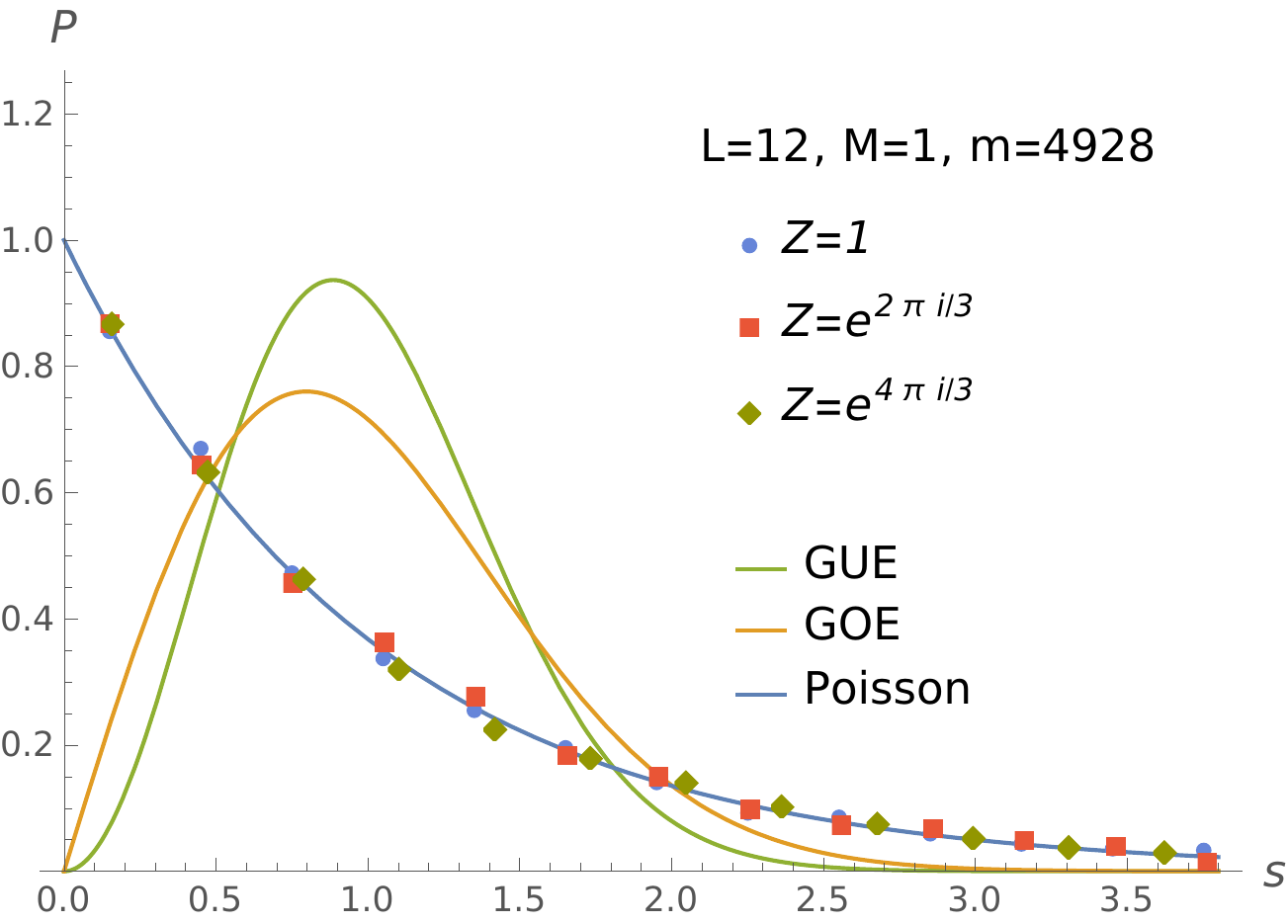}$ 
	\caption{Left: Distribution of level spacings for (top) $L=11$, $M=0$ with positive and negative parity and (bottom) $L=12$, $M=1$ in all three sectors with generic values of the parameters resulting in a chaotic spectrum. Right: Distribution of level spacings with integrable values of the parameters for (top) $L=11$, $M=0$ both parities and (bottom) $L=12$, $M=1$. }
	\label{fig:su3levelspace}
\end{figure}

The computation of the level-spacing distribution for spectra in the SU$(3)$ sector is essentially the same as in the previously discussed SU$(2)$ and twisted SU$(2)$ cases: We numerically compute the energy eigenvalues by direct diagonalisation, then de-symmetrise the spectrum by focussing on specific sectors of states, and finally we carry out the appropriate unfolding and binning of the data. We first consider the spin chain with $L=11$ and $M=0$ and take the deformation parameters to have the numerical values
\<
h=\tfrac{1}{3}\text{exp}\left(i \tfrac{11}{13}\right)~,~~~q=\text{exp}\left(i\tfrac{7}{10}+\tfrac{5}{13}\right)~.
\>
These do not correspond to any of the integrable points \eqref{eq:qhInt} found by \cite{Bundzik:2005zg}. The corresponding spin-chain system has a total number of 5369 states which can be organised into 2806 states with even and 2563 states with odd generalised parity. The distribution of level spacings is shown in the top left panel of Fig.\ \ref{fig:su3levelspace}.

Using the generalised Brody distribution \eqref{eq:gBrody} and assuming $\omega=1$ for a Wigner-Dyson distribution, we find the best fit is for $\alpha=1.98$ for the positive-parity states and $\alpha=2.12$ for the negative-parity states. If instead we assume $\alpha=2$, we find the best fit is for $\omega=1.00$ and $\omega=1.05$, respectively. 
We can attempt to fit both parameters and in this case the results are less convincing, in particular we find $\{\alpha, \omega\}$ is $\{1.71,1.10\}$ for positive parities and $\{1.32,1.37\}$ for negative, but this seems to be a feature of the sensitivity of the fitting process. Thus this distribution appears to be well described by the GUE Wigner-Dyson distribution 
with $\alpha=2$. Similar results are found for $L=11$ with $M=1$ and $M=2$. The case where $L$ is a multiple of three has different symmetries but, once complete de-symmetrisation is carried out, the results are again similar, cf.\ bottom left panel of Fig.\ \ref{fig:su3levelspace} for the case of $(L=12, M=1)$ for the three values of $Z$. 

Thus, in contrast to the chaotic SU$(2)$ spin-chain models studied above, here we find that the level-spacing distribution is well described by the unitary instead of the orthogonal ensemble. This is explained by the fact that generally there does not seem to be a  time-reversal operator for the deformed SU$(3)$ Hamiltonian. Although there is the cyclic symmetry transforming the flavours at each site as in \eqref{eq:flcyc}, this cannot be combined with complex conjugation to form a symmetry. Additionally, in the case of non-vanishing complex $h$ the combination of spin-chain parity and complex conjugation is not a symmetry of the Hamiltonian \eqref{eq:Hsu3} as
\<
E^{\ell+2,\ell}_i 
E^{\ell+2,\ell+1}_{i+1} 
\neq 	E^{\ell+1,\ell+2}_i E^{\ell,\ell+2}_{i+1}~,
\>
while for $q\bar{q} \neq 1$ the term $E_i^{\ell+1,\ell+1}E_{i+1}^{\ell,\ell}$ breaks parity symmetry as it is not mapped into the corresponding $E_i^{\ell,\ell}E_{i+1}^{\ell+1,\ell+1}$ term. 

The distribution of spacings is largely independent of the values of $(q,h)$, however we can see that for particular cases which are known to be integrable the distribution becomes Poisson. For example if we take
\<
h=\tfrac{1}{3} \text{exp}\left( i \tfrac{2\pi}{3}\right)~,~~~q=\tfrac{4}{3} \text{exp}\left(i \tfrac{4\pi}3\right)~,
\>
then we find the distribution shown in the top right panel of Fig.\ \ref{fig:su3levelspace} for $(L=11,  M=0)$ and bottom right for $(L=12, M=1)$, both of which follow the Poisson curve.

\begin{figure}
	\centering
	$
	\includegraphicsbox[scale=0.5]{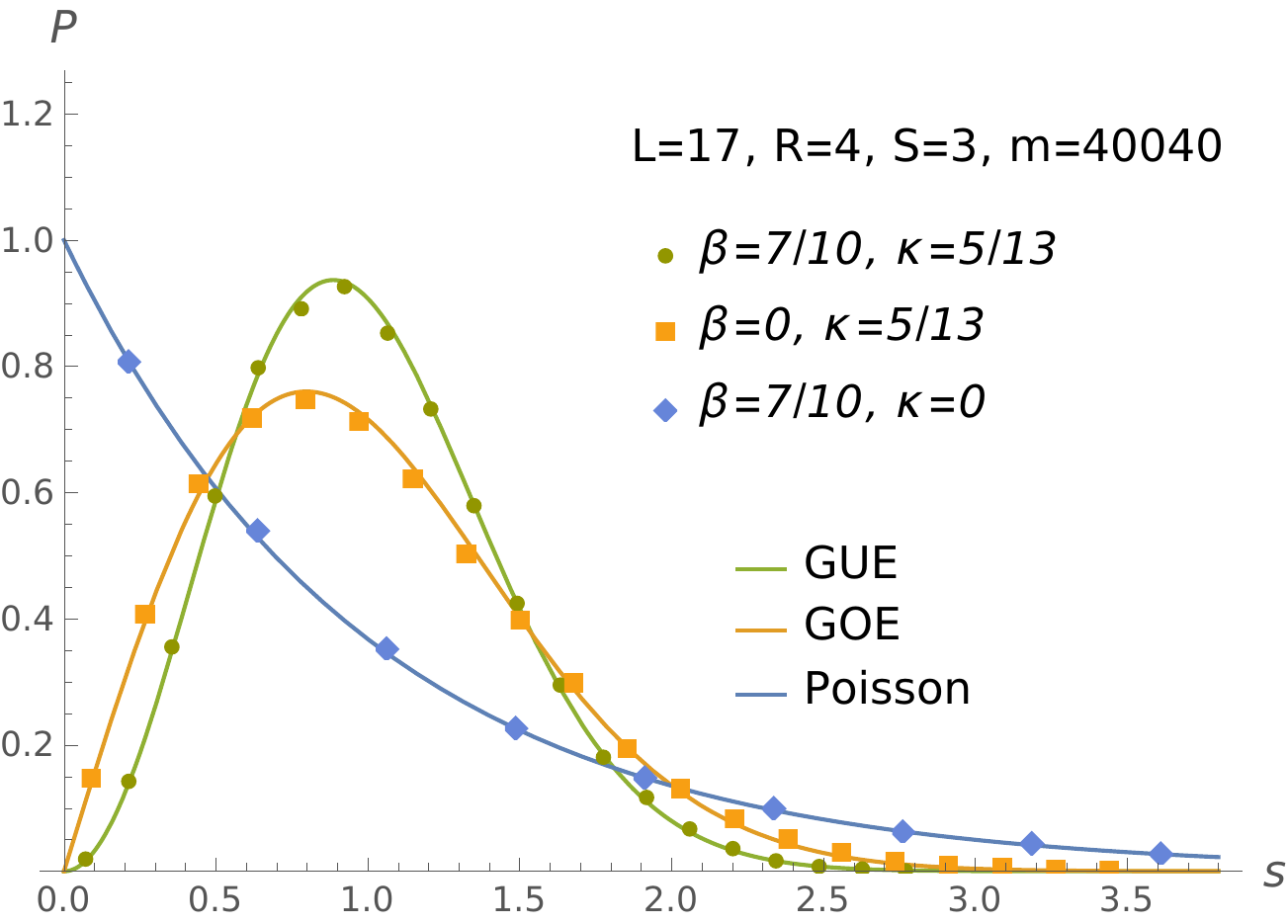}$ 	$
	\includegraphicsbox[scale=0.5]{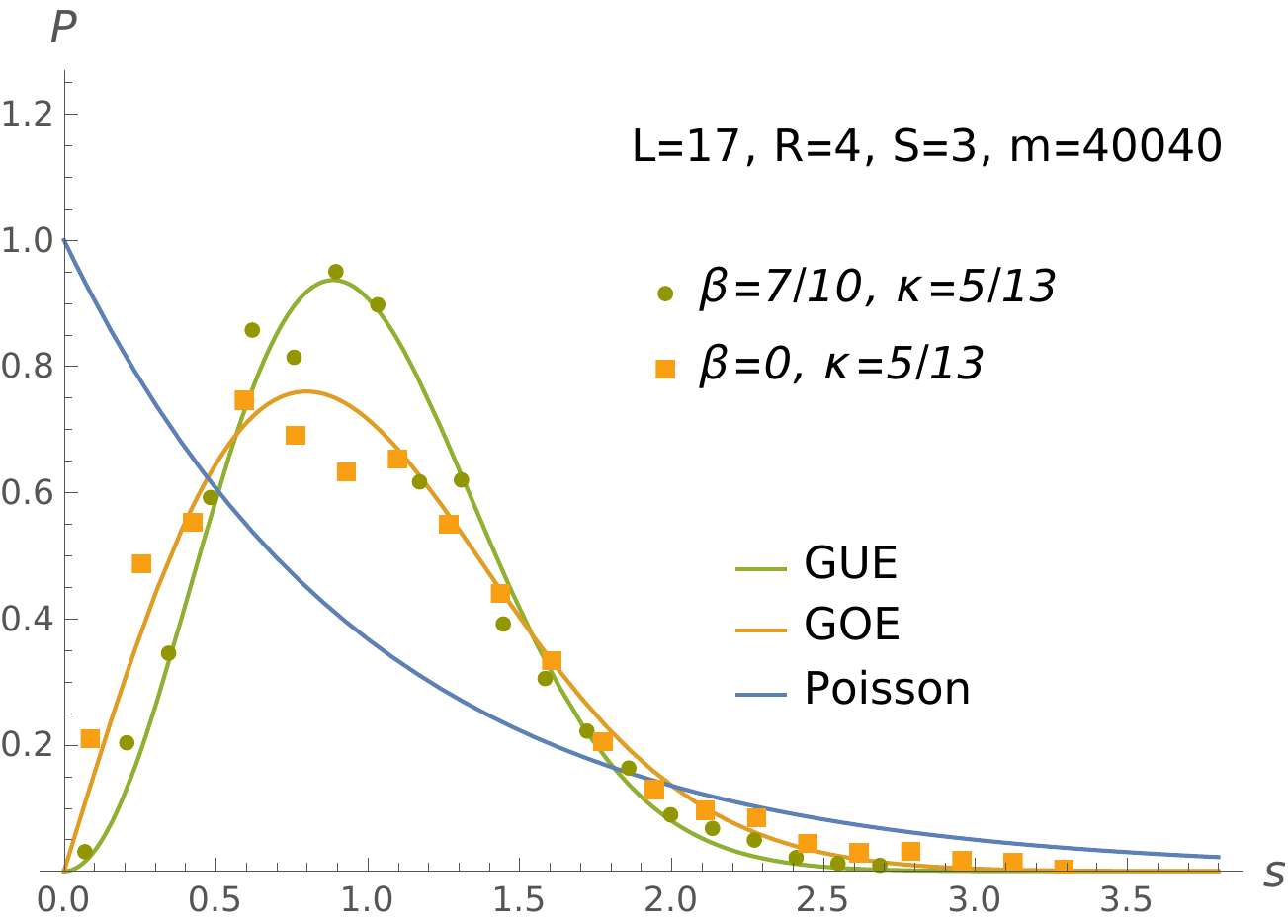}$ 
\caption{Distribution of level spacings for $(L=17, R=4, S=3)$ with $(\beta\neq 0, \kappa \neq 0)$, $(\beta=0, \kappa\neq 0)$ and $(\beta\neq 0, \kappa=0)$. Right: distribution with the highest and lowest 10\% of energy  states removed. Left: distribution for the lowest 6\% of energy states.}
	\label{fig:su3L17R4S3}
	\end{figure}

Another special region of parameter space is when $h=0$, where we can label a state by its length $L$ and the number of sites, $R$, with state $\ket{1}$, and the number $S$ with state $\ket{2}$. In this case there is no additional generalised parity charge which makes computing the spectrum relatively straightforward. For example if we consider $(L=17, R=4, S=3)$ chains, there are 40,040 states. Taking
\<
q=\text{exp}\left(i\tfrac{7}{10}+\tfrac{5}{13}\right)~,
\>
i.e.\ $\beta\neq 0$ and $\kappa\neq 0$, we find the distribution is again well described by the GUE Wigner-Dyson distribution as shown in Fig.\ 
\ref{fig:su3L17R4S3}. If we take $\beta=0$ but keep $\kappa\neq0$, we find that the spectrum is chaotic but now GOE.  In this last case the orthogonal ensemble's reappearance is natural as the Hamiltonian for $q\in \mathbb{R}$ is real, and so we expect the ensemble to be GOE. Alternatively if we take $\kappa=0$ and $\beta\in\mathbb{R}$, we find the Poisson distribution corresponding to an integrable Hamiltonian. In \tabref{tab:su3L17R4S3} we show the best fit values for the parameters $\omega$ and $\alpha$ of the Brody distribution.
\begin{table}[h]
		\centering
		\begin{tabular}{|c c || c c c |} 
		\hline
		$\beta$ & $\kappa$  & $\omega$ & $\alpha$ & $\{\omega, \alpha\}$ \\ [0.5ex] 
		\hline\hline
		0.7 & 5/13 & 0.98  & 1.9  & \{0.97, 2.0\} \\ 
				\hline
		0 & 5/13 & 0.92  & 0.95  & \{0.9, 1.1\} \\ 
		\hline
		0.7 & 5/13 & 0.89$^\star$  &  1.71$^\star$ &  \\ 
		\hline
		0 & 5/13 & 0.51$^\star$  & 0.73$^\star$ &  \\ 
		\hline
		0.7 & 0 & 0.01  & N/A  & N/A \\ 
		\hline
	\end{tabular}
\caption{Best-fit parameters in the Brody distribution for $(L=17, R=4, S=3)$ with $(\beta\neq 0, \kappa \neq 0)$, $(\beta=0, \kappa\neq 0)$ and $(\beta\neq 0, \kappa=0)$. The value of $\omega$ is first fitted with $\alpha$ fixed to either 1 or 2, then $\alpha$ is fitted with $\omega$ fixed in the chaotic cases, and finally both values are fitted. The values are calculated with the $10\%$ of highest and lowest energies removed. Those marked by $\star$ are calculated for only the lowest $6\%$ of energies. }
\label{tab:su3L17R4S3}
\end{table}		
		
In Sec.\ \ref{sec:LL} we will be interested in studying the dynamics of spin chains in the limit of infinite length and low energy. This is a difficult regime to access by direct diagonalisation methods due to the rapidly increasing size of the Hamiltonian mixing matrices. In general, see e.g. \cite{zelevinsky1996nuclear,d2016quantum}, one expects the low-energy states to be less chaotic than those from the bulk of the spectrum. Nonetheless, we can analyse the level-spacing distribution for the lowest 6\% of energy states, see the right panel of Fig.\ \ref{fig:su3L17R4S3} and the starred data in Tab.\ \ref{tab:su3L17R4S3}. Even for these low-energy states, the distributions are quite close to the Wigner-Dyson distribution although the case of $\beta=0$ is further from GOE. One might suspect that for sufficiently large $L$ even the low-energy states are chaotic and this is consistent with what we will find in the LL model, though the cross-over from integrable to chaotic is non-trivial. 

\subsection{Spectral rigidity}

\begin{figure}
\centering
\includegraphicsbox[scale=0.5]{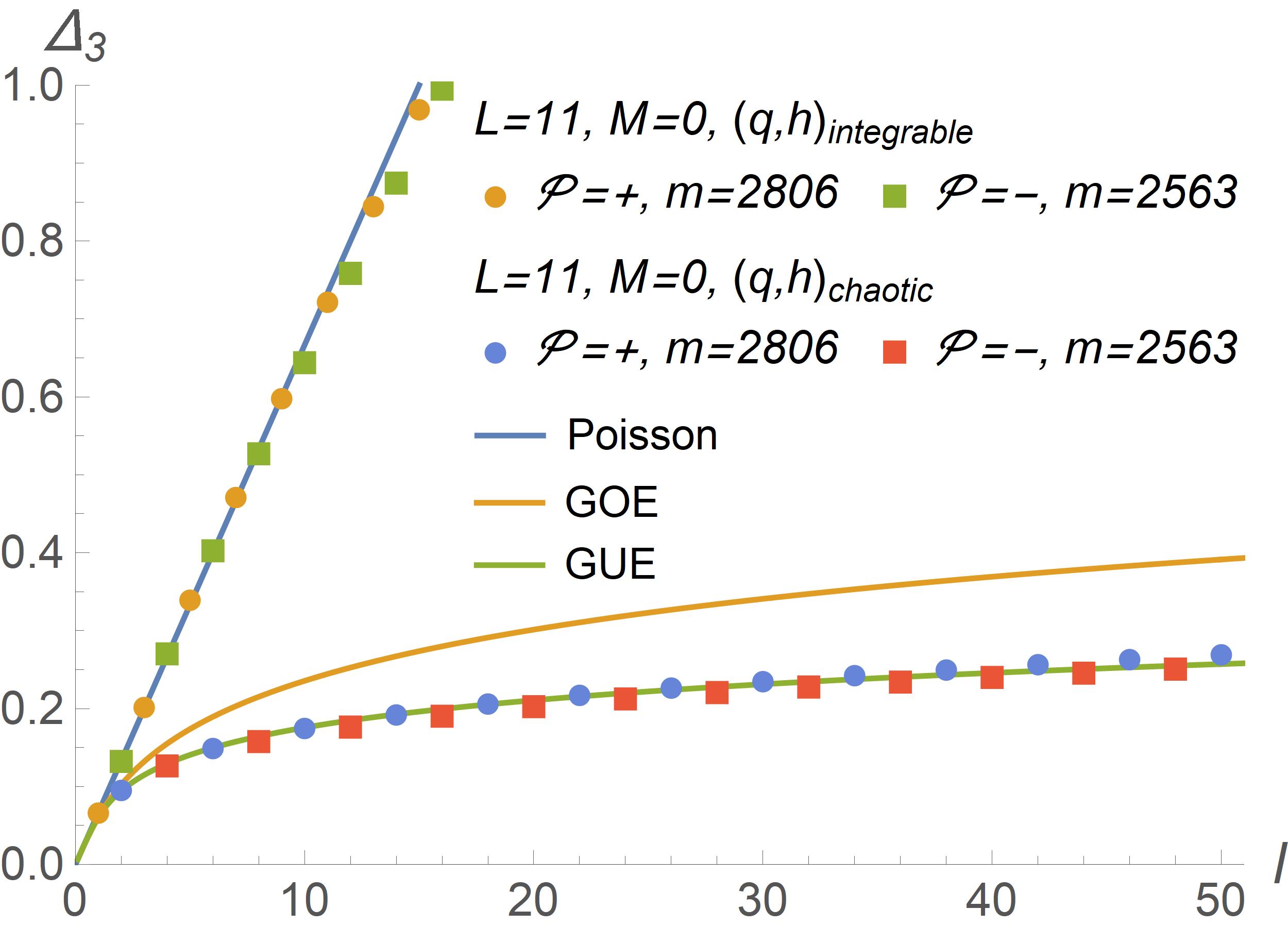}
\quad
\includegraphicsbox[scale=0.5]{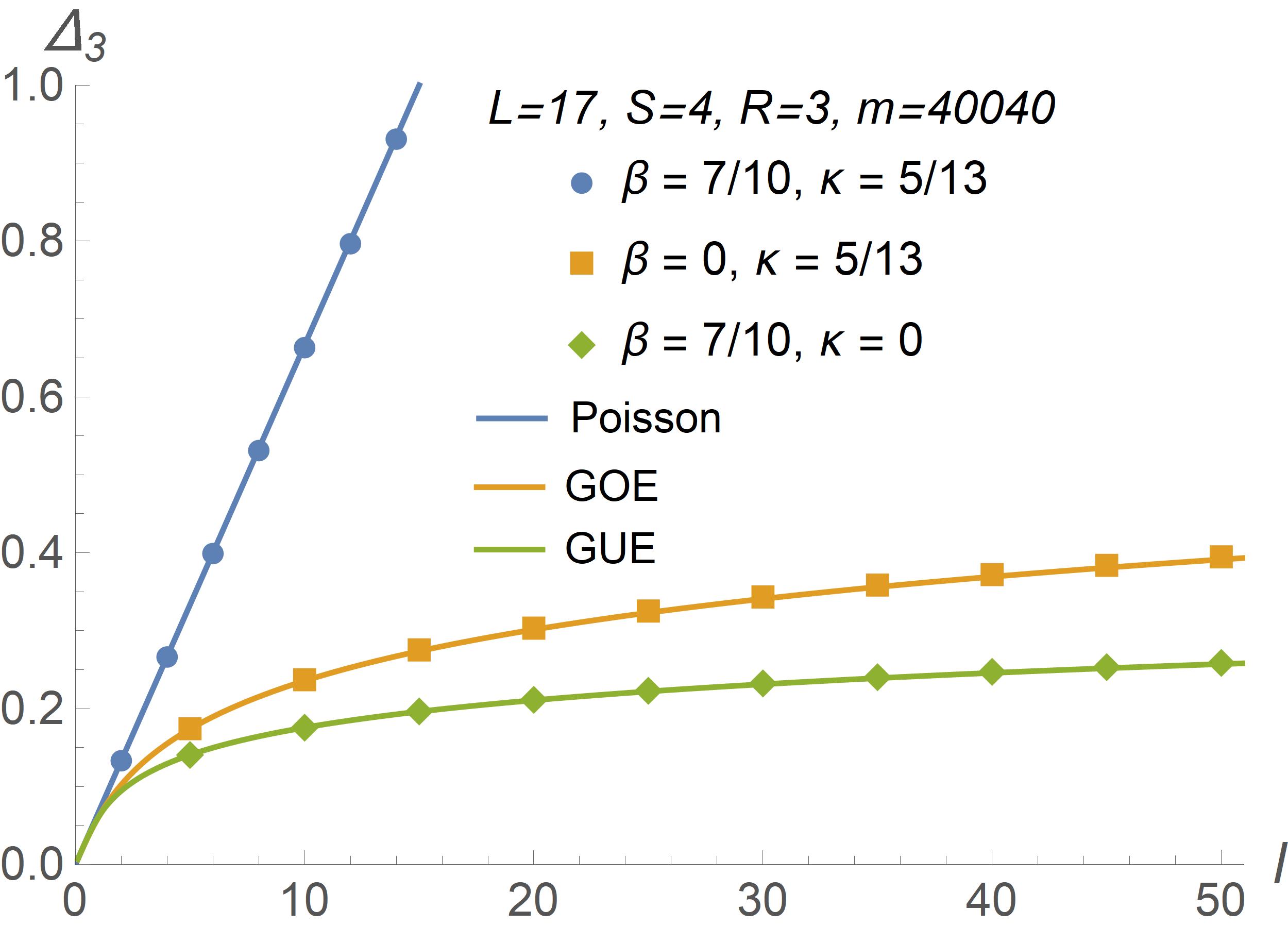}
\caption{Left: Dyson-Mehta statistic for $(L=11, M=0)$ spectra at an integrable and a chaotic point. Right: Dyson-Mehta statistic for $(L=17, R=4, S=3)$ and different configurations of $\beta$ and $\kappa$ (with $h=0$).}
\label{fig:SR}
\end{figure}

The different RMT Gaussian ensembles not only describe the nearest-neighbour correlations of the unfolded spin-chain spectra, but also long-range spectral properties and even the statistics of eigenvectors. In this section we consider the spectral rigidity which is a measure for the regularity, or rigidity, of a spectrum resulting from level repulsion. It can be measured by the Dyson-Mehta statistic $\Delta_3(l)$ \cite{DysonMehtaStatistic} defined as
\begin{align}
\Delta_3(l)=\frac 1l\Big\langle\min_{A,B}\int_{\varepsilon_0}^{\varepsilon_0+l}d\varepsilon(\hat n(\varepsilon)-A\varepsilon-B)^2\Big\rangle_{\varepsilon_0}\ .
\end{align}
Here $\hat n(\varepsilon)$ denotes the cumulative level number, cf.\ \eqref{eq:staircase}, of the unfolded and de-symmetrised spectrum, and the expression inside the angle brackets measures its least-square deviation from a straight line $A\varepsilon+B$ in an interval of length $l$. $\langle...\rangle_{\varepsilon_0}$  indicates the averaging over different interval starting points $\varepsilon_0$ taken from a discretisation of the unfolded spectral range, excluding a number of points at the boundaries. Increasing the interval length $l$ increases the number of probed levels and thus $\Delta_3$ is a measure for long-range correlations.
The solid lines in Fig.~\ref{fig:SR} show the expected behaviour of the Dyson-Mehta statistic for matrices in the Gaussian orthogonal and unitary ensemble, as well as for an uncorrelated spectrum, see \cite{mehta2004random} for further details and e.g.\ \cite{McLoughlin:2020zew} for an efficient way to perform the extreme value problem.

The Dyson-Mehta statistic is related to other statistical measures like the number variance and spectral form factor, and these relations can be made explicit by realising that all of these measures essentially probe the two-level correlations in a spectrum. For example the spectral form factor $g(t)$, which has been studied e.g.\ in the context of quantum chaos in the SYK model \cite{Cotler_2017}, is related to the Dyson-Mehta statistic via \cite{stockmann_1999}
\begin{align}
\Delta_3(l)=\frac 1\pi\int_0^\infty\frac{dt}{t^2}g(t)G(lt/2)
\end{align}
with
\begin{align}
G(x)=1-\left(\frac{\sin(x)}{x}\right)^2-3\left(\frac{x\cos(x)-\sin(x)}{x^2}\right)^2~.
\end{align}

In Fig.\ \ref{fig:SR}  we plot spectral-rigidity data for the $L=11$ and $L=17$ deformed SU$(3)$ spin chains for various parameter configurations $(q,h)$. At integrable points the Dyson-Mehta statistic follows the linear behaviour of uncorrelated spectra, reflecting the absence of level repulsion in these spectra. At non-integrable points it follows the logarithmic growth of the GOE and GUE prediction, respectively, according to the discrete symmetry properties of the spin-chain model. Together with the results for the short-range nearest-neighbour statistics from the previous section, these findings for the Dyson-Mehta statistics imply that fluctuations in spectra of the considered planar supersymmetric theories are consistently described by the Poisson or Wigner-Dyson RMT distribution according to their symmetry structure.

\subsection{Transition from integrable to chaotic dynamics}
We now examine the transition between integrable and chaotic statistics in the SU$(3)$ spin-chain model. 
In particular, it is integrable in the case where $h=0$ and $\kappa=0$, and transitions to a chaotic model as either $h$ or $\kappa$ become non-vanishing. We can thus view $h\neq 0$ and $\kappa \neq 0$ as perturbations about an integrable line parametrised by $\beta$. In the case of $h\neq 0$, the perturbation breaks the global symmetry which separately preserves the number of $\ket{0}$, $\ket{1}$ and $\ket{2}$ sites, and so we cannot smoothly turn on this deformation in the level-statistics analysis as the size of de-symmetrised subsectors changes. However for the case of $h=0$ and $\beta \neq 0$, the subspace remains the same as we turn on $\kappa$ and so we can study the transition from Poisson to Wigner-Dyson more clearly. 
  
\begin{figure}
	\centering
	$
	\includegraphicsbox[scale=0.5]{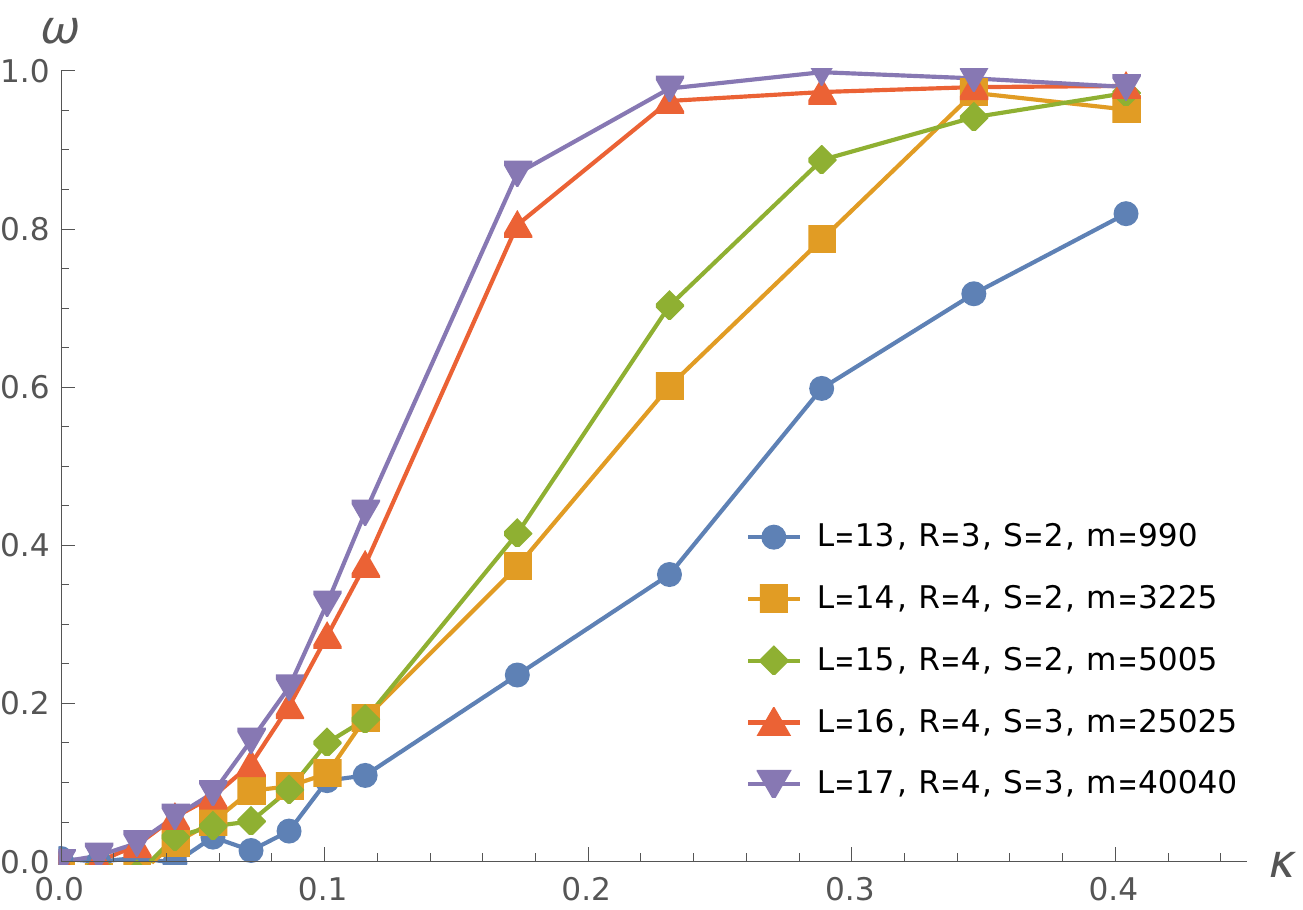}$ 
	$
	\includegraphicsbox[scale=0.5]{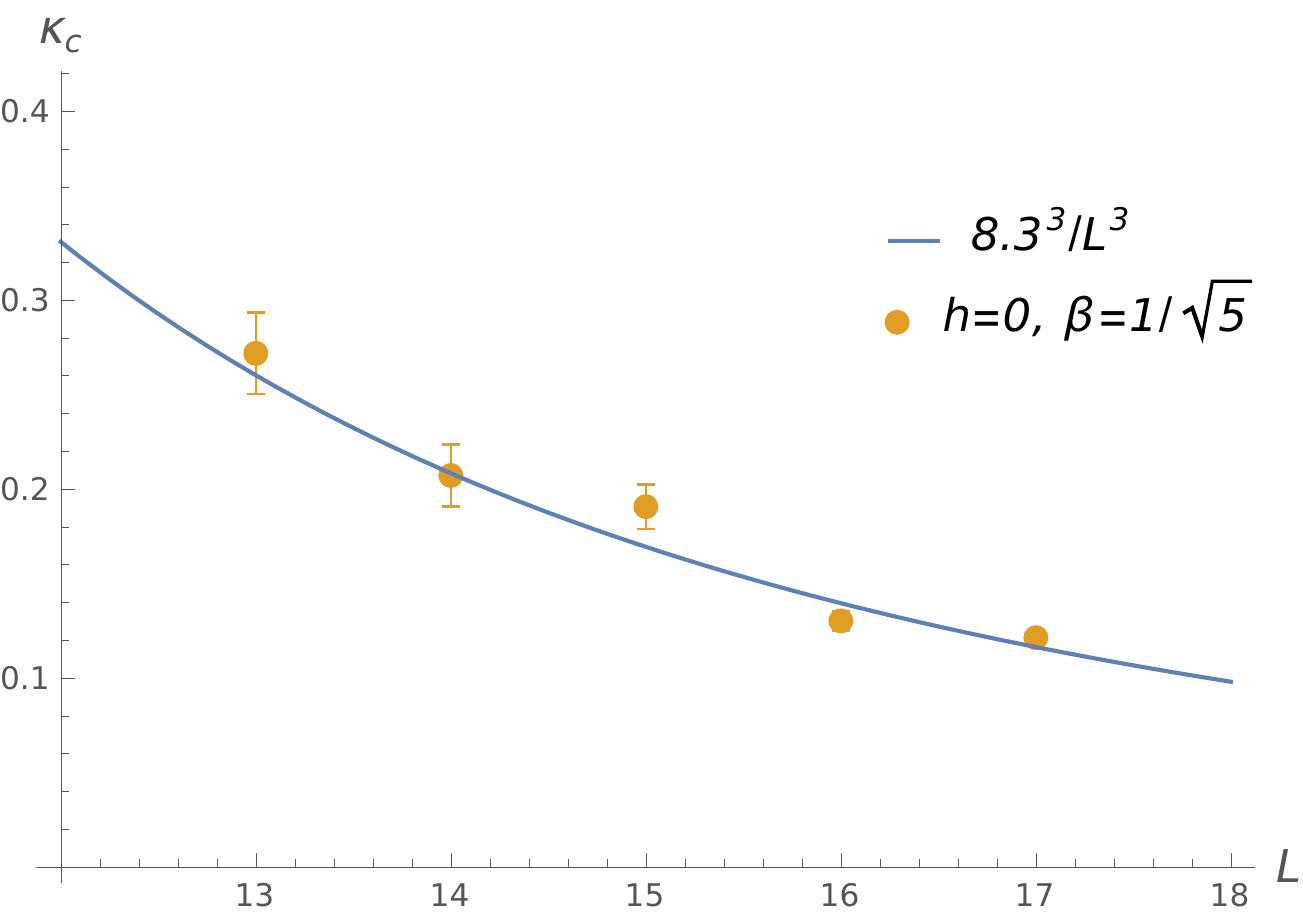}$ 
	\caption{Left: Best fit $\omega$ in Brody distribution of level spacings for $\beta=1/\sqrt{5}$, as $\kappa$ varies from 0 to $0.4$ for $L=13$ to $17$. Right: Cross-over coupling as a function of length $L$. The curve $\sim 8.4^3/L^3$ is shown for comparison. }
	\label{fig:su3h0b1oversqrt5}
\end{figure}

We study chains of length $L\in[13,17]$ and consider sectors with the excitation densities $R/L\sim 1/4$ and $S/L\sim 1/6$ held as close to constant as possible as we vary $L$. We take $\beta=1/\sqrt{5}$ so that the chaotic distribution is GUE. For a range of values of $\kappa \in[0,0.5]$ we compute the value of $\omega$ for which the Brody distribution best fits the data, cf.\ left panel in Fig.\  \ref{fig:su3h0b1oversqrt5}. For larger values of the length, and so larger values of the dimension of the subspace, the value of $\omega$ increases more rapidly toward the Wigner-Dyson value of 1 with increasing $\kappa$. Additionally, for larger $L$, i.e.\ $L=16$ and $L=17$, it can be seen that $\omega$ is greater than zero even for very small $\kappa$ and so there seems to be no region of integrable dynamics at finite $\kappa$. This is consistent with the hypothesis that at infinite $L$ the dynamics are chaotic for any finite value of the deformation. 

The curves of increasing $\omega$ can be fitted by the function $\tfrac{1}{2}(\tanh( a \kappa-b)+1)$ which can then be used to determine the cross-over parameter $\kappa_c$ for which $\omega=0.5$. The value of this parameter for different values of $L$ is shown in the right panel of Fig.\ \ref{fig:su3h0b1oversqrt5} where it can be seen to scale as $\sim 1/L^3$. Fitting the cross-over values to $\kappa_c=N_c/L^{b_c}$, we find the best fit is for 
\<
b_c=3.0 \pm 0.4~,~~~N_c=(6\pm 7)\times 10^2~
\>
 which suggests that the deformation is strongly chaotic and which is consistent with the fact that there is no known integrable completion of this deformed spin chain. 
 
\begin{figure}
 	\centering
 	$
 	\includegraphicsbox[scale=0.5]{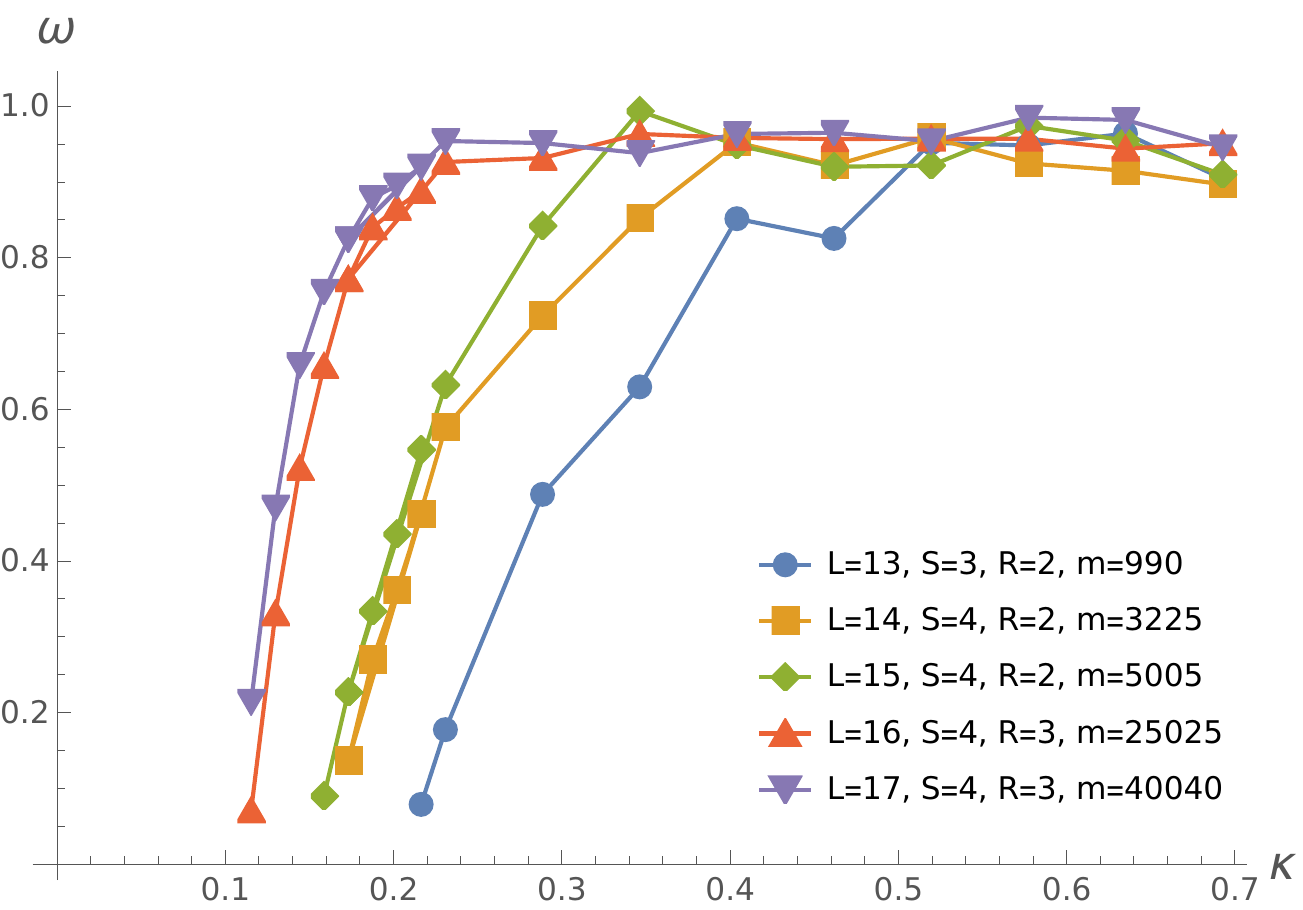}$ 
 	$
 	\includegraphicsbox[scale=0.5]{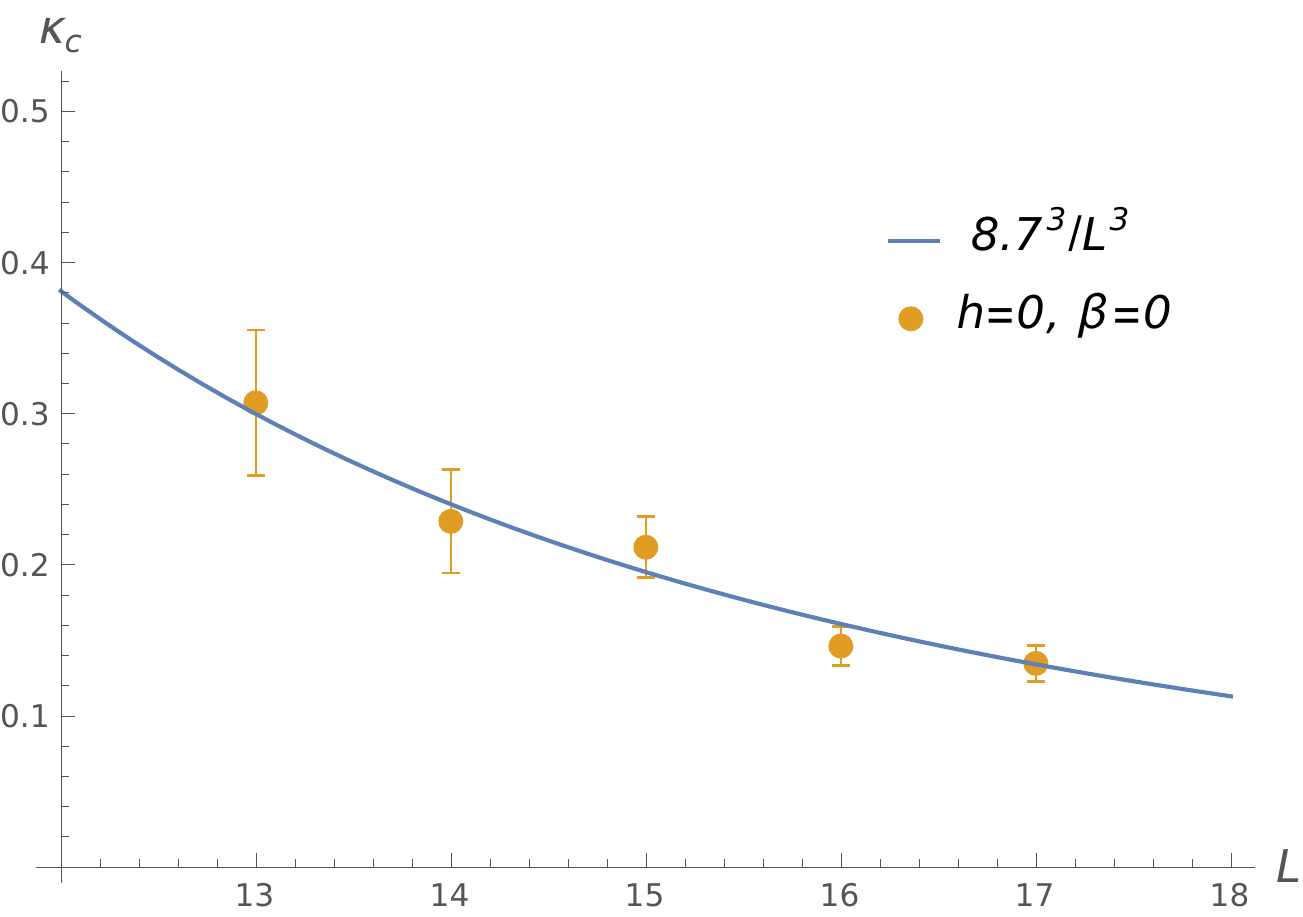}$ 
 	\caption{Left: Best fit $\omega$ in Brody distribution of level spacings for $\beta=0$, as $\kappa$ varies from 0 to $0.7$ for $L=13$ to $17$. Right: Cross-over coupling as a function of length $L$. The curve $\sim 1/L^3$ is shown for comparison. }
 	\label{fig:su3h0b0}
 \end{figure}

For later comparison with the Landau-Lifshitz model, it is also of interest to compute the cross-over for the case of $\beta=0$ where the chaotic dynamics are of GOE type. In this case, when $\kappa=0$, the global SU$(3)$ symmetries are restored and the limit is no longer smooth as the subsector splits up into smaller sectors. As a consequence, for small values of $\kappa$, the best fit for $\omega$ in the Brody distribution becomes negative. Nonetheless, we can examine the cross-over behaviour for a range of values $\kappa \in [0, 0.5]$, where we keep only those values for which $\omega>0$.  With this caveat, one can again see, cf.\ Fig.\ \ref{fig:su3h0b0}, that the onset of chaotic dynamics occurs earlier for larger $L$. The cross-over parameter is, as previously, computed by fitting it to a tanh-function and the scaling with $L$ is shown in the right panel of Fig.\  \ref{fig:su3h0b0}. The scaling is still $1/L^3$ with the best fit values being
 \< 
 b_c=3.2 \pm 0.5~,~~~N_c=(1.0\pm 1.3)\times 10^3~,
 \>
 indicating that the deformation is again strongly chaotic. 
 
These results imply that for any finite $\kappa$, or any $\kappa=\kappa_0/L^b$ with $\kappa_0$ constant and $b<3$,  the system will be chaotic as $L\to \infty$. In Sec.\ \ref{sec:LL} we will study the large $L$ limit with $b=1$ which gives rise to a Landau-Lifshitz model. In this limit we additionally focus on the low-energy states and one can ask whether the cross-over behaviour is different for these states. Although this is difficult to analyse in the spin chain, if we restrict to the low-energy states there is a clear increase in the cross-over coupling. For example, with $(L=17, R=4, S=3)$ and parameters $h=\beta=0$, the cross-over coupling computed with the highest and lowest deciles of energies removed is approximately $\kappa_c=0.13$, while if we consider just the lowest 6\% of energies the cross-over coupling is $\kappa_c=0.24$.

\section{Landau-Lifshitz Models}
\label{sec:LL}
In the context of the AdS/CFT correspondence, studying the long-wavelength limit about the ferromagnetic vacuum of the spin chain proved a useful approach in understanding the connection between the spectrum of planar anomalous dimensions and the string energies. The generalised Landau-Lifshitz  model, which emerges as an effective two-dimensional action in the limit of long chains, $L\to \infty$ with $\lambda/L^2$ fixed, was shown to agree with the string sigma model action expanded in the same limit \cite{Kruczenski:2003gt, Kruczenski:2004kw}.

Let us briefly review the derivation of the Landau-Lifshitz action from the continuum limit for  SU$(2)$ spin-$1/2$ chains, for details see e.g.\ \cite{Fradkin:1991nr}. The starting point is the definition of the coherent state for a single spin-$\tfrac{1}{2}$ particle as
\<
\ket{\vec{n}(\psi,\phi)}= e^{i \phi}\cos\psi \ket{1}+e^{-i \phi}\sin \psi \ket{2}~,
\>
where the vacuum, spin-up, state is denoted by $\ket{1}$ and the spin-down state is $\ket{2}$ (see \appref{app:CS} for more details on the construction of coherent states). For the full spin chain we can then define a coherent state as the tensor product over each lattice site
\<
\ket{\vec{n}}= \bigotimes_{i=1}^L\ket{\vec{n}(\psi_i,\phi_i)}~.
\>

 We now consider the time-evolution operator for a spin chain with Hamiltonian $H=\sum_{i=1}^L H_{i,i+1}$ and the transition amplitude $\bra{\Psi_a}e^{-i H T}\ket{\Psi_b}$ between two coherent states $\bra{\Psi_{a,b}}$. We begin by breaking the time interval $T$ into $N$ steps of duration $\delta$ and look at the limit $N\to \infty$ and $\delta\to 0$, with $N\delta=T$ kept fixed. At each intermediate time step $t_j$ we insert the resolution of the identity \eqref{eq:su2iden} so that 
\<
\bra{\Psi_a}e^{-i H T}\ket{\Psi_b}=\lim_{\substack{N\to \infty\\ \delta\to 0}}\left(\prod_{j=1}^N \int d\mu_2(t_j)\right)\left(\prod_{j=0}^N \bra{\vec{n}(t_j)}e^{-i \delta H}\ket{\vec{n}(t_{j+1})}\right)~,
\>
where we have made the identification $\ket{\Psi_a}=\ket{\vec{n}(t_0)}$ and $\ket{\Psi_b}=\ket{\vec{n}(t_{N+1})}$. Expanding in $\delta$ and using the identity
\<
\braket{\vec{n}(t_j)}{\vec{n}(t_{j+1})}=1+i \delta \sum_{i=1}^L  \cos (2\psi_i(t_j)) \dot \phi_i(t_j) +\mathcal{O}(\delta^2)~,
\>
 which follows from the definition of the scalar product \eqref{eq:su2prod} combined with the assumption that the trajectories are sufficiently smooth, we can write the transition function as
\<
\bra{\Psi_a}e^{-i H T}\ket{\Psi_b}&=&
\\
& &\kern-60pt
\lim_{\substack{N\to \infty\\ \delta\to 0}}\left(\prod_{j=1}^N \int d\mu_2(t_j)\right) \Big[1+i\delta \sum_{j=0}^N  \sum_{i=1}^L \big[\cos (2 \psi_i(t_j)) \dot{\phi}_i(t_j) - \bra{\vec{n}(t_j)} H_{i, i+1}\ket{\vec{n}(t_{j})}\big]+\mathcal{O}(\delta^2)\Big]\nn
\>
where $\dot \phi=\partial_t \phi$.
In the usual fashion this can be considered as defining a path integral over all trajectories of the vector $\vec{n}$ on the unit two-sphere with appropriate boundary conditions
\<
\bra{\Psi_a}e^{-i H T}\ket{\Psi_b}&=&\int \mathcal{D}n~ e^{i S[\vec{n}]}~,
\>
where we have introduced the action 
\<
\label{eq:Action1}
S[\vec{n}]=\sum_{i=1}^L\int_0^T  dt  \big( \cos (2 \psi_i) \dot{ \phi}_i- \bra{\vec{n}(t)}H_{i,i+1}\ket{\vec{n}(t)}\big)~.
\>
The first term is the topological Wess-Zumino term that follows from the definition of the coherent states, while the second encodes the specific dynamics of the system. 
For the spin chain with 
\<
H_{i,i+1}=\frac{\lambda}{16\pi^2} (\unit_{i,i+1}-\vec{\sigma}_i \cdot \vec{\sigma}_{i+1})~,
\>
corresponding to the undeformed planar SU$(2)$ one-loop dilatation operator where we re-introduced the overall coupling $\frac{\lambda}{16\pi^2}$,
this second term is given by 
\<
 \sum_{i=1}^L\int_0^T dt ~\bra{\vec{n}(t)}H_{i,i+1}\ket{\vec{n}(t)}=\frac{\lambda}{16\pi^2}\sum_{i=1}^L\int_0^T dt~(1-\vec{n}_i\cdot \vec{n}_{i+1})
 \>
 where $\vec{n}_i=(\cos 2\phi_i \sin 2\psi_i, \sin 2\phi_i \sin 2\psi_i, \cos 2 \psi_i)$ so that $\vec{n}_i^2=1$. 
If we now make the additional assumption that the states are such that $\vec{n}_i$ varies smoothly as a function of the lattice position, then we can replace the discrete index $i=1, \dots, L$ with a continuous parameter $\sigma\in [0,1]$ in the limit of the large $L$. 
Expanding the lattice fields so that
 \<
 \sum_{i=1}^L~(1-\vec{n}_i\cdot \vec{n}_{i+1})=\frac{2}{L}\int_0^1 d\sigma ~ \big[(\acute{\psi}^2+\sin^2 (2\psi) \acute{\phi}^2)+\mathcal{O}(1/L)\big]~,
 \>
 where the derivative with respect to $\sigma$ is denoted by the primed fields, and taking the same continuum limit for the first term in \eqref{eq:Action1}, we find the  Landau-Lifshitz action
\<
\label{eq:LL_udef}
S= L \int d\sigma dt~ \cos( 2 \psi) \dot{\phi}-\frac{\lambda}{8\pi^2 L} \int d\sigma dt (\acute{\psi}^2+\sin^2(2\psi) \acute{\phi}^2)~.
\>
The fields $(\phi(t,\sigma),\psi(t,\sigma))$, or equivalently $\vec{n}(t,\sigma)$, can be viewed as describing the long wavelength modes of the spin chain about the ferromagnetic vacuum which corresponds to setting $\vec{n}(t,\sigma)$ to a constant vector. 
Holding $\lambda/L^2$ fixed while taking $L\to \infty$, we see that this limit corresponds to a semi-classical limit of the spin chain where classical solutions describe long spin-chain energy eigenstates. 
Restricting to closed periodic spin chains corresponds to imposing the boundary conditions
\<
\phi(\sigma=1,t)=\phi(\sigma=0,t)~~~\text{and}~~~ \psi(\sigma=1,t)=\psi(\sigma=0,t)
\>
on the coherent state. 
Furthermore, to restrict to zero-momentum spin chains, we impose the condition $P=0$ with the generator of world-sheet translations given by
\<
P=\int d\sigma (p_\psi \acute{\psi}+p_\phi \acute{\phi})= L\int d\sigma \cos (2\psi) \acute{\phi}~.
\>
. 
\subsection{Higher-derivative LL model}

At higher loop orders one can map the planar dilatation operator to a spin chain involving longer-range interactions. A description in terms of long-wavelength modes is still possible, but the effective action includes terms with higher numbers of derivatives \cite{Kruczenski:2004kw}. For a spin chain with SU$(2)$ symmetry and next-to-nearest-neighbour interactions, the effective action is fixed by symmetries to be of the form
\<
\label{eq:hdea}
S=L \int dt \int_0^{2\pi} \frac{d\sigma}{2\pi}  (\cos(2\psi) \dot{\phi}-H[\vec{n}])~,~~~\text{with}~~~H[\vec{n}]=a_0 (\partial_\sigma \vec{n})^2+a_1 (\partial^2_\sigma \vec{n})^2+a_2(\partial_\sigma \vec{n})^4~,
\>
where we have re-scaled $\sigma$ by $2\pi$ compared to the previous section.
The Wess-Zumino term is fixed by kinematic considerations and so is the same for all spin-1/2 chains, while the coefficients $a_0, a_1, a_2$
depend on the detailed structure of the Hamiltonian. The derivation of the effective action is in general more complicated than simply taking the continuum limit as above, and one must carefully include the quantum corrections before taking the continuum limit \cite{Kruczenski:2004kw}. For the next-to-nearest neighbour Hamiltonian \eqref{eq:NNNHam} with the choice of parameters corresponding to the two-loop planar dilatation operator, the coefficients were computed in \cite{Kruczenski:2004kw} and found to be\footnote{The time coordinate is rescaled here compared to \eqref{eq:LL_udef} by $\frac{\lambda}{4\pi^2L}$.}
\<
\label{eq:LL2l}
a_0=\frac{1}{8}~,~~~a_1=-\frac{\lambda}{32L^2}~,~~~a_2=\frac{3\lambda}{128L^2}~.
\>
Again we only keep the leading term in the limit $L\to \infty$ while holding $\lambda/L^2$ fixed.

The equations of motion following from \eqref{eq:hdea} can be written as
\<
\label{eq:LLhdeom}
\partial_t \vec{n}=2 a_0~\vec{n}\times \partial_\sigma^2 \vec{n} -2 a_1~\vec{n}\times \partial^4_\sigma \vec{n}+4a_2 \big[(\vec{n}\times \partial_\sigma^2 \vec{n})(\partial_\sigma \vec{n})^2+2 (\vec{n}\times \partial_\sigma \vec{n})(\partial_\sigma \vec{n} \cdot \partial_\sigma^2 \vec{n})\big]~,
\>
where we have also expressed the contribution from the Wess-Zumino term using $\vec{n}$.  
We are interested in understanding the dynamics of this model and in particular the presence (or absence) of classical dynamics analogous to the quantum chaos found in the spin-chain system. To this end we wish to explore different regions of the space of solutions and, as a complete categorisation of solutions for this non-linear field theory is not feasible, we make a simplifying ansatz
\<
\label{eq:LLHD1ph}
\phi(\sigma, t)= \omega t~,~~~\psi(\sigma, t)=\tfrac{1}{2}q(\sigma)~,
\>
where $\omega$ is a constant. Note that in this section $q$ corresponds to a phase-space coordinate and not the Leigh-Strassler parameter. The ansatz \eqref{eq:LLHD1ph} automatically satisfies the zero-momentum constraint $P=0$. It also includes the constant vacuum solution $\psi=$ const. and the LL limit of the folded spinning string solution considered in e.g.\ \cite{Kruczenski:2003gt, Kruczenski:2004cn} and thus we refer to it as the spinning string ansatz. 
The equations of motion imply that 
\<
\label{eq:HDeom}
a_0 q^{(2)} +6(a_1+a_2) (q^{(1)})^2q^{(2)}-a_1 q^{(4)}-\omega \sin q=0
\>
where $q^{(n)}$ is the $n$-th derivative of $q$.

If we first consider the case of the undeformed LL-model, where $\lambda=0$, the equation is essentially that of an exact pendulum and can be explicitly solved in terms of elliptic integrals. As described in \cite{Kruczenski:2003gt}, there are essentially two classes of solutions depending on the initial conditions: those where the angle $\psi$ oscillates over some finite interval and those where the angle increases without limit.
\begin{figure}
	\centering
	$
	\includegraphicsbox[scale=0.4]{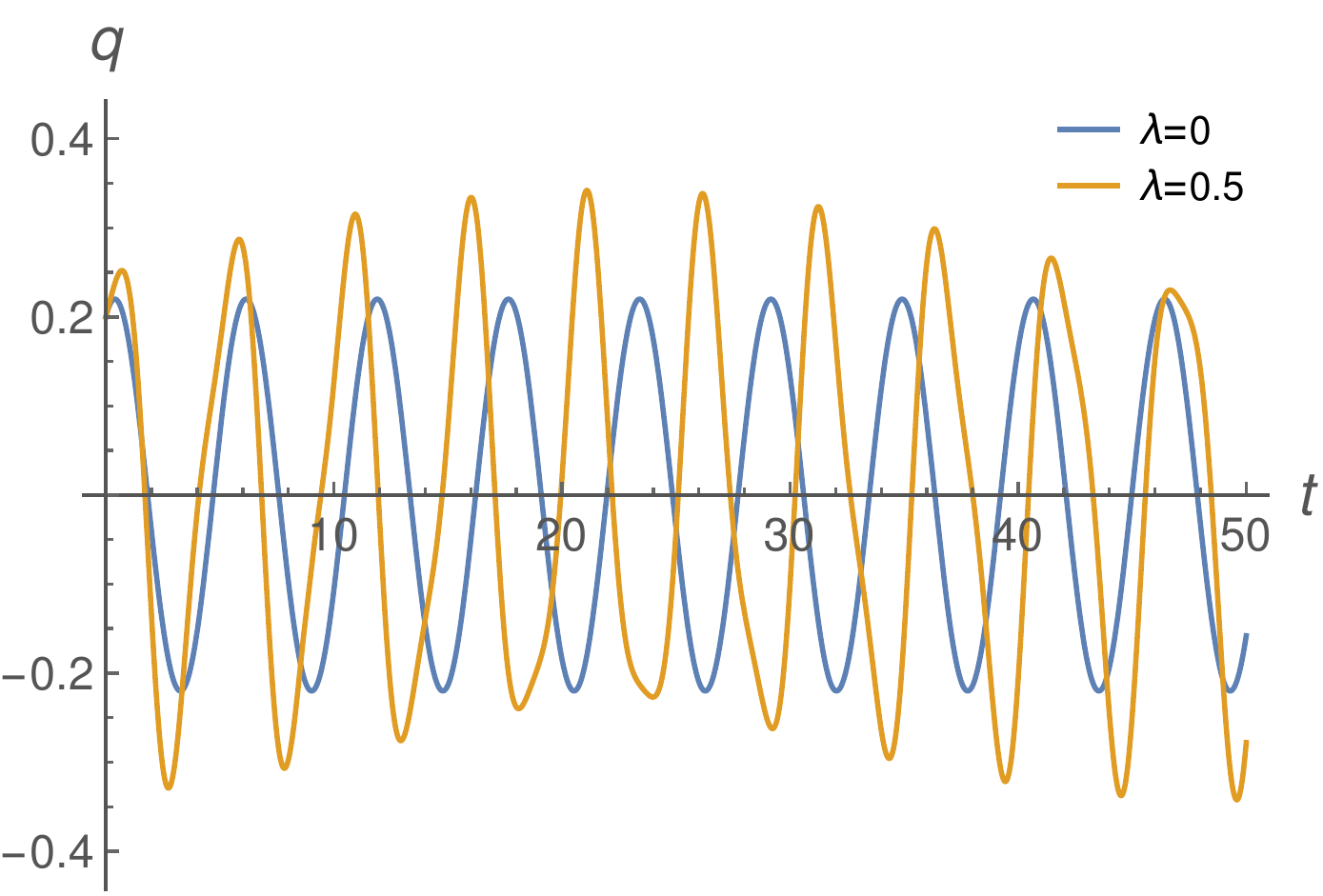}$ ~~~\includegraphicsbox[scale=0.45]{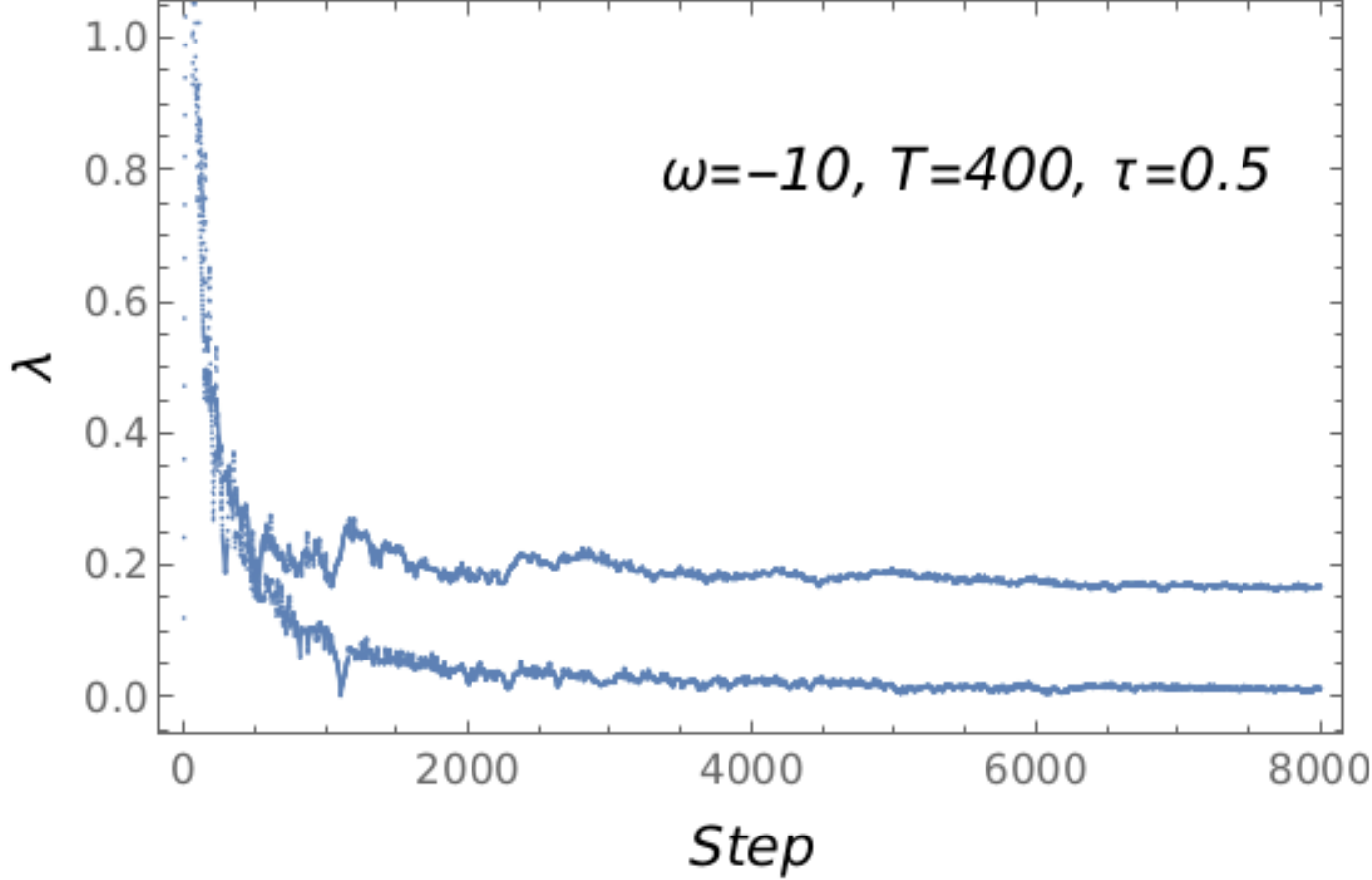}
	\caption{Left: Spinning string ansatz for undeformed, $\lambda=0$, and deformed, $\lambda=1/2$, LL model. We plot the solution with $\omega=-0.15$, initial conditions $q(0)=0.2$, $q^{(1)}(0)=0.1$ (and $q^{(2)}(0)=0$, $q^{(3)}(0)=0$ in the deformed case). Right: Plot of the two largest LCEs for LL model with parameters (\ref{eq:hdLL_ch}) and $\omega=-10$.
		The initial conditions are  $q(0)=0=p_0(0)=p_1(0)$, $q^{(1)}(0)=\tfrac{1}{10}$. 
		The total evolution time is $T=400$ which is broken into $8000$ iterative steps. }
	\label{fig:LLmotion}
\end{figure}
When we include the higher-derivative corrections, the equations are non-linear and we must generally resort to numerical methods. To gain some insight we can set $\omega$ to zero, integrate \eqref{eq:HDeom} and introduce the variable $u=q^{(1)}$ which satisfies
\<
\ddot{u}=\frac{a_0}{a_1} u+2\left(1 +\frac{a_2}{a_1}\right) u^3+c~
\>
where $\dot u$ denotes the derivative w.r.t. $\sigma$ and $c$ is an integration constant. 
This is simply the equation for a particle moving in a double-well potential combined with a constant force term given by $c$.  

For the choice of parameters corresponding to the two-loop dilatation operator \eqref{eq:LL2l}, the potential well is inverted and so one again finds both unbounded and bounded motion. To this we must add the $\sin q$-driving term which, depending on the sign of $\omega$, will either be a restoring force or tend to drive the particle away from the $q$-origin. If the initial conditions are such that the particle is close to the $u$-origin, with small initial $u$-velocity and acceleration, and the sign of $\omega$ is such that the driving $\sin q$-term provides a restoring force, the motion will be bound. Alternatively for large initial values or when the driving force is sufficiently large, the motion appears unbounded though numerical issues make following the trajectory difficult. For initial conditions on $q$, $q^{(1)}$, $q^{(2)}$  and $q^{(3)}$ giving bound motion, we plot the time evolution of $q(t)$ in both the undeformed and deformed theories, cf.\ left panel in \figref{fig:LLmotion}. This illustrates that in the theory deformed by the two-loop higher-derivative terms, the motion is somewhat more irregular than the undeformed case.

To obtain a more quantitative characterisation, it is useful to switch to a Hamiltonian formulation of the reduced dynamics. Substituting the ansatz \eqref{eq:LLHD1ph} into the action yields the Lagrangian 
\<
\mathcal{L}=-a_1 (\dot{q}^{(1)})^2-a_0 (q^{(1)})^2-(a_1+a_2)(q^{(1)})^4+ \omega \cos q~.
\>
Following the general procedure of Ostrogradsky (briefly reviewed in App.~\ref{app:OstHD}), the phase-space variables are $q$, $q^{(1)}$, $p_0$ and $p_1$, and we define the Hamiltonian 
\<
\label{eq:Ham_hdLL}
H=q^{(1)} p_0-\frac{p_1^2}{4 a_1}+a_0 (q^{(1)})^2+(a_1+a_2)(q^{(1)})^4-\omega \cos q~.
\>
It is easy to check that the resulting Hamiltonian equations of motion are equivalent to those of the original model \eqref{eq:LLhdeom} upon substitution of the ansatz \eqref{eq:LLHD1ph}. Note that while the reduced theory is one-dimensional, due to the higher derivatives the phase space is four-dimensional and so the theory is not trivially integrable and there is a possibility of chaotic motion. 

One quantitative approach to chaotic dynamics is to compute the Lyapunov characteristic exponents (LCEs)  \cite{oseledec1968multiplicative, benettin1976kolmogorov, benettin1980lyapunov} (see \cite{skokos2010lyapunov} for a review and further references) which measure the rate of change of the separation between initially nearby trajectories. For chaotic motion at least one LCE is positive, while for regular motion all LCEs are zero and so the maximal LCE (mLCE) is an indicator of the chaotic nature of orbits in a dynamical system. More specifically for an autonomous Hamiltonian system the spectrum of LCEs of order 1, $\{\lambda_i\}$ with $i=1,\dots, 2N$, where $2N$ is the dimension of phase space, and ordered so that $\lambda_1 \geq \lambda_2\geq  \dots \geq \lambda_{2N}$,  consist of pairs having opposite signs, $\lambda_j=-\lambda_{2N-j+1}$ for $j=1, \dots, N$, with at least two LCEs vanishing. For each additional independent integral of motion, one more pair of LCEs will vanish. There are well-established methods for numerically evaluating the LCEs of a given dynamical system including an implementation in Mathematica \cite{sandri1996numerical} which we followed closely (see \appref{app:numLCE} for additional details of our numerical methods for computing LCEs). Taking $q^{(1)}(0)\in[0,1]$ and $\omega=-1/10$ we find that the mLCE is less than $0.01$, and similarly fixing $q^{(1)}(0)=0.1$ and letting $\omega\in[-0.2,0.03]$ we find that the mLCE is less than $0.08$. Such values are relatively small and suggest the motion of the higher-derivative theory corresponding to the two-loop dilatation operator is not chaotic. 
 
One might worry that this contradicts the finite-$L$ level statistics, however the lack of chaotic motion should perhaps not be too surprising. In similar systems, e.g.\ the Duffing oscillator or double pendulum, the motion is regular at low energies and only becomes chaotic when the energy is sufficiently large. In the LL model following from the spin chain, we cannot increase the energy too much without the motion becoming unbound/unstable. This is also not unexpected from the spin-chain description as the LL model corresponds to taking $\tilde{g}^2=g^2/L^2$ fixed while $L$ is large. The ratio of the next-to-nearest-neighbour to nearest-neighbour terms \eqref{eq:effgc} becomes in this limit 
\<
\delta=\frac{\tilde{g}^2L^2}{1 -4 \tilde{g}^2L^2}\to-\frac{1}{4} ~,
\>
a value for which the spectral statistics are intermediate between Wigner-Dyson and Poisson. 
Relatedly, the S-matrix of the deformed LL theory computed about the $q=0$ vacuum is known to factorise \cite{Gerotto:2017sat}. However it should be noted that this calculation was perturbative in $g$ and $q$, and so is strictly  valid only near $g^2=0$ and $q=0$.  

For comparison, we can study the dynamics of the deformed LL model for values of the parameters $a_0, a_1$ and $a_2$ where, in the $\omega=0$ limit, the double-well potential in the $u$-variable results in bound motion for arbitrary initial conditions. For example, we can repeat the phase-space analysis with 
\<
\label{eq:hdLL_ch}
a_0=a_1=-1~~~ \text{and}~~~ a_2=3/2~.
\>
 In this case the dynamics appear to be chaotic as can be seen in the mLCE which, for $\omega=-10$ and initial conditions $q(0)=0=p_0(0)=p_1(0)$, as well as $q^{(1)}=\tfrac{1}{10}$, is found to be $\lambda_1=0.17$. In \figref{fig:LLmotion} we show the numerical evaluation of the two largest LCEs for times up to $T=400$. There is one non-zero value and one vanishing LCE as is expected for an autonomous Hamiltonian system with a four-dimensional phase space. The two remaining LCEs are simply, to within numerical accuracy, the negatives of those shown in \figref{fig:LLmotion}. Thus we see that the higher-derivative model is chaotic for generic parameters $a_0,a_1,a_2$, but not for the specific values \eqref{eq:hdLL_ch} corresponding to the two-loop dilatation operator which is compatible with the weak chaos seen in the level statistics. 

\subsection{SU(3) LL model}

We now turn to consider the deformed one-loop SU$(3)$ sector described in \secref{sec:su3}, where the Hamiltonian density is given in \eqref{eq:Hsu3} and where the coherent state is 
\<
\ket{n}=\cos \theta  e^{i \phi} \ket{1}+\cos \psi \sin \theta e^{ i \varphi} \ket{2} +\sin \psi \sin \theta e^{-i \varphi} \ket{3}~,
\>
cf.\ \eqref{eq:su3cs}. For simplicity we consider only the case with $h=0$ and $q=\exp(i \beta+\kappa)$, and when taking the $L\to \infty$ limit we keep $\tilde{\beta}=\beta L$ and $\tilde{\kappa}=\kappa L$ fixed (see \cite{Frolov:2005ty} for a discussion in the SU$(2)$ case). 
Repeating the procedure described above we find for the action
\<
S=L\int dt \int ^{2\pi}_0 \frac{d\sigma}{2\pi}(\cos^2 \theta~ \dot \phi+\cos 2 \psi \sin^2 \theta~ \dot \varphi-H[\vec{n}])
\>
with 
\begin{align}
\label{eq:def_su3_ham}
H[\vec{n}]=&~(\partial_\sigma \theta-\tfrac{\tilde{\kappa}}{2}\cos 2 \psi \sin 2 \theta)^2\nn\\
 &
+\sin^2 \theta \big[(\partial_\sigma \psi+\tfrac{\tilde{\kappa}}{4}(1+3 \cos 2\theta \sin 2 \psi)^2+ \cos^2 \theta (\partial_\sigma \phi - \cos 2 \psi (\partial_\sigma \varphi+\tilde{\beta}))^2\nn\\
 &+ \sin^2 2 \psi (\partial_\sigma \varphi+\tfrac{\tilde{\beta}}{4}(1+3 \cos 2 \theta))^2\big]+\tfrac{9}{4}(\tilde{\beta}^2+\tilde{\kappa}^2)\cos^2 \theta \sin^4 \theta \sin^2 2 \psi~.
\end{align}
The undeformed version of this action was studied in \cite{Hernandez:2004uw, Stefanski:2004cw}, while the deformed version was studied in \cite{Frolov:2005iq}. 
If we introduce the variables
\begin{align}
&\rho_1= \cos \theta~,&&\rho_2=\sin \theta \sin \psi~,&&\rho_3= \sin \theta \cos \psi\nn\\
&\phi_1 =\phi~,&&\phi_2=\varphi~,&&\phi_3=\varphi~,
\end{align}
we can write the Hamiltonian as 
\<
H=\sum_{i=1}^3 (\epsilon_{ijk} \rho_i \partial_\sigma \rho_j-\tilde \kappa |\epsilon_{ijk}| \rho_i \rho_j)^2+\sum_{i<j}(\partial_\sigma \phi_i-\partial_\sigma \phi_j -\epsilon_{ijk} \tilde \beta) \rho_i^2 \rho_j^2~.
\>
This is not quite the deformed action found in \cite{Frolov:2005iq} as it is missing a term
\<
H_{sextic}=-9(\tilde \beta^2+\tilde \kappa^2)\rho_1^2 \rho_2^2 \rho_3^2
\>
which is a result of the coherent state missing the vacuum configuration consisting of totally symmetrised products of all three spin-chain states. As a consequence of this, the Landau-Lifshitz action does not match the ``fast-motion'' limit of the world-sheet theory for strings in the deformed AdS geometry. This limitation can in principle be remedied by either considering generalised coherent states or by making an appropriate change of basis \cite{Frolov:2005iq}. 

In order to study the dynamics of the model, we again restrict to a specific class of solutions which effectively reduce the model to one dimension. We start from the ansatz
\begin{align}
\label{eq:spinanz}
&\phi(\sigma, t)=w_\phi t~,&&\varphi(\sigma, t)=w_\varphi t~,&&
\theta(\sigma, t)=\theta(\sigma)~,&&\psi(\sigma, t)=\psi(\sigma)
\end{align}
which implies that the continuum limit of the spin-chain momentum vanishes
\begin{align}
P=&~\int d\sigma \big[p_\theta \partial_\sigma \theta+p_\psi \partial_\sigma \psi+p_\phi \partial_\sigma \phi+ p_\varphi \partial_\sigma \varphi\big]\nn\\
=&~ L \int \frac{d\sigma}{2\pi} \big[\cos^2 \theta \partial_\sigma \phi+\cos 2 \psi \sin^2 \theta \partial_\sigma \varphi\big]=0 
\end{align}
and so corresponds to a zero-momentum spin-chain state. 
Starting with the undeformed theory, $\tilde \beta=\tilde \kappa=0$, and substituting the ansatz into the equations of motion, one sees that the $\phi$ and $\varphi$ equations are automatically satisfied while those of $\theta$ and $\psi$ follow from the reduced Hamiltonian
\<
H^{(0)}_{red}=p_\theta^2+ p_\psi^2\csc^2 \theta -4 w_\phi \cos^2 \theta-4 w_{\varphi} \cos 2 \psi \sin^2 \theta~.
\>
More generally, in the case $\tilde \beta=0$, $\tilde \kappa \neq 0$ the equations of motion follow from the reduced Hamiltonian with the additional terms
\<
H_{\tilde \kappa}=\tilde \kappa \big[
-2p_\theta \sin 2\theta \cos 2 \psi 
	 + p_\psi \sin 2 \psi (3 \cos 2 \theta+1)
	  +
9 \tilde \kappa (\alpha-1) \sin ^4\theta \cos ^2\theta \sin ^2 2 \psi \big]~,
\>
where setting the coefficient $\alpha=0$ corresponds to the naive coherent-state limit of the spin chain and $\alpha=1$ corresponds to adding the contribution $H_{sextic}$. In general, the $\phi$ and $\varphi$ equations are not satisfied by the ansatz \eqref{eq:spinanz} when $\tilde \beta\neq 0$ so the equations of motion do not reduce to a one-dimensional Hamiltonian problem in the same way and we will not consider this case further. 

The reduced dynamics, in both the undeformed and deformed cases considered, are thus given by a Hamiltonian system with a four-dimensional phase space. We can compute the Lyapunov exponents for different parameter values and initial conditions. 

\begin{figure}
	\centering
	\includegraphicsbox[scale=0.75]{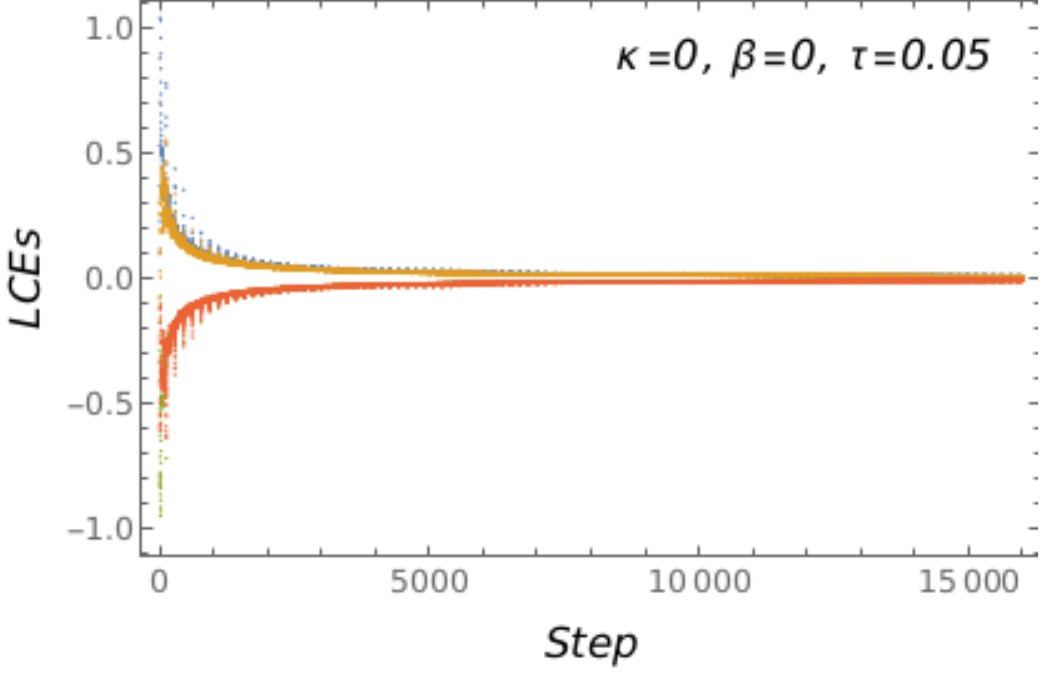}
	\caption{Plot of the full LCE spectrum for the undeformed SU$(3)$ Landau-Lifshitz model. 
		The initial conditions are  $p_\theta(0)=3/10$, $p_\psi(0)=11/10$, $\theta(0)=1$, $\psi(0)=1$. 
		The total evolution time is $T=800$ which is broken into $16000$ iterative steps.  }
	\label{fig:LLSU3LCEundef}
\end{figure}

\paragraph{Undeformed Model}
Starting with the undeformed theory and taking, for example, the initial conditions 
\<
\label{eq:su3_init}
p_\theta(0)=3/10, ~~p_\psi(0)=11/10, ~~\theta(0)=1, ~~\psi(0)=1
\>
 with $w_\phi=w_\varphi=1$ and numerically integrating for $T=1600$, we find that the mLCE is less than $0.005$. Similar values were found for different initial conditions and values of $w_\phi$ and $w_\varphi$ see \tabref{table:mLCESU3un}. That all such calculations can be done for such long times at relatively low numerical precision (only 30 digits) is another signature of the non-chaotic nature of the system. Using the iterated method to calculate the full LCE spectrum, we similarly find that
all exponents rapidly converge to zero, cf.\ \figref{fig:LLSU3LCEundef}.
\begin{table}
	\centering
	\begin{tabular}{|c c c |} 
		\hline
		$w_\phi$  & $w_\varphi$ & mLCE \\ [0.5ex] 
		\hline\hline
		1 & 1 & 0.0042 \\ 
		\hline
		-1 & 1 & 0.0044 \\ 
		\hline
		1 & -1 & 0.0035 \\ 
		\hline
		2 & 1 & 0.0045 \\
		\hline
		1 & 2 & 0.0048 \\
		\hline
		3/4 & 7/11 & 0.0045 \\
		\hline
	\end{tabular}
	\caption{mLCE for $\tilde \kappa=\tilde \beta=0$ integrated for $T=1600$ with initial conditions $p_\theta(0)=3/10$, $p_\psi(0)=11/10$, $\theta(0)=1$, $\psi(0)=1$.}
	\label{table:mLCESU3un}
\end{table}

\begin{figure}
	\centering
	\<
	\includegraphicsbox[scale=0.65]{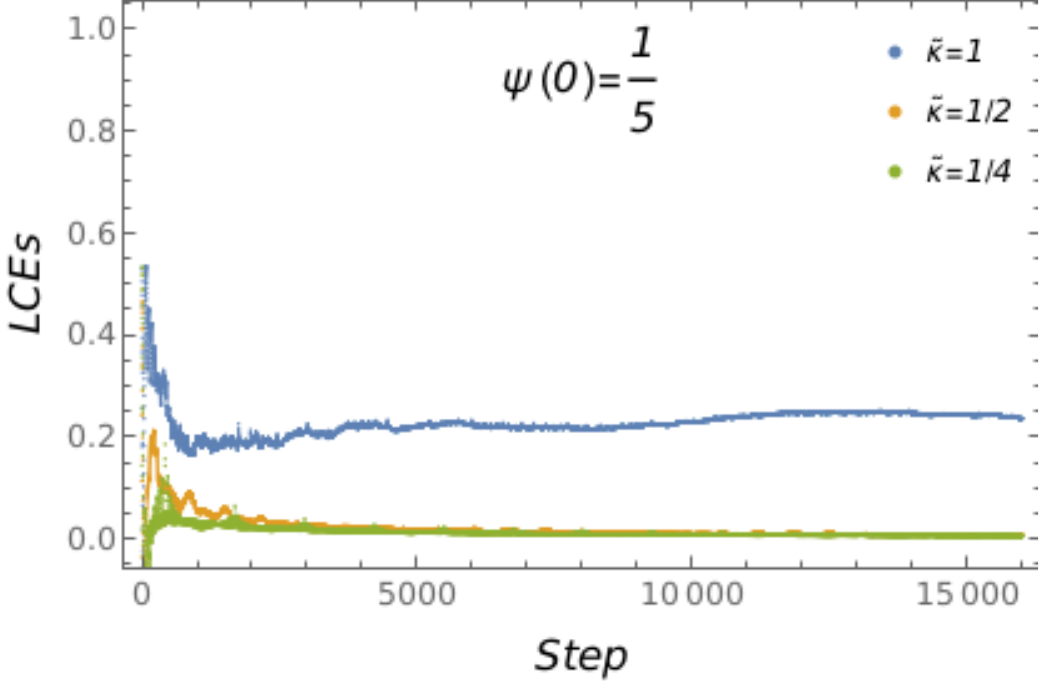}& & 	\includegraphicsbox[scale=0.65]{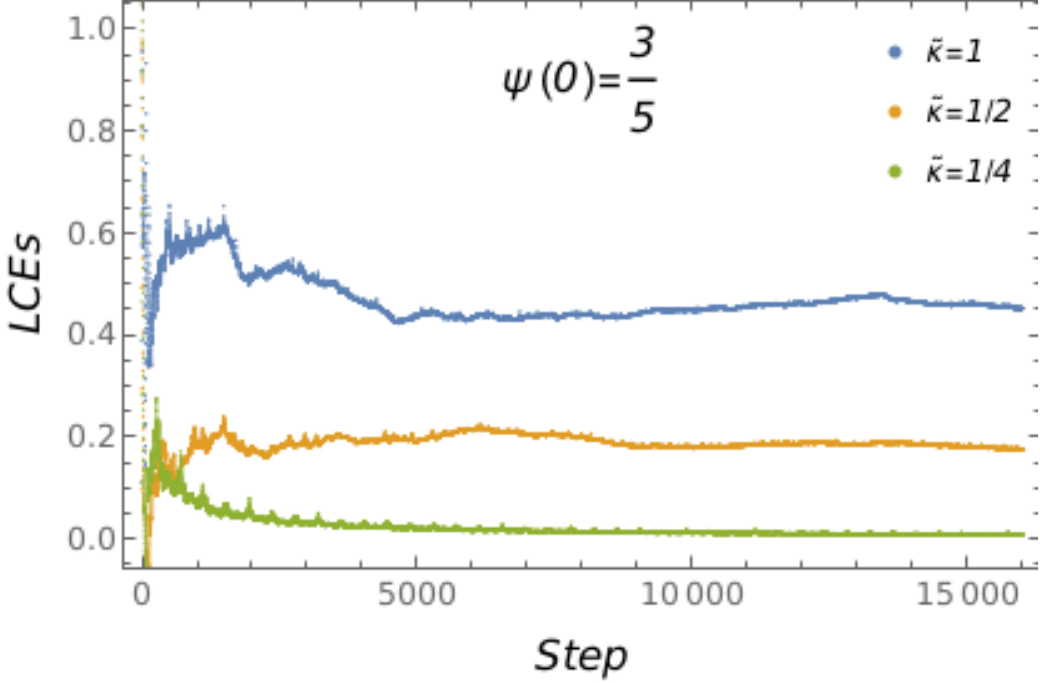}\nn \\
	\includegraphicsbox[scale=0.65]{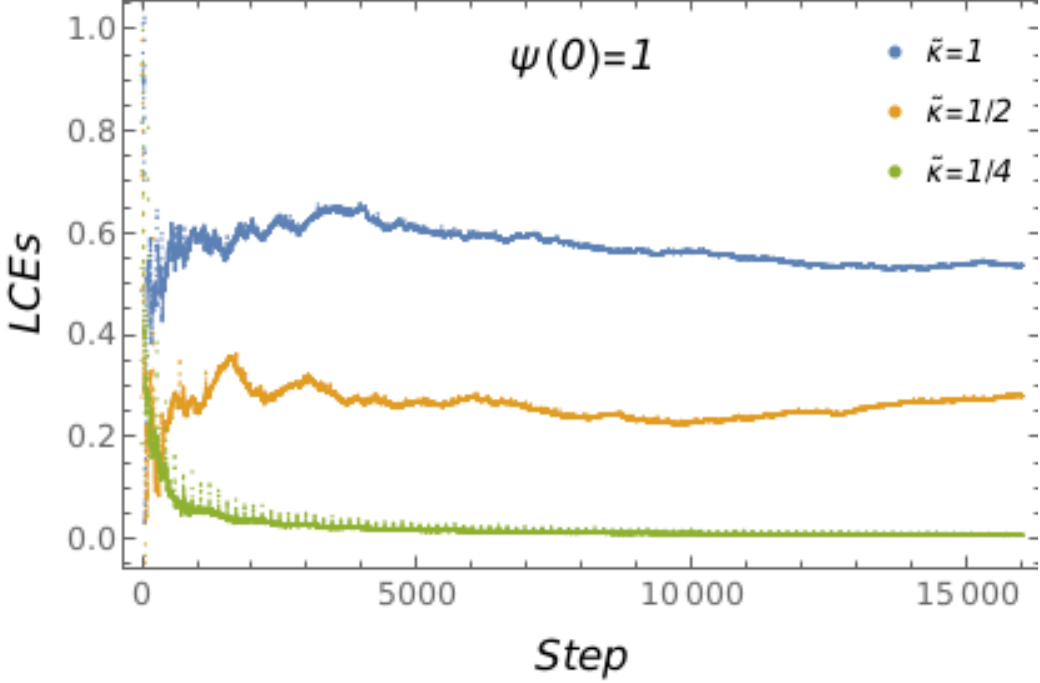}& &	\includegraphicsbox[scale=0.65]{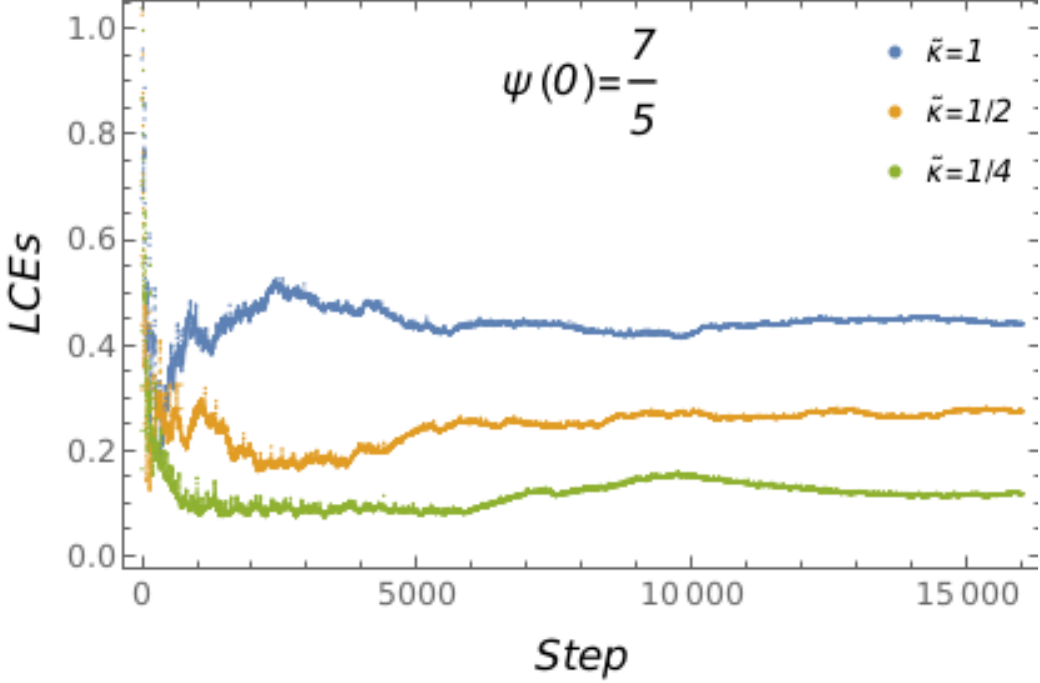}\nn
	\>
	\caption{Plot of the mLCE for the deformed SU$(3)$ Landau-Lifshitz model. 
		The initial conditions are  $p_\theta(0)=3/10$, $p_\psi(0)=11/10$, $\theta(0)=1$ and for different $\psi(0)$. The parameter values are $\alpha=w_\phi=w_\varphi=1$ with $\tilde\kappa$ varying.
		The total evolution time is $T=800$ which is broken into $16000$ iterative steps.  }
	\label{fig:LLSU3LCEspec}
\end{figure}
\paragraph{Deformed Model} Computations in the deformed theory are more difficult as, due to the chaotic nature of the system, greater numerical precision is required to compute an accurate solution. For example, with the same reference initial conditions \eqref{eq:su3_init}, we can integrate the equations of motion with $\tilde \kappa=w_\phi=w_\varphi=\alpha=1$ to $T=100$ to an accuracy of $10^{-9}$ at working precision $55$. That is, integrating forward and back gives solutions which differ by no more than $10^{-9}$ in any phase-space coordinate which are generally of order $1$. However, by $T=110$ the accuracy has decreased to $10^{-3}$ and by $T=150$ solution is completely unreliable. In order to get a reliable solution for this time interval it is necessary to increase the working precision to $75$. Nevertheless, even after only $T=100$ we have reasonably good results for the mLCE with the value $0.59$ compared to the $T=150$ value of $0.58$. There is an inherent error in this method of computing the mLCE in that a random direction in phase space is used, but the use of finite-time intervals is the largest source of error in our estimates. In general the results seem to be reliable to within at least $\pm 0.05$ after $T=100$. 

Using the iterated method to compute the full LCE spectrum, we can study the convergence over time. In \figref{fig:LLSU3LCEspec} we plot the largest LCE for each of the considered configurations, two of the remaining LCEs are zero and the fourth is just the negative of the first to within numerical accuracy. It should be noted that in the iterated calculation the solution is reliable over each iteration, but generally not over the whole time interval. This potentially results in a certain averaging of the LCEs over different phase-space points in an uncontrolled fashion. Nonetheless,  the results for the LCEs appear to be robust, the iterated method gives $0.54$ for the mLCE with the same initial conditions and parameter values, \eqref{eq:su3_init}, $\tilde \kappa=w_\phi=w_\varphi=\alpha=1$. This is possibly due to the LCEs being constant over at least small regions of phase space.

\subsection{Transition from integrable to chaotic dynamics}
\begin{figure}
	\centering
	
	\includegraphicsbox[scale=0.45]{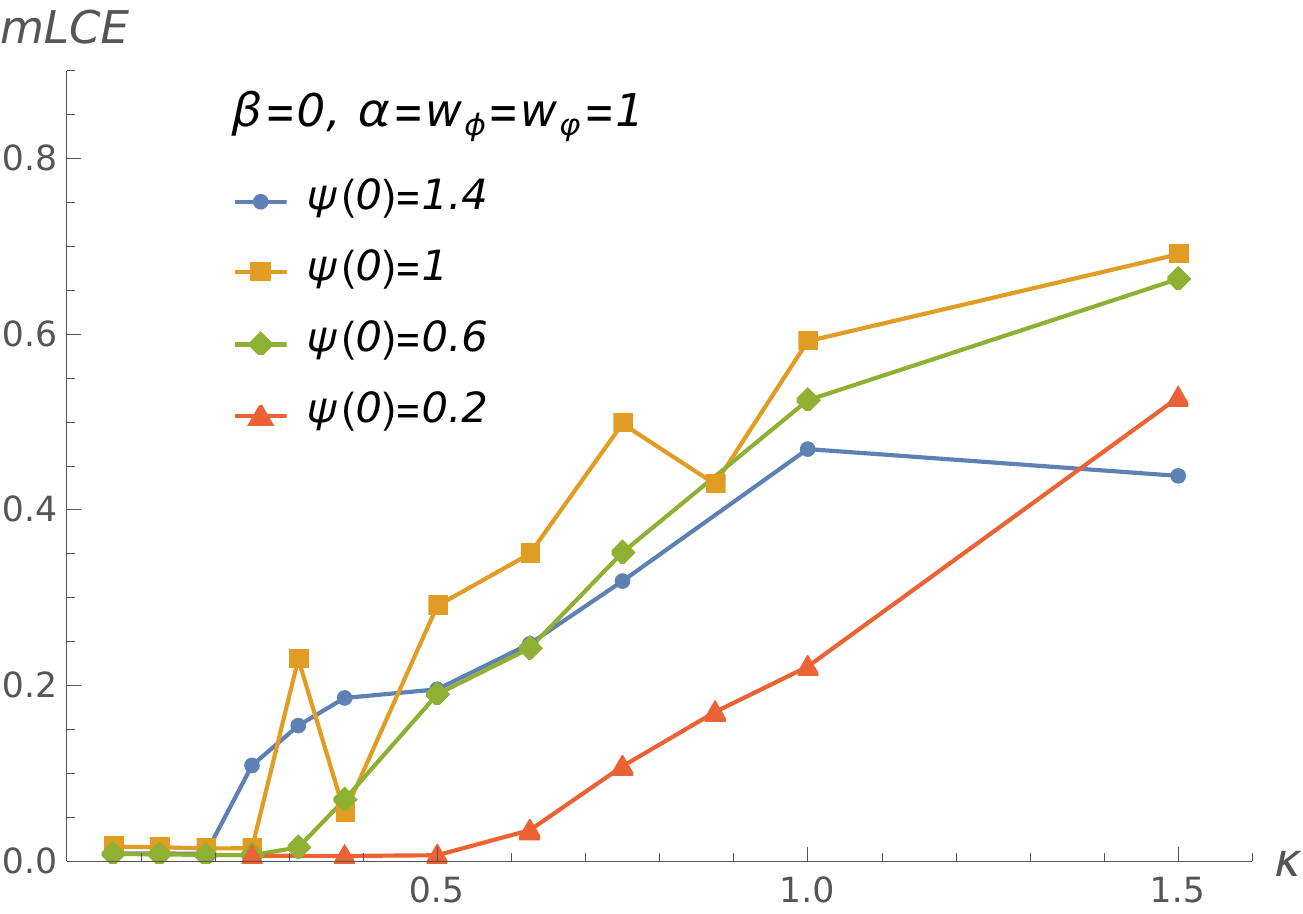}~~~\includegraphicsbox[scale=0.45]{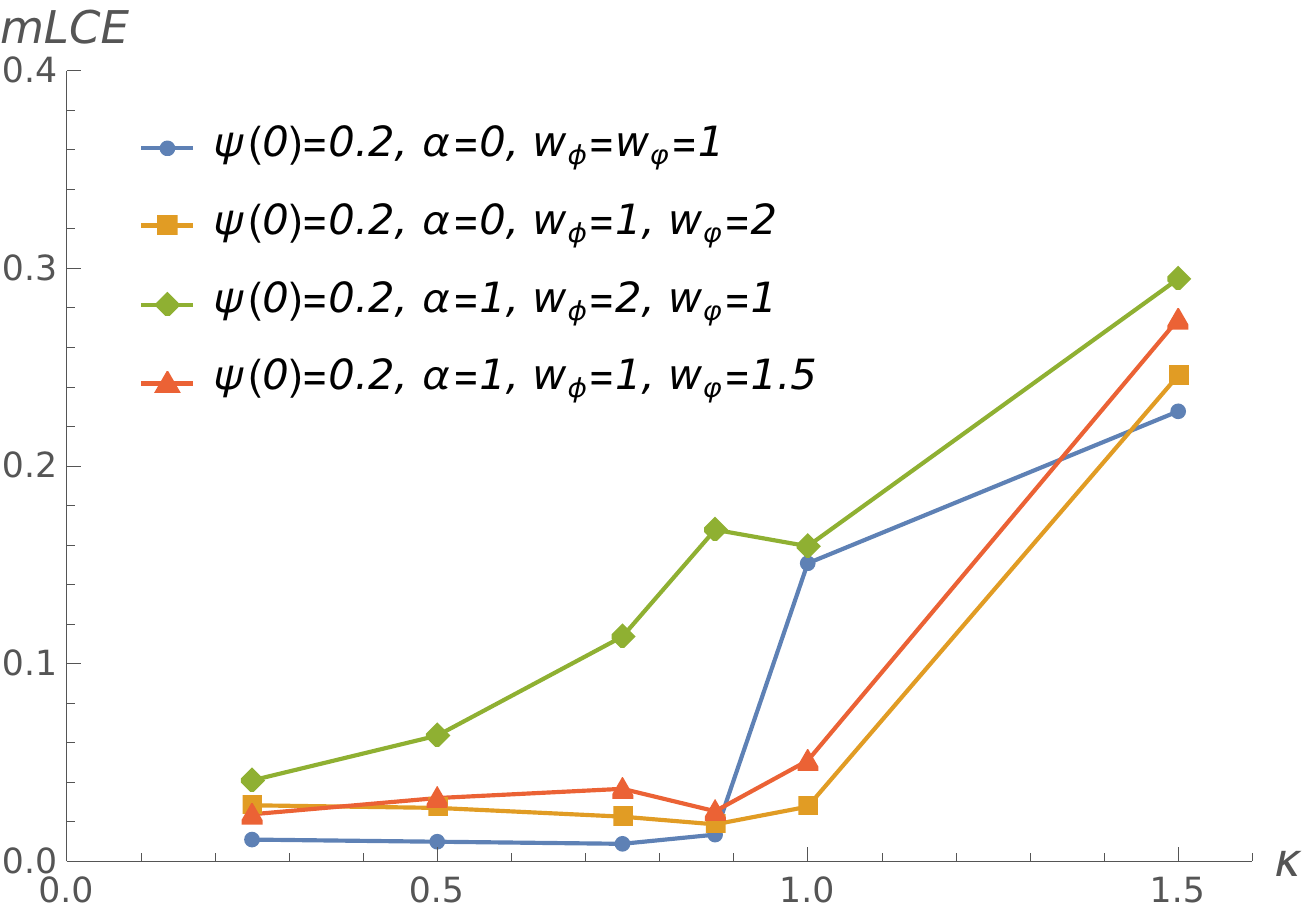}
	
	\caption{Left: mLCE for the deformed SU$(3)$ Landau-Lifshitz model with $\tilde \beta=0$ and $w_\phi=w_\varphi=\alpha=1$ but different $\tilde \kappa$ for the
		initial conditions  $p_\theta(0)=3/10$, $p_\psi(0)=11/10$, $\theta(0)=1$ and a range of $\psi(0)$. Right: mLCE for with $\tilde \beta=0$ and $p_\theta(0)=3/10$, $p_\psi(0)=11/10$, $\theta(0)=1$, $\psi(0)=1/5$ and different parameter values.
	}
	\label{fig:mLCEkrange}
\end{figure}

In the final part of this section, we study the transition from integrable to chaotic dynamics in the deformed LL model. \figref{fig:mLCEkrange} shows the variation of the mLCE with changing $\tilde \kappa$. One notable feature is that for small, but finite values of $\tilde \kappa$ the mLCE appears to vanish. For example, with the initial conditions \eqref{eq:su3_init} and  $\tilde \kappa=1/4$, $w_\phi=w_\varphi=\alpha=1$, the mLCE is less than $5\times 10^{-2}$ when we integrate out to $T=100$, and $5\times 10^{-3}$ when we integrate out to $T=1600$. This suggests it converges to zero, see \tabref{table:mLCESU3small}. By comparison, for $\tilde\kappa=9/32$, we find the mLCE is $0.08$ at $T=100$ and $0.1$ at $T=400$, suggesting it will converge to a small but non-vanishing number. 
This implies that the dynamics at this phase space point only become chaotic when the deformation parameter crosses some critical value close to $\tilde\kappa=1/4$. Further increasing the value of $\tilde\kappa$ generally, though not monotonically, results in an increasing value of the mLCE, cf.\  \figref{fig:mLCEkrange}.

\begin{table}
	\centering
	\begin{tabular}{|c |c c c c c |c|} 
		\hline
				 $\tilde \kappa$  &  &  &  T&  & & Soln. \\ [0.5ex]
				  
		 {} & $100$ & $200$ & $400$ & $800$ & $1600 $& Error \\ [0.5ex] 
		\hline\hline
		1/4 & 0.047 & 0.025 & 0.015 &0.0083 & 0.0048& $4\times 10^{-8}$\\ 
		\hline
		9/32 & 0.84 & 0.86 & 0.11 & - & - &$7\times 10^{-9}$\\ 
		\hline
	\end{tabular}
	\caption{mLCE for $\tilde \beta=0$ integrated for different times and different values of $\tilde \kappa$ with initial conditions (\ref{eq:su3_init}). The error is that of the numerical solution for the largest-$T$ solution quoted, not the mLCE,  with the accuracy of shorter $T$ solutions being significantly greater. }
	\label{table:mLCESU3small}
\end{table}

Note that different regions of the phase space generally give different Lyapunov exponents for fixed values of $\tilde \kappa$. For example, the initial conditions $p_\theta(0)=3/10$, $p_\psi(0)=11/10$, $\theta(0)=1$, $\psi(0)=1/5$ are not chaotic for $\tilde\kappa=1/2$ with the mLCE being less that $7\times 10^{-3}$, however they become chaotic by $\kappa=3/4$, see left panel in \figref{fig:mLCEkrange}. Thus it seems that different points in phase space have different critical values, but for a sufficiently large value of $\tilde \kappa$ it seems reasonable to suppose that all regions of phase space are chaotic. 

We also studied the behaviour of the model for different values of the parameters $w_\phi$, $w_\varphi$ and $\alpha$, see the right panel in \figref{fig:mLCEkrange}. While the values of the LCEs differ, generally being lower than for $w_\phi=w_\varphi=\alpha=1$, the behaviour is generically the same. In particular, the dependence on $\tilde \kappa$ is the same. There does not seem to be a qualitative difference when we remove the additional sextic term and so chaotic behaviour for non-zero values of $\tilde \kappa$ is found in both the naive coherent-state limit and when the potential term needed to match the string model is included.

\section{Conclusions}

In this work, we have considered the statistical properties of anomalous dimensions in planar $\mathcal{N}=4$ SYM theory and its marginal deformations. We have found that the spectral statistics for the one-loop marginally-deformed theory, in an SU$(3)$ sector with generic parameters, $h\neq0 $, $\beta\neq 0$, $\kappa \neq0$, is that of GUE RMT. This implies that the model is chaotic according to the usual definition following from the BGS conjecture. Additionally, for the case of the imaginary-$\beta$ theory, where $h=\beta=0$ and $\kappa \neq 0$, the ensemble becomes GOE with the statistics still being those of a chaotic system. This provides a weak-coupling analogue of the classical chaos seen in the holographic string dual \cite{Giataganas:2013dha}. 

To further characterise the dynamics, we study the distribution of level spacings  as the parameters transition from values where the system is known to be integrable to those where it is chaotic. For infinite-dimensional quantum many-body systems, it is expected that the transition to chaotic dynamics occurs for any non-zero value of the deformations, while for finite systems there is a smooth cross-over. We consider the scaling of the cross-over coupling, defined in terms of the Brody distribution, as a function of the spin-chain length $L$. For the case of the imaginary-$\beta$ theory, the scaling is $1/L^3$ which is consistent with generic quantum chaos, e.g.\ \cite{modak2014finite}. For comparison, we study the transition behaviour in $\mathcal{N}=4$ SYM theory and its twisted analogue, where the integrable  one-loop nearest-neighbour XXX spin chain is deformed by a two-loop NNN-interaction. In this case the statistics at finite coupling are those of the GOE Wigner-Dyson distribution. The cross-over coupling at fixed $L$ is higher than for generic deformations and, in the twisted case, scales as $1/L$. We thus find that the two-loop theory is weakly chaotic in the sense of \cite{Szasz-Schagrin:2021pqg}. 

This is consistent with the fact that the two-loop deformation, while it breaks integrability at finite coupling, is perturbatively integrable and can be completed into an integrable long-range theory, namely quantum strings in AdS$_5 \times$S$^5$. Viewed from a purely spin-chain perspective, this result is quite interesting as it suggests that the SU$(2)$-invariant NNN-deformation of the XXX model, and the twisted versions, are only weakly chaotic. 
This is a perhaps surprising result, however it is consistent with the realisation that the NNN deformation is  weaker than an interchain perturbation \cite{hsu1993level}, and that for the NNN deformation there are quasi-conserved quantities while for the next-to-NNN deformation there are no quasi-conserved charges at all \cite{malikis2020quasi, Kurlov:2021pxl}. 

Finally, in order to further our understanding of the relationship between the spin-chain statistics and string-theory dynamics, we studied the behaviour of generalised Landau-Lifshitz models. These models are known to capture the dynamics of both string solutions with large angular momenta and long-wavelength states of the spin chain. In the integrable case, they give a very explicit mapping between string solutions and spin-chain states, and thus provide a potential approach to understanding how the chaotic dynamics interpolate between weak and strong coupling. For the higher-derivative LL model corresponding to the two-loop dilatation operator, we find that the classical dynamics has vanishing Lyapunov exponents. This is consistent with the classical integrability of the string theory, but non-obvious from the spin-chain perspective. It is potentially explained by the weakly chaotic nature of the two-loop dilatation operator combined with the fact that the LL model captures only the low-energy states which are generally the least chaotic. For the imaginary-$\beta$ theory, we find non-vanishing Lyapunov exponents consistent with classically chaotic dynamics. This  provides further qualitative evidence for a matching between the string and spin-chain chaos. One notable feature of the imaginary-$\beta$ LL dynamics is that the transition to non-vanishing Lyapunov exponents occurs at finite values of the deformation. This is familiar in classically non-integrable systems, but is different from that found in the spin-chain description. It suggests that there may be interesting transition dynamics in spin chains if one combines the thermodynamic limit with a low-energy limit.

\section*{Acknowledgements}

  We would like to particularly thank Raul Pereira for many contributions and earlier collaboration on which this work builds. We thank B. Pozsgay, D. Sz\'asz, G. Takacs, L. Pando Zayas, and  A. Tseytlin for useful discussions and comments. Some of the calculations reported here were performed on the Lonsdale cluster maintained by the Trinity Centre for High Performance Computing. This cluster was funded through grants from Science Foundation Ireland. This work was supported by the Science Foundation Ireland through grant 15/CDA/3472, by the ERC starting grant 757978, by the research grant 00025445 from Villum Fonden, and it has received funding from the European Union's Horizon 2020 research and innovation programme under the Marie Sk\l{}odowska-Curie grant agreement No.\ 764850 ``SAGEX''.

\appendix
\section{Coherent states}
\label{app:CS}
A general group element of SU$(2)$ can be written as 
\<
g_2(\psi, \varphi_1,\varphi_2)=\begin{pmatrix} e^{i \varphi_1} \cos \psi & - e^{-i \varphi_2} \sin \psi \\ e^{i \varphi_2} \sin \psi & e^{- i \varphi_1} \cos \psi
		\end{pmatrix}~,
	\>
where $0\leq \psi \leq \pi/2$ and $0\leq \varphi_1, \varphi_2\leq 2 \pi$. In order to construct the SU$(2)/\text{U}(1)$ coherent states, we rewrite this as  
\<
g_2(\theta, \varphi_1,\varphi_2)
&=&\begin{pmatrix}  \cos {\theta}/2 & -e^{-i \varphi} \sin {\theta}/2 \\ e^{i \varphi} \sin {\theta}/2 &  \cos {\theta}/2
	\end{pmatrix}\begin{pmatrix} e^{i \varphi_1} & 0 \\ 0 & e^{-i \varphi_1} \end{pmatrix}~,
\>
where ${\theta}=2\psi$, $\varphi=\varphi_2-\varphi_1$. The first matrix, which we denote as $x(g)$, is the coset representative, while the second is an element of the stationary group  of the highest-weight state
\<
\ket{\psi_0}= \ket{1}= \begin{pmatrix} 1 \\ 0 \end{pmatrix}~.
\>
Coherent states can be written in terms of the coset representative as 
\<
\ket{\vec{n}}=x(g) \ket{\psi_0}=\begin{pmatrix} \cos {\theta}/2 \\ e^{i \varphi} \sin {\theta}/2
\end{pmatrix}
\>
which can be viewed as a point on the two-sphere. Alternatively, we can construct these states by simply acting with the group element $g_2$ on the highest-weight state
\<
\ket{\vec{n}'}=\begin{pmatrix} e^{i \varphi_1} \cos {\psi} \\ e^{i \varphi_2} \sin {\psi}
\end{pmatrix}~,
\>
which  describes a point on the three-sphere and we can recover the original coherent state by fixing the value of $\varphi_1$ such that $e^{i\varphi_1}=1$. Alternatively, we can also choose $\varphi_1+\varphi_2=0$ and $\phi=\tfrac{1}{2}(\varphi_1-\varphi_2)$ with $-\pi\leq\phi\leq+\pi$, which corresponds to the definition of the coherent state as
\<
\ket{\vec{n}}=e^{i \phi \sigma^z}e^{-i \psi \sigma^y} \ket{1}~.
\>
The inner product of two states in this latter construction is 
\<
\label{eq:su2prod}
\braket{\vec{n}(\psi_1, \phi_1)}{\vec{n}(\psi_2,\phi_2)}=e^{i(\phi_2-\phi_1)}\cos \psi_1\cos \psi_2+e^{i(\phi_1-\phi_2)}\sin \psi_1 \sin \psi_2~,
\>
while we can define a resolution of the identity 
\<
\label{eq:su2iden}
\unit =\frac{2}{\pi}\int d\mu_2 \ket{\vec{n}}\bra{\vec{n}}
\>
by integrating over all such states using the usual expressions for the volume measure on the three-sphere 
$d\mu'_2 =\cos \psi \sin\psi d\psi d\varphi_1 d \varphi_2$
but restricting the integration by $\varphi_1+\varphi_2=0$ so that 
\<
d\mu_2 =\cos \psi \sin\psi d\psi d \phi~.
\>

We can generalise this to SU$(3)$ by making use of the following parametrisation of group elements, e.g.\  \cite{nemoto2000generalized},
\<
g_3(\theta, \xi_1, \xi_2, \phi, \varphi_1,\varphi_2,  \varphi_3,\varphi_4)&=&\\
& & \kern-160pt\begin{pmatrix} 1 & 0 & 0 \\
0 &	e^{i \varphi_1} \cos \xi_1 & - e^{-i \varphi_2} \sin \xi_1 \\ 0 & e^{i \varphi_2} \sin \xi_1 & e^{- i \varphi_1} \cos \xi_1
\end{pmatrix}
\begin{pmatrix} 
		e^{i \phi} \cos \theta & - \sin \theta & 0  \\  \sin \theta & e^{- i \phi} \cos \theta & 0 \\ 0 & 0 & 1
\end{pmatrix}
\begin{pmatrix} 1 & 0 & 0 \\
	0 &	e^{i \varphi_3} \cos \xi_2 & - e^{-i \varphi_4} \sin \xi_2 \\ 0 & e^{i \varphi_4} \sin \xi_2 & e^{- i \varphi_3} \cos \xi_2
\end{pmatrix}~,\nn
\>
where $0\leq \varphi_i\leq 2\pi$, $0\leq \xi_i\leq \pi/2$, $0\leq \phi\leq 2\pi$ and $0\leq \theta \leq \pi/2$. 
The coherent states corresponding to the coset SU$(3)/\text{SU}(2)\times \text{U}(1)$ are found by acting on the highest-weight vector $\ket{\psi_0}=(1,0,0)^T$ so that
\<
\ket{\vec{n}'_3}=g_3 \ket{\psi_0}=\begin{pmatrix} e^{i \phi} \cos \theta \\ e^{i \varphi_1} \cos \xi_1 \sin \theta \\ e^{i \varphi_2} \sin \xi_1 \sin \theta \end{pmatrix}~.
\>
The metric is that of the five-sphere
\<
ds^2=d\theta^2+\cos^2\theta d\phi^2+\sin^2 \theta(d\xi_1^2+\cos^2 \xi_1 d\varphi_1^2+\sin^2 \xi_1 d\varphi_2^2)
\>
and the corresponding measure is
\<
d\mu'_3=\cos \theta\sin^3 \theta\sin \xi_1\cos \xi_1 d\theta d\xi_1 d\phi d\varphi_1d \varphi_2~.
\>
As in the SU$(2)$ case we can use the remaining U$(1)$ to fix $\varphi_1+\varphi_2=0$ and $\varphi=\tfrac{1}{2}(\varphi_1-\varphi_2)$ and for convenience we relabel $\xi_1\to \psi$ so that
\<
\label{eq:su3cs}
\ket{\vec{n}_3}=\begin{pmatrix} e^{i \phi} \cos \theta \\ e^{i \varphi} \cos \psi \sin \theta \\ e^{-i \varphi} \sin \psi \sin \theta \end{pmatrix}~.
\>

\section{Higher-derivative theories}
\label{app:OstHD}
Let us consider a dynamical system described by a Lagrangian
\<
\mathcal{L}\equiv\mathcal{L}\left(x_i; \dot x_i;\ddot{x}_i; \dots; x^{(m_i)}_i\right)
\>
which is functional of the coordinates $x_i(t)$, $i=1,\dots, n$, which depend on the time parameter $t$ and the derivatives, $x_i^{(r)}=\tfrac{d^r x_i}{dt^r}$ up to some maximal order $m_i$ for each coordinate $x_i$.  The Euler-Lagrange equations are 
\<
\sum_{r=0}^{m_i} (-1)^r \frac{d^r}{dx_i^r} \left(\frac{\partial \mathcal{L}}{\partial x_i^{(r)}}\right)=0~,~~~i=1,2,\dots, n~.
\>
In the Ostrogradsky approach to higher-derivative dynamics we define the generalised momenta
\<
\label{eq:momrec}
p_{i,r_i-1}=\sum_{k=r_i}^{m_i} (-1)^{k-r_i} \frac{d^{k-r_i}}{dt^{k-r_i}}\frac{\partial \mathcal{ L}}{\partial x^{(k)}}~,~~~r_i=1,\dots,m_i~,
\>
and we take the phase-space variables to be $q^{i,r_i}=x^{(r_i)}_i$ and $p_{i,r_i}$ with $r_i=1, \dots, m_i-1$. A key assumption is that we can invert the expressions for $p_{i,m_i-1}$  to write $\dot{q}^{i,m_i-1}=x_i^{(m_i)}$ in terms of $p_{i,m_i-1}$ and $q^{i,r_i}$ with $r_i<m_i$ for each $i=1, \dots, n$. However we do not express the remaining variables $q^{i,r_i}$ with $r_i\leq m_i-1$ in terms of the momenta. We now define the Hamiltonian
\<
H=\sum_{i=1}^n\left[\sum_{r_i=0}^{m_i-2} p_{i,r_i}{q}^{i,r_i+1}+p_{i,m_i-1}\dot{q}^{i,m_i-1}\right]-\mathcal{L}(q^{i,0},\dots, q^{i,m_i-1},\dot{q}^{i,m_i-1})
\>
depending on the independent variables $(q^{i,r_i},p_{i,r_i})$ for $r_i=1,\dots m_i-1$, so that 
the evolution equations are 
\<
\dot{q}^{i,r_i}=\frac{\partial H}{\partial p_{i,r_i}}~,~~~\dot{p}_{i,r_i}=-\frac{\partial H}{\partial q_{i,r_i}}~, ~~~r_i=0,\dots,m_i-1~.
\>
It is clear that for each $i$ the first $m_i-1$ of these equations imply 
\<
\dot{q}^{i,r_i}=q^{i,r_i+1}~, ~~~r_i=0,\dots, m_i-2
\>
as the corresponding momenta only occur linearly and not in the Lagrangian. The last coordinate equation is equivalent to $\dot{q}^{i,m_i-1}=\dot{q}^{i,m_i-1}(q^{i,r_i<m_i-1},p_{i,m_i-1})$.  Additionally the last $m_i-1$ of the momenta equations give the recursion relations
\<
p_{i,r_i-1}=\frac{\partial \mathcal{L}}{\partial x^{(r_i)}}-\frac{dp_{i,r_i}}{dt}~,~~~r_i=1,\dots,m_i-1~,
\>
which have \eqref{eq:momrec} as a solution, while the first equation gives
\<
\dot{p}_{i,0}=\frac{\partial \mathcal{L}}{\partial x_i}~.
\>
This last equation is just the Euler-Lagrange equation of motion for $x_i$.  

\section{Numerical computation of LCEs}
\label{app:numLCE}
We briefly summarise our approach to the numerical computation of LCEs which follows closely the methods described in \cite{sandri1996numerical, skokos2010lyapunov}. Given the $n$-dimensional phase-space variables $\vec{x}$, to compute the LCEs we numerically solve the equations of motion
\<
\label{eq:gen_eom}
\frac{d\vec{x}}{dt}=\vec{f}(\vec{x}(t))
\>
with initial conditions $\vec{x}(t=0)=\vec{x}_0$, simultaneously with the variational equation
\<
\label{eq:var_eq}
\frac{d\Phi}{dt}=J \Phi
\>
where $J$ is the matrix Jacobian
\<
J^i_j=\frac{\partial f^i}{\partial x^j}
\>
 and we take $\Phi(t=0)$ to be the $n\times n$ identity matrix. The matrix $\Phi(t)$ provides a linear map on the space of tangent vectors which can be used to evolve the deviation vectors. That is, given some initial configuration $\vec{x}_0$ which corresponds to a flow with coordinates $\vec{F}_t(\vec{x}_0)$ at time $t$, we can consider a system with infinitesimally shifted initial conditions $\vec{x}_0+\epsilon \vec{w}(0)$. The deviation vector at time $t$ is then given by 
 \<
 \label{eq:dev_ev}
 \vec{w}(t) =\Phi(t) \vec{w}(0)~,
 \> 
where $\Phi^i{}_j(t)=\frac{\partial {F}^i_t(\vec{x}_0)}{\partial {x}^j_0}$ satisfies the variational equations \eqref{eq:var_eq}~. For almost all vectors $\vec{w}(0)$, which can thus be chosen randomly, we can find the mLCE of order one, $\lambda_1$, by computing
\<
\lambda_1(\vec{x}_0)=\lim_{t\to \infty}\frac{1}{t}\ln \frac{\| \vec{w}(t)\|}{\|\vec{w}(0)\|}~.
\>
In practice, one numerically solves \eqref{eq:gen_eom} and \eqref{eq:var_eq} for a large time $T$. One issue is that for chaotic systems it can be difficult to maintain sufficient numerical precision when integrating over large times as, almost by definition, small errors rapidly grow in the flow trajectory. To ensure that our numerical integration is sufficiently precise, we first integrate for time $T$, then take the end point as our new initial condition and integrate backward in time to check that we find the original trajectory. For example, if we consider the non-chaotic case of the higher-derivative LL model \eqref{eq:Ham_hdLL} with spin-chain parameters \eqref{eq:LL2l} and $\omega=-\tfrac{1}{10}$, we can consistently numerically integrate the equations of motion for $T=400$ using the Mathematica function NDSolve with the ``ExplicitRungeKutta'' method with working precision of only 14 digits. By comparison, for the chaotic case where the parameters are as in \eqref{eq:hdLL_ch} and $\omega=-10$, we need to have a working precision of 60 digits so that the forward and backward integrations are everywhere consistent to $\pm 0.003$.

A second practical issue is that the vectors $\vec{w}(t)$ grow very quickly. To address this, one can split the integration range into $k$ steps of length $\tau=T/k$ 
\<
\label{eq:NmLCE}
\lambda_1(\vec{x}_0))=\lim_{k\to \infty} 
\frac{1}{k \tau}\sum_{r=1}^k \ln \alpha_r~,
\>
with expansion coefficient $\alpha_r=\frac{\| \vec{w}(r \tau)\|}{\|\vec{w}((r-1)\tau)\|}$. We can use the linearity of \eqref{eq:dev_ev} to rescale $\vec{w}((r-1)\tau)$ to have unit length at each step, and compute the expansion coefficient for the evolution of the unit vector over the interval $[(r-1)\tau,r\tau]$. 

More generally one can compute the full spectrum of LCEs. To do this one starts with the definition of the LCE of order $p$, which is given in terms of the volume spanned by $p$ linearly independent vectors $\vec{w}_1, \dots,\vec{w}_p$  whose evolution is given by the variational equation as in \eqref{eq:dev_ev}
\<
\lambda^{(p)}(\vec{x}_0)=\lim_{t\to\infty} \frac{1}{t} \ln \frac{\text{vol}^{(p)}(\vec{w}_1(t), \dots, \vec{w}_p(t))}{\text{vol}^{(p)}(\vec{w}_1(0), \dots, \vec{w}_p(0))}~.
\>
Further, one uses the known relation between the LCE of order $p$ for a suitably chosen subspace and the $p$ largest LCEs of order $1$
\<
\lambda^{(p)}(\vec{x}_0)=\sum_{j=1}^p \lambda_j (\vec{x}_0)~.
\>
In our numerical calculations, we start from a set of $n$ orthonormal vectors $\vec{w}_1(0), \dots,\vec{w}_n(0)$ and we again split the integration into $k$ steps of length $\tau$. After numerically integrating \eqref{eq:gen_eom} and \eqref{eq:var_eq} we compute the evolved vectors 
$\vec{{w}}_1(\tau), \dots,\vec{{w}}_n(\tau)$ and use the Gram-Schmidt procedure to compute an orthogonal, but not normalised, basis  $\tilde{\vec{w}}_1(\tau), \dots,\tilde{\vec{w}}_n(\tau)$. The evolved volume is given by the product of norms
\<
\text{vol}^{(n)}(\vec{w}_1(\tau), \dots, \vec{w}_p(\tau))=\| \tilde{\vec{w}}_1(\tau) \| \dots\| \tilde{\vec{w}}_n(\tau)\|~.
\>
As above in the case of the mLCE we can rescale the $n$ vectors so they are again orthonormal and then take these vectors to provide our initial conditions. Iterating the calculation we have that 
\<
\lambda^{(n)}=\lim_{k\to \infty} \frac{1}{k\tau}\sum_{r=1}^\infty \ln \| \tilde{\vec{w}}_1(r \tau) \| \dots\| \tilde{\vec{w}}_n(r \tau)\|~.
\>
We could of course have similarly computed $\lambda^{(n-1)}$ by considering an $(n-1)$-dimensional subspace and so we have that the $n$-th largest LCE is
\<
\lambda_n=\lambda^{(n)}-\lambda^{(n-1)}=\lim_{k\to \infty} \frac{1}{k\tau}\sum_{r=1}^\infty \ln \| \tilde{\vec{w}}_n(r \tau) \|~.
\>
 Repeating this we find the full spectrum of LCEs
\<
\lambda_1=\lim_{k\to \infty} \frac{1}{k\tau}\sum_{r=1}^\infty \ln \| \tilde{\vec{w}}_1(r \tau) \|~,\dots~,~ \lambda_n=\lim_{k\to \infty} \frac{1}{k\tau}\sum_{r=1}^\infty \ln \| \tilde{\vec{w}}_n(r \tau) \|~,
\>
which can be approximated by computing for some large $T=k\tau$ until convergence is found. 

\subsection{Lorenz Model}
\begin{figure}
	\centering
	\includegraphicsbox[scale=0.6]{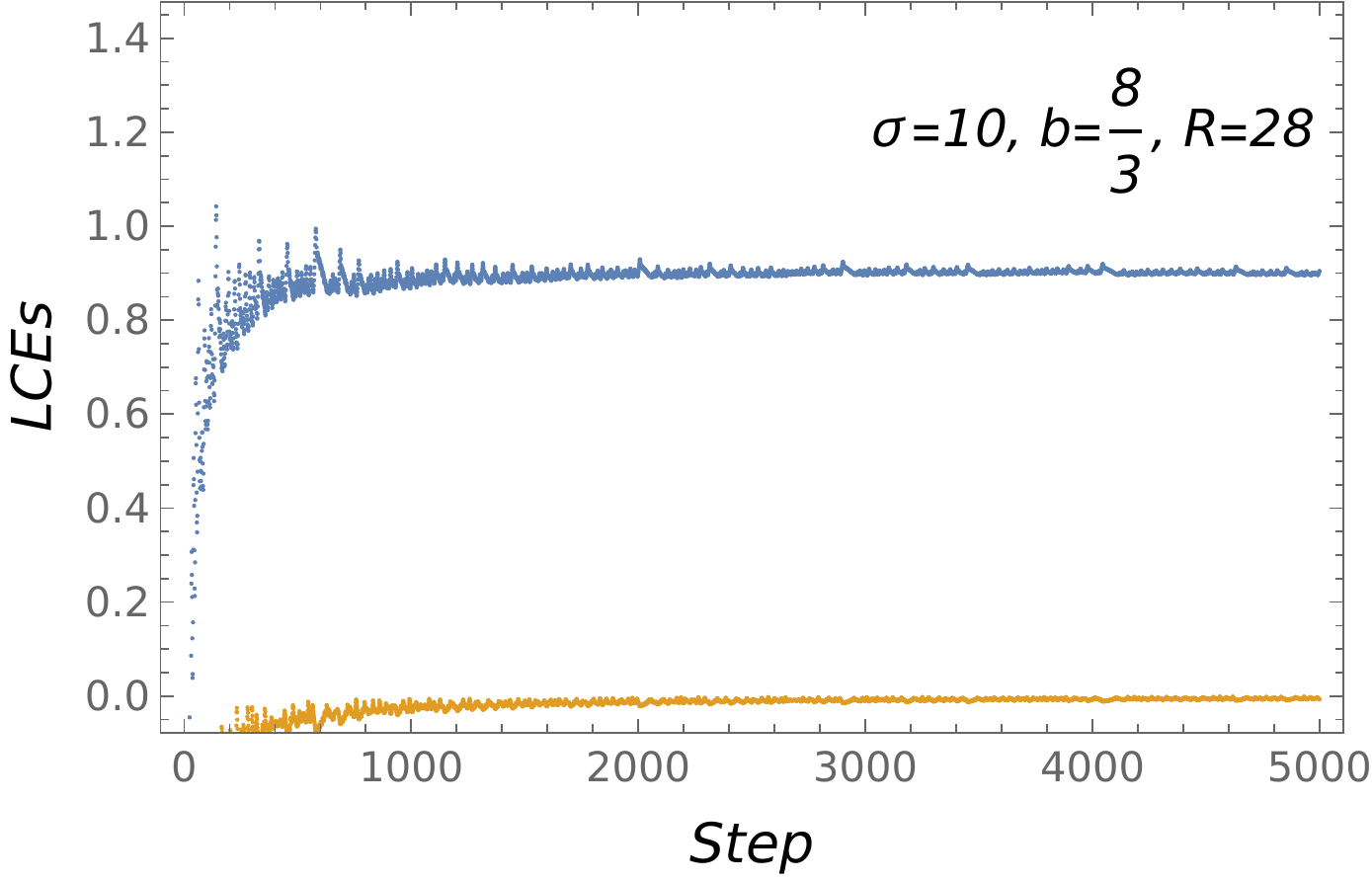}
	\caption{Plot of the two largest LCEs for the Lorenz model.
		The initial conditions are  $x(0)=29$, $y(0)=2$, $z(0)=20$. 
		The total evolution time is $T=400$ which is broken into $5000$ iterative steps.  }
	\label{fig:LorenzLCE}
	\end{figure}

A very well-studied system, originating in atmospheric science, is the Lorenz model described by the variables $(x,y,z)$ with the equations of motion
\<
\dot x=\sigma (y-x)~,~~~\dot y =x(R-z)-y~,~~~\dot z= x y -b z~.
\>
 We use this relatively simple system to check our numerical methods. For $\sigma=10$ and $b=8/3$ the model is known to be chaotic whenever the parameter $R$, the Rayleigh number, is greater than a critical value $R\simeq 24.74$. The trajectories of the system generally have one positive Lyapunov exponent and it follows from Liouville's formula that the sum of the Lyapunov exponents is $-(\sigma+1+b)=-13.667$. Numerical estimates of the LCEs for $R=28$ are $\{0.905 \pm 0.005, 0.0, -14.57\pm 0.01\}$ see for example \cite{alligood1996chaos}. 
Numerically computing the LCEs for $T=400$, $k=5000$ using Mathematica's NDSolve with the ExplicitRungeKutta method at Working Precision 40, we find $\{0.904, -0.005, -14.57\}$ though the results depend on the exact initial conditions to at least $\pm 0.02$. We plot the estimates of the two largest LCEs in \figref{fig:LorenzLCE} for specific initial conditions where the good convergence to the quoted results can be seen.  
This theory is non-Hamiltonian and so we cannot check our numerical integration solution by integrating forward and back, however we can alternatively check the numerical solution by integrating at different precisions and seeing that the solution is unchanged to a required accuracy. 
\bibliographystyle{nb}
\bibliography{QuantumStats}

\end{document}